\documentclass[a4paper,12pt]{article}

\usepackage{amsmath,amssymb}
\usepackage{graphicx}
\usepackage{psfrag}
\usepackage[usenames]{color}
\usepackage{epsf}
\usepackage{cite}

\makeatletter

\def\ten{{\rm A}}

\def\EllipticF#1#2{{\bf F}\Bigl({#1}\Big|{#2}\Bigr)}
\def\EllipticE#1#2{{\bf E}\Bigl({#1}\Big|{#2}\Bigr)}

\def\deltabD{\boldsymbol{\delta}^{\text{D}}}
\def\floor#1{{\lfloor{ #1 }\rfloor}}

\def\half{{\frac{1}{2}}}

\def\p{\partial}
\def\pb{\bar{\partial}}
\def\unit{{1\kern-.65ex {\rm l}}}
\def\1{{1\kern-.65ex {\rm l}}}
\def\ap{{\alpha'}}

\def\slash#1{\ooalign{\hfil/\hfil\crcr$#1$}} 

\def\dim{\mathop{\mathrm{dim}}\nolimits}

\def\Im{\mathop{\mathrm{Im}}\nolimits}

\def\rank{\mathop{\mathrm{rank}}\nolimits}

\def\tr{\mathop{\mathrm{tr}}\nolimits}

\def\zb{{\overline{z}}}

\def\taub{{\overline{\tau}}}
\def\rhob{{\overline{\rho}}}

\def\CC{{\cal C}}

\def\CF{{\cal F}}

\def\CH{{\cal H}}

\def\CM{{\cal M}}
\def\CN{{\cal N}}
\def\CO{{\cal O}}

\def\CQ{{\cal Q}}
\def\CR{{\cal R}}

\def\CT{{\cal T}}

\def\cC{{\cal C}}

\def\cF{{\cal F}}
\def\cG{{\cal G}}

\def\cN{{\cal N}}

\def\cQ{{\cal Q}}

\def\cT{{\cal T}}

\def\bbP{{\mathbb{P}}}

\def\bbR{{\mathbb{R}}}

\def\bbZ{{\mathbb{Z}}}

\def\dtt{\widetilde{dt}}
\def\EllipticF#1#2{{\bf F}\Bigl({#1}\Big|{#2}\Bigr)}
\def\EllipticE#1#2{{\bf E}\Bigl({#1}\Big|{#2}\Bigr)}

\newcount\hour \newcount\minute
\hour=\time \divide \hour by 60
\minute=\time
\def\now{%
\ifnum \hour<13
  \ifnum \hour=0 \advance \hour by 12 \number\hour:\else \number\hour:\fi%
     \ifnum \minute<10 0\fi%
     \number\minute%
\ A.M.%
\else \advance \hour by -12 \number\hour:%
  \ifnum \minute<10 0\fi%
  \number\minute%
  \ P.M.%
\fi%
}

\makeatother

\begin{document}

\baselineskip=18pt  
\numberwithin{equation}{section}  




\thispagestyle{empty}


\vspace*{2.5cm} 
\begin{center}
 {\LARGE Exotic Branes in String Theory}\\
 \vspace*{1.7cm}
 Jan de Boer$^1$ and Masaki Shigemori$^2$\\
 \vspace*{1.0cm} 
 $^1$ Institute for Theoretical Physics, University of Amsterdam\\
Science Park 904, P.O. Box 94485, 1090 GL Amsterdam, the Netherlands 
 \\[1ex]
 $^2$ Kobayashi-Maskawa Institute for the Origin of Particles and the Universe,\\
 Nagoya University, Nagoya 464-8602, Japan
\end{center}
\vspace*{1.5cm}

\noindent

Besides the familiar D-branes, string theory contains a vast number of
other non-per\-tur\-ba\-tive objects. While a complete classification is
lacking, many of these objects are related to each other through various
dualities.  Codimension two objects play a special role, because their
charges are no longer additive but are instead expressed in terms of
holonomies of scalar fields, which is given by an element of the
relevant duality group.  In this paper we present a detailed exposition
of these ``exotic'' objects, the charges they carry, and their
connection to non-geometric compactifications.  Despite the name
``exotic branes,'' these objects are in fact ubiquitous in string
theory, as they can automatically appear when describing bound states of
conventional branes, and as such may be of particular importance in
describing the microscopic degrees of freedom of black holes.

\newpage
\setcounter{page}{1} 



\tableofcontents


\section{Introduction}

String theory includes various extended objects as collective
excitations, such as D-branes. The $U$-duality symmetry
\cite{Hull:1994ys} which maps these objects into one another has played
a pivotal role in the development of string theory and provided crucial
insights into its non-perturbative behavior.
When string/M-theory is compactified to lower dimensions, the
$U$-duality group gets enhanced, relating objects that were not related
in higher dimensions.  For example, when M-theory is compactified on
$T^k$, the lower $(d=11-k)$ dimensional theory has an $E_{k(k)}(\bbR)$
symmetry as the $U$-duality group, at the level of classical
supergravity.  This continuous $E_{k(k)}(\bbR)$ symmetry is believed to
be broken to a discrete $E_{k(k)}(\bbZ)$ symmetry in the quantum theory
 \cite{Hull:1994ys}.

As the torus dimension $k$ increases, the number of gauge fields in the
lower ($d=11-k$) dimensional theory increases, and so does that of
associated charged particles.  The spectrum of these charged particles
represent the orbit of the $U$-duality group.  For $d\ge 4$, the
11-dimensional origin of such charged particles is easily understood;
they are ordinary branes partially wrapped on $T^k$.  For $d\le 3$,
however, the lower dimensional theories contain particles, called
\emph{exotic states}, whose higher-dimensional origin is less obvious
\cite{Elitzur:1997zn, Blau:1997du, Hull:1997kb, Obers:1997kk, Obers:1998fb}. In Type
II language, most of them have tension proportional to $g_s^{-3}$ or
$g_s^{-4}$, clearly indicating that they cannot be explained in terms of
ordinary branes whose tension can at most be $\sim g_s^{-2}$.

For example, in Type II superstring compactified on $T^2$, consider an
NS5-brane extending along six of the eight remaining non-compact
directions, not wrapping the internal $T^2$ (Table
\ref{table:NS5to522}).
\begin{table}
\begin{quote}
 \begin{center}
 \begin{tabular}{|c|ccccccc|cc|}
  \hline
      & 1 & 2 & 3 & 4 & 5 & 6 & 7 & 8 & 9 \\
  \hline
  NS5 & $\cdot$ & $\cdot$ & $\bigcirc$ & $\bigcirc$ & $\bigcirc$ & $\bigcirc$ & $\bigcirc$ & $\cdot$ & $\cdot$\\
  \hline
  \multicolumn{10}{c}{$\downarrow T_{89}\rule{0pt}{3ex}$}\\[1ex]
  \hline
  ~$5^2_2$~ & $\cdot$ & $\cdot$ & $\bigcirc$ & $\bigcirc$ & $\bigcirc$ & $\bigcirc$ & $\bigcirc$ & $\circledcirc$ & $\circledcirc$\\
  \hline
 \end{tabular}
 \caption{\sl Two transverse $T$-duality transformations on an NS5-brane
 produce an exotic $5^2_2$-brane.  ``$\,\cdot\,$'' (``$\,\bigcirc\,$'')
 indicates that the brane is localized (extended) in that direction. The
 mass of the $5^2_2$ is proportional to the squared radii of the directions
 with ``$\,\circledcirc\,$''.\label{table:NS5to522}}
 \end{center}
\end{quote}
\end{table}
It is well-known that, if we perform a $T$-duality along one of the
$T^2$ directions, we obtain a Kaluza--Klein (KK) monopole.  However, it
is much less known that, if we $T$-dualize further along the remaining
direction of the $T^2$, we obtain a codimension-2 exotic state called
$5^2_2$; see Table \ref{table:NS5to522}.  As was noted in
\cite{deBoer:2010ud} and will be reviewed later, the $5^2_2$ solution is
a \emph{non-geometric} background\footnote{For reviews on non-geometric
backgrounds, see \cite{Wecht:2007wu, Andriot:2011uh}.}
known as a $T$-fold \cite{Hull:2004in}; namely, as we go around it, the
internal $T^2$ is non-trivially fibered and does not come back to
itself, but rather to a $T$-dual version.  Other exotic states can be
obtained by further application of $U$-duality transformations on this
$5^2_2$.  They have non-trivial $U$-duality monodromies around them,
{\it i.e.}, the spacetime is glued together around them by $U$-duality
twists, and they are hence non-geometric $U$-folds
\cite{Hull:2004in}. These exotic states can be thought of as
codimension-2 branes and we will call them \emph{exotic branes}.

Being dual to the standard branes such as D-branes, exotic branes are as
essential ingredients of string theory as standard branes are, and are
worth studying on its own.  In particular, the fact that their charge is
characterized by the non-trivial monodromy around them is a novel and
peculiar feature and is expected to lead to interesting structures; this
was indeed the case with F-theory 7-branes, of which exotic branes can
be thought of as generalizations.  For example, the monodromic structure
of F-theory 7-branes were crucial for realizing gauge theories with
exceptional gauge groups \cite{Johansen:1996am, Gaberdiel:1997ud,
Gaberdiel:1998mv, DeWolfe:1998eu, DeWolfe:1998pr, Bergshoeff:2009th}.
Furthermore, the non-geometric nature of the monodromy is of much
interest in view of the recent developments in the double field theory
\cite{Hull:2009mi, Hohm:2010jy, Hohm:2010pp} (for reviews, see
\cite{Hohm:2011gs, Zwiebach:2011rg}) and generalizations thereof
\cite{Hillmann:2009ci, Berman:2010is, Berman:2011pe, Berman:2011jh,
Berman:2011cg}, which is a framework to incorporate stringy
non-geometric nature of spacetime.

One might think that such codimension-2 objects are problematic due to
logarithmic divergences \cite{Greene:1989ya}; generally, codimension-2
objects will backreact on the spacetime very badly and destroy the
asymptotics.  Also, exotic branes were found in three dimensions whereas
our universe is four dimensional.  Based on these, one might conclude
that exotic branes are irrelevant as long as we are concerned with
physics of ordinary branes in realistic spacetimes.
However, these are naive and incorrect because of the supertube
effect \cite{Mateos:2001qs}---the spontaneous polarization phenomenon
that occurs when we bring a particular combination of charges together.
Let us briefly recall what this phenomenon is.  A basic example of the supertube
effect is
\begin{align}
 \rm D0+F1(1)\to d2(1\psi)+p(\psi),
 \label{supertube_D0+F1}
\end{align}
in which D0-branes and fundamental strings along $x^1$ spontaneously
polarize into a D2-brane extending along $x^1$ and an arbitrary closed
curve in the transverse eight directions parametrized by $\psi$.  On the
D2-brane worldvolume, there is also a density of momentum charge along
$\psi$.  Note that the D2 charge did not exist in the original
configuration.  However, this does not violate charge conservation,
because the D2 is along a closed curve and there is no net D2 charge but
only a D2 dipole charge. We wrote the lowercase ``d2'' on the right hand
side of \eqref{supertube_D0+F1} to clarify that the D2 is a
dipole. Similarly, the momentum density $\rm p(\psi)$ does not give a
net momentum but only angular momentum.  The microscopic entropy of the
D0-F1 system can be recovered by counting the possible $\psi$ curves
that the system can polarize into \cite{Palmer:2004gu, Bak:2004kz}.

What does this have to do with exotic branes?  The point is that, the
phenomenon \eqref{supertube_D0+F1} implies that \emph{ordinary branes
can polarize into exotic branes}.  Namely, by taking $U$-duality of the
process \eqref{supertube_D0+F1}, one can show that, even if we start
with a configuration only of ordinary branes, the supertube effect can
produce exotic charges, as was first noted in \cite{deBoer:2010ud} and
will be discussed in detail later.  Because the exotic charges thus
produced are dipole charges, there is no net exotic charge at infinity.
So, there is no problem with charge conservation or of log divergences.
This means that, even in asymptotically flat spacetime in $d\ge 4$
dimensions, if we consider a system involving various ordinary branes,
exotic branes are spontaneously generated by the supertube effect 
generically and become crucial for understanding the physics.
So, exotic branes are ubiquitous and must play an important role for
generic physics of string theory.  Exotic branes are not exotic at all!

One particularly interesting situation in view of this is the black
hole, which is typically constructed in string theory as a bound state
of multiple (ordinary) branes.  Because the component branes can
polarize into exotic branes by the supertube effect, exotic branes are
expected to
be of great relevance for our understanding of black hole physics
in string theory.
More concretely, it was argued in \cite{deBoer:2010ud, Bena:2011uw} that
the microstates of black holes involve codimension-2 (exotic) branes
along arbitrary surfaces, dubbed superstrata.  This is an interesting
possibility especially in view of the fuzzball conjecture
\cite{Mathur:2005zp, Bena:2007kg, Skenderis:2008qn,
Balasubramanian:2008da, Chowdhury:2010ct}, which claims that the
microstates of black holes are made of fuzzballs, a mess of stringy
sources extending over the naive horizon scale.  Superstrata, if they
exist, may be giving a concrete realization of some or all of the fuzzballs.

In summary, exotic branes are basic ingredients of
string theory which can appear in various situations and relevant for
diverse aspects of string theory.  The purpose of the current paper is
to introduce this fascinating subject and to start exploring it, by
studying basic properties of exotic branes and examining their
implications for black hole physics as a particular example.
One main take-home message is that non-geometric exotic branes are not
the exception but the rule; they are simply \emph{inevitable}, if we are
to consider generic situations in string theory.  This is exactly
analogous to the state of affairs in flux compactification, for which it
has become clear by now that the conventional geometric compactification
with fluxes is a very tiny (probably measure zero) portion of all
generic compactifications in string theory, and the generic
compactifications involve non-geometric internal space (see, {\it e.g.},
\cite{Wecht:2007wu}).  Because string theory goes beyond the standard
notion of geometry, non-geometries and exotic branes are expected to be
generically present in the theory.

The plan of the rest of the paper is as follows.  In section
\ref{sec:exotic_branes_higher-D_origin}, we review how exotic branes
arise in the context of three dimensional supergravity and then discuss
their higher dimensional origin.  We summarize the duality relations
among different exotic branes, and also discuss exotic branes in $d>3$
dimensions. Section \ref{sec:aspects} involves analyses of some aspects
of exotic branes, such as how to define their charge by monodromies.  In
particular, we discuss the apparent non-conservation of brane charge
when it is moved around an exotic brane.  The resolution lies in
choosing the appropriate notion of charge, {\it i.e.}, Page charge,
which is shown to be conserved in all cases we study.  Section
\ref{sec:sugra_description} discusses how exotic branes are described
within supergravity, taking the $5^2_2$-brane as the main example. We
demonstrate that, around the $5^2_2$-brane, a torus direction undergoes
a $T$-duality and hence the solution represents a non-geometric
spacetime.  In this section, we will consider infinitely long straight
exotic branes.  This is not well-defined as a stand-alone object and
should be thought of as an effective description near the brane core.
Better defined solutions are discussed in section
\ref{sec:supertube_effect_and_exotic_branes}.  There, we discuss the
exotic supertube effect in which two stacks of D4-branes polarize into a
$5^2_2$-brane along a closed curve.  These solutions can be regarded as
non-geometric microstates of the D4-D4 system.  In section
\ref{sec:toward_truly_non-geometric_config}, we discuss in what sense
exotic brane solutions are non-geometric and how non-geometric we can
make them.  In section \ref{sec:exotic_branes_and_black_holes}, we
discuss the implications of exotic branes and the supertube effect for
black hole microphysics.  Section
\ref{sec:discussion_and_future_directions} is devoted to a discussion on
the results and possible future directions.  Appendices \ref{app:conv},
\ref{app:10D-8D}, and \ref{app:derive_int_db,int_dgamma} discuss
conventions and some detailed calculations used in the main text.
Appendix \ref{app:page_chg} is an extended discussion on the notions of
charge in string theory. It is known \cite{Marolf:2000cb} that there are
multiple different notions of charge in string theory and one has to be
careful to use the appropriate one depending on the purpose.  We clarify
the notion of brane charge and Page charge for D-branes in the presence
of NS5-brane source.

\section{Exotic branes and their higher dimensional origin}
\label{sec:exotic_branes_higher-D_origin}

\subsection{Exotic states in three dimensions}
\label{ss:exotic_states_3D}

Since exotic states (or branes) were first discovered in three
dimensions as a consequence of the $U$-duality of string theory
\cite{Elitzur:1997zn, Blau:1997du, Hull:1997kb, Obers:1997kk, Obers:1998fb}, it is perhaps the most
appropriate to start our discussion by reviewing how they arise in three
dimensional supergravity.

If we compactify M-theory on $T^8$ or Type IIA/B string theory on $T^7$
down to three dimensions, we obtain maximally supersymmetric ($\CN=16$)
supergravity with $E_{8(8)}(\bbR)$ as the $U$-duality group
\cite{Marcus:1983hb}.  This theory has 128 scalars parametrizing the
moduli space $\CM=SO(16)\backslash E_{8(8)}(\bbR)/E_{8(8)}(\bbZ)$. In
three dimensions, gauge fields (1-forms) can be Hodge dualized into
scalars,\footnote{This is a statement in the ungauged theory; in the
gauged theory in which (a subgroup) of the $U$-duality is promoted to a
local symmetry, we have both scalars and 1-forms at the same time and
the 1-forms cannot Hodge dualized into scalars.\cite{Nicolai:2000sc,
Nicolai:2001sv}} and the moduli space $\CM$ and the $E_{8(8)}(\bbR)$
symmetry are manifest only after such dualization. The classical
$E_{8(8)}(\bbR)$ symmetry is broken to the discrete subgroup
$E_{8(8)}(\bbZ)$ in string theory \cite{Hull:1994ys}, which is generated
by $S$- and $T$-dualities along the internal torus.

For example, let us consider Type IIB and take a D7-brane wrapped on the
$T^7$.  From the 3D viewpoint, this is a point particle with mass
\begin{align}
 M_{\rm D7(3456789)}={R_3R_4\cdots R_9\over g_s l_s^8},\label{D7mass}
\end{align}
where $R_3,R_4,\dots,R_9$ are the radii of the $T^7$ and
$l_s=\sqrt{\ap}$ is the string length. If we act on this point particle
with $U$-duality transformations, we obtain an orbit of the $U$-duality
group, called the ``particle multiplet'' \cite{Obers:1998fb}.  The mass
of the other states in the multiplet can be easily found by repeatedly
applying the $T$- and $S$-duality transformation rules,
\begin{align}
 T_y&:\quad R_y\to {l_s^2\over R_y}, 
 \quad g_s\to {l_s\over R_y}g_s;&
 S:&\quad g_s\to {1\over g_s},\quad l_s\to g_s^{1/2}l_s,\label{TSformulas}
\end{align}
to the original mass \eqref{D7mass}.  Here, $y$ is the direction along
which we take $T$-duality.  From the expression
for the mass, we can identify what the state corresponds to in 10
dimensions.

\begin{table}
\begin{center}
   \begin{tabular}{|l|l|}
   \hline
   Type IIA&  
       P ({\bf 7}), F1 ({\bf 7}), D0 ({\bf 1}), D2 ({\bf 21}), D4 ({\bf 35}), D6 ({\bf 7}),\\
   & NS5 ({\bf 21}), KKM ({\bf 42}),
       $5_2^2$ ({\bf 21}), 
       $0^7_3$ ({\bf 1}), $2^5_3$ ({\bf 21}), \\
   & $4^2_3$ ({\bf 35}), $6^1_3$ ({\bf 7}), $0^{(1,6)}_4$ ({\bf 7}), $1^6_4$ ({\bf 7})\\
   \hline
 Type IIB&
       P ({\bf 7}), F1 ({\bf 7}), D1 ({\bf 7}), D3 ({\bf 35}), D5 ({\bf 21}), D7 ({\bf 1}),\\
   &NS5 ({\bf 21}), KKM ({\bf 42}) ,
       $5_2^2$ ({\bf 21}), 
       $1^6_3$ ({\bf 7}), $3^4_3$ ({\bf 35}),\\
   &$5^2_3$ ({\bf 21}), $7_3$ ({\bf 1}), $0^{(1,6)}_4$ ({\bf 7}), $1^6_4$ ({\bf 7})\\
   \hline
   M-theory&  
   P ({\bf 8}), M2 ({\bf 28}), M5 ({\bf 56}), KKM ({\bf 56}),\\
   &
   $5^3$ ({\bf 56}), $2^6$ ({\bf 28}), $0^{(1,7)}$ ({\bf 8})\\
   \hline
  \end{tabular}
 \caption{\sl The 240 point particle states in 3D $\CN=16$ supergravity
 and their brane interpretations in string theory and M-theory. Their
 multiplicities are displayed in boldface numbers. These multiplicities
 can be interpreted as those of $SL(7)$ representations for Type IIA/B
 and those of $SL(8)$ representations for M-theory.  For the notation
 for exotic branes, see the text. \label{table:exotic_states}}
\end{center}
\end{table}

If we follow this procedure, we find 240 possible states in total,
including various states of ordinary branes partially wrapped on $T^7$,
as well as some peculiar states whose mass formula cannot be interpreted
in terms of any of ordinary branes \cite{Elitzur:1997zn, Blau:1997du,
Hull:1997kb, Obers:1997kk, Obers:1998fb}.  The latter states are called \emph{exotic
states} .  In Table \ref{table:exotic_states}, we listed all the 240
states in the particle multiplet, including the exotic ones.  The
notation used in the table for ordinary states is standard; {\it e.g.},
P denotes a gravitational wave and KKM denotes a Kaluza--Klein (KK)
monopole.
For exotic branes, on the other hand, we follow \cite{Obers:1998fb}
and denote them by how their mass $M$ depends on the radii of the
internal torus.  For Type IIA/B exotic states, the mass $M$ of a brane
denoted by $b^c_n$ depends linearly on $b$ radii and quadratically on
$c$ radii.  For $b^{(d,c)}_n$, $M$ also depends cubically on $d$ radii.
Moreover, $M$ is proportional to $g_s^{-n}$.  In equations,
\begin{align}
\begin{array}{lcl}
 \text{$b_n^c$-brane}&:&
 M=
 \dfrac{R_{i_1}\cdots R_{i_b}\, (R_{j_1}\cdots R_{j_c})^2
 }{g_s^n l_s^{b+2c+1}},
 \\[3ex]
 \text{$b_n^{(d,c)}$-brane}&:&
 M=\dfrac{R_{i_1}\cdots R_{i_b}\, (R_{j_1}\cdots R_{j_c})^2(R_{k_1}\cdots R_{k_d})^3
 }{g_s^n l_s^{b+2c+3d+1}}.
\end{array}
\end{align}
For example, the mass of $5^2_2$ mentioned in the introduction is
$M=R_3\cdots R_7(R_8 R_9)^2/g_s^2l_s^{10}$. We often display how the
brane ``wraps'' the internal $T^7$ as $5^2_2(34567,89)$.  In M-theory,
we use a similar notation except that we do not have the subscript $n$.

For illustration, let us work out the $T$-duality between NS5 and
$5^2_2$ displayed in Table \ref{table:NS5to522}.  The NS5-brane in Type
II theory wrapped on $x^{3,\dots,7}$ has mass
\begin{align}
 M_{\rm NS5(34567)}={R_3\cdots R_7\over g_s^2 l_s^6}.
\end{align}
If we $T$-dualize this configuration along $x^{8}$ using \eqref{TSformulas}, the mass turns into
that of a KK monopole as
\begin{align}
 M_{\rm NS5(34567)} 
 ~~\xrightarrow{T_8}~~
 {R_3\cdots R_7\over (g_sl_s/R_8)^2 l_s^6}
 ={(R_3\cdots  R_7) R_8^2\over g_s^2 l_s^8}
 =M_{\rm KKM(34567,8)}.\label{oyn12Mar12}
\end{align}
Further $T$-duality along $x^{9}$ gives a $5^2_2$-brane as
\begin{align}
 M_{\rm KKM(34567,8)}
 ~~\xrightarrow{T_9}~~
 {(R_3\cdots  R_7) R_8^2\over (g_s l_s/R_9)^2 l_s^8}
 ={(R_3\cdots  R_7) (R_8 R_9)^2\over g_s^2 l_s^{10}}
 =M_{5^2_2(34567,89)}.\label{fxxw20Mar12}
\end{align}
Similarly, one can readily work out other states in the multiplet.

\subsection{Duality rules for exotic branes}

Using the procedure explained above, it is straightforward to find how
the exotic branes map into one another under $T$- and $S$-dualities, as
well as under M-theory lift.  Such duality rules have already appeared
explicitly and implicitly in various papers including
\cite{Elitzur:1997zn, Blau:1997du, Hull:1997kb, Obers:1997kk, Obers:1998fb,
Eyras:1999at, LozanoTellechea:2000mc, Kleinschmidt:2011vu,
Bergshoeff:2011se, Kikuchi:2012za}, although notations may be different
.  In this subsection, we give a summary of such duality rules, for the
convenience of the reader and for future reference in the current paper.

In order to display the duality rules, it is convenient to introduce
another notation for exotic branes:\footnote{This notation is identical
to the ones introduced in \cite{Bergshoeff:2011se}, except that we flip
the sign of the subscript $n$ relative to theirs.}
\begin{align}
 b^c_n&=(abc)_n,\qquad a=9-b-c,\\
 b^{(d,c)}_n&=(abcd)_n, \qquad a=9-b-c-d.
\end{align}
Namely, a $(abcd)_n$-brane has mass which is independent of $a$ spatial
directions, linearly dependent on $b$ radii, quadratically dependent on
$c$ radii, cubicly dependent on $d$ radii, and so on.  We omit entries
after the last non-vanishing entry.  In this notation, ${\rm
NS5}=5_2=(45)_2$, the Type II KK monopole is $5^1_2=(351)_2$, and
$0^{(1,6)}_4=(2061)_4$.  We use a similar notation for the states in
M-theory, except that we do not have a subscript; for example, ${\rm
M2}=(82)$ and the M-theory KK monopole is $(361)$. Also notice that
when wrapping an $(abcd)_n$-brane on a $p$-torus, there are 
$\left(\begin{smallmatrix} & p & \\ b & c & d
\end{smallmatrix} \right)$ ways to do so.

In this notation, the ordinary and exotic branes that belong to the
particle multiplet in three dimensions are, in the Type IIA picture,
\begin{align}
\begin{split}
  g_s^{0}&:\quad \rm P, ~~ F1=1_0=(81)_0\\
 g_s^{-1}&:\quad \rm D0=0_1=(90)_1,~~ D2=2_1=(12)_1,~~ D4=4_1=(54)_1,~~ D6=6_1=(36)_1\\
 g_s^{-2}&:\quad \rm NS5=5_2=(45)_2, ~~  KKM=5_2^1=(351)_2, ~~  5_2^2=(252)_2\\
 g_s^{-3}&:\quad \rm 0^7_3=(207)_3, ~~ 2^5_3=(225)_3,~~ 4^3_3=(243)_3,  ~~6^1_3=(261)_3 \\
 g_s^{-4}&:\quad  0^{(1,6)}_4=(2061)_4, ~~1^6_4=(216)_4
\end{split}\label{IIAexotics}
\end{align}
where we classified the states according to how their mass depends on
$g_s$.  In the Type IIB picture, we have
\begin{align}
\begin{split}
  g_s^{0}&:\quad \rm P, ~~ F1=1_0=(81)_0\\
 g_s^{-1}&:\quad \rm D1=1_1=(81)_1,~~ D3=3_1=(63)_1,~~ D5=5_1=(45)_1,~~ D7=7_1=(27)_1\\
 g_s^{-2}&:\quad \rm NS5=5_2=(45)_2, ~~  KKM=5_2^1=(351)_2, ~~  5_2^2=(252)_2\\
 g_s^{-3}&:\quad \rm 1^6_3=(216)_3, ~~ 3^4_3=(234)_3, ~~5^2_3=(252)_3, ~~ 7_3=(27)_3\\
 g_s^{-4}&:\quad \rm 0^{(1,6)}_4=(2061)_4, ~~1^6_4=(216)_4
\end{split}\label{IIBexotics}
\end{align}
In M-theory, we have
\begin{align}
\begin{split}
 &\rm P,~~
 M2=2=(82),~~
 M5=5=(55),~~
 KKM=6^1=(361),\\
 &\rm 
 5^3=(253),~~ 
 2^6=(226),~~ 
 0^{(1,7)}=(2071).
\end{split}\label{Mexotics}
\end{align}

\def\DIR#1{{\underline{#1}}}
In order to specify the direction acted by $T$-duality, we put an
underscore at the corresponding position. For example, the $T$-duality
relation in \eqref{oyn12Mar12}  can be written as
\begin{align}
(\DIR{4}5)_2\to (35\DIR{1})_2.
\end{align}
With this notation, the $T$-duality relations among various exotic branes
are as in Table \ref{table:T-duality_IIA/B}.
%
\begin{table}[htbp]
\begin{align*}
\begin{array}{|r@{~}l|}
 \hline
  T:\rm IIA&\to \rm IIB\\
 \hline
 \rm NS5=5_2=(\DIR{4}5)_2&\to (35\DIR{1})_2=5^1_2=\rm KKM \\
         (4\DIR{5})_2    &\to (4\DIR{5})_2=5_2=\rm NS5 \\[0.5ex]
 \rm KKM=5^1_2= (\DIR{3}51)_2&\to(25\DIR{2})_2=5_2^2 \\
        (3\DIR{5}1)_2&\to(3\DIR{5}1)_2=5^1_2=\rm KKM\\
        (35\DIR{1})_2&\to(\DIR{4}5)_2=5_2=\rm NS5\\[0.5ex]
 5_2^2= (2\DIR{5}2)_2&\to(2\DIR{5}2)_2=5^2_2\\
        (25\DIR{2})_2&\to(\DIR{3}51)_2=5^1_2=\rm KKM\\[0.5ex]
 \hline
 0^7_3= (20\DIR{7})_3&\to(2\DIR{1}6)_3=1^6_3\\[0.5ex]
 2^5_3= (2\DIR{2}5)_3&\to(21\DIR{6})_3=1^6_3\\
        (22\DIR{5})_3&\to(2\DIR{3}4)_3=3^4_3\\[0.5ex]
 4^3_3= (2\DIR{4}3)_3&\to(23\DIR{4})_3=3^4_3\\
        (24\DIR{3})_3&\to(2\DIR{5}2)_3=5^2_3\\[0.5ex]
 6^1_3= (2\DIR{6}1)_3&\to(25\DIR{2})_3=5^2_3\\
        (26\DIR{1})_3&\to(2\DIR{7})_3=7_3\\[0.5ex]
 \hline
 0^{(1,6)}_4= (20\DIR{6}1)_4&\to(20\DIR{6}1)_4=0^{(1,6)}_4\\
              (206\DIR{1})_4&\to(2\DIR{1}6)_4=1^6_4\\[0.5ex]
 1^6_4=  (2\DIR{1}6)_4&\to(206\DIR{1})_4=0^{(1,6)}_4\\
         (21\DIR{6})_4&\to(21\DIR{6})_4=1^6_4\\[0.5ex]
	  \hline
\end{array}
\quad
\begin{array}{|r@{~}l|}
 \hline
  T: \rm IIB&\to\rm IIA\\
 \hline
 \rm NS5=5_2=
 (\DIR{4}5)_2&\to(35\DIR{1})_2=5^1_2=\rm KKM \\
 (4\DIR{5})_2&\to(4\DIR{5})_2=5_2=\rm NS5 \\[0.5ex]
  \rm KKM=5^1_2=
 (\DIR{3}51)_2&\to(25\DIR{2})_2=5_2^2 \\
 (3\DIR{5}1)_2&\to(3\DIR{5}1)_2=5^1_2=\rm KKM\\
 (35\DIR{1})_2&\to(\DIR{4}5)_2=5_2=\rm NS5\\[0.5ex]
 5_2^2=
 (2\DIR{5}2)_2&\to(2\DIR{5}2)_2=5^2_2\\
 (25\DIR{2})_2&\to(\DIR{3}51)_2=5^1_2=\rm KKM\\[0.5ex]
 \hline
 1^6_3=
 (2\DIR{1}6)_3&\to(20\DIR{7})_3=0^7_3\\
 (21\DIR{6})_3&\to(2\DIR{2}5)_3=2^5_3\\[0.5ex]
 3^4_3=
 (2\DIR{3}4)_3&\to(22\DIR{5})_3=2^5_3\\
 (23\DIR{4})_3&\to(2\DIR{4}3)_3=4^3_3\\[0.5ex]
 5^2_3=
 (2\DIR{5}2)_3&\to(24\DIR{3})_3=4^3_3\\
 (25\DIR{2})_3&\to(2\DIR{6}1)_3=6^1_3\\[0.5ex]
 7_3=
 (2\DIR{7})_3&\to(26\DIR{1})_3=6^1_3\\[0.5ex]
 \hline
 0^{(1,6)}_4=
 (20\DIR{6}1)_4&\to(20\DIR{6}1)_4=0^{(1,6)}_4\\
 (206\DIR{1})_4&\to(2\DIR{1}6)_4=1^6_4\\[0.5ex]
 1^6_4= 
 (2\DIR{1}6)_4&\to(206\DIR{1})_4=0^{(1,6)}_4\\
 (21\DIR{6})_4&\to(21\DIR{6})_4=1^6_4\\[0.5ex]
  \hline
\end{array}
\end{align*}
\caption{\sl The $T$-duality relations among exotic branes in Type
IIA/B\@. The underscore specifies the direction along which $T$-duality
is taken.\label{table:T-duality_IIA/B}}
\end{table}

We also list the $S$-duality relations in Table \ref{table:S-duality_IIB}.
%
\begin{table}[htbp]
\begin{align*}
\begin{array}{|r@{~}l|}
 \hline
  S: \rm IIB&\leftrightarrow \rm IIB\\
 \hline
 ~~~~~~~ 5_2^2=(252)_2 &\leftrightarrow 5^2_3=(252)_3\\
 7_3=(27)_3 &\leftrightarrow \rm D7=7_1=(27)_1\\
 1^6_3=(216)_3 &\leftrightarrow 1^6_4=(216)_4\\
 \multicolumn{2}{|c|}{
 3^4_3=(234)_3,~~ 0^{(1,6)}_4=(2061)_4 : \text{self-dual}}\\
 \hline
\end{array}
\end{align*}
\caption{\sl The $S$-duality relations among exotic branes in Type
IIB. \label{table:S-duality_IIB}}
\end{table}

The relation between M-theory exotic branes and their type IIA reduction
can be read off from the mass formula using the standard relation
between 10D and 11D quantities,
\begin{align}
 l_s= R_{10}^{-1/2}l_{11}^{3/2},\qquad g_s= R_{10}^{3/2}l_{11}^{-3/2},
\end{align}
where $R_{10}$ is the radius of the 11th direction and $l_{11}$ is the
11D Planck length.  We list the relation in Table \ref{table:M2IIA}.
We displayed the direction of the M-theory circle by an underscore.

\begin{table}
 \begin{align*}
 \begin{array}{|r@{~}l|}
 \hline
  \rm M&\to \rm IIA\\
 \hline
 \rm KKM=6^1=
 (\DIR{3}61) &\to (261)_3=6^1_3\\
 (3\DIR{6}1) &\to (351)_2=5^1_2=\rm KKM\\
 (36\DIR{1}) &\to (36)_1=\rm D6\\
 5^3=
 (2\DIR{5}3) &\to (243)_3=4^3_3\\
 (25\DIR{3}) &\to (252)_2=5^2_2\\
 2^6=
 (2\DIR{2}6) &\to (216)_4=1^6_4\\
 (22\DIR{6}) &\to (225)_3=2^5_3\\
 0^{(1,7)}=
 (20\DIR{7}1) &\to (2061)_4=0_4^{(1,6)}\\
 (207\DIR{1}) &\to (207)_3=0^7_3\\
  \hline
 \end{array}
 \end{align*}
 \caption{\sl The rule for the reduction of exotic branes from M-theory
 to Type IIA\@.  The underscore specifies the direction of the
 dimensional reduction.\label{table:M2IIA}}
\end{table}

\subsection{10D/11D origin of exotic states}

The fact that most of the exotic states have mass proportional to
$g_s^{-3}$ or $g_s^{-4}$ clearly indicates that they cannot be
interpreted in terms of ordinary branes, whose mass is proportional to
$g_s^{-2}$ at most. Here, we argue that the exotic states are
interpreted in higher dimensions as non-geometric backgrounds, or
$U$-folds \cite{Hellerman:2002ax, Hull:2004in}.  This connection between
exotic branes and non-geometric $U$-folds was pointed out first in
\cite{deBoer:2010ud}.

The argument \cite{deBoer:2010ud} is simple.  As an example, consider a
D7-brane wrapped on $T^7$, which is magnetically coupled to the RR
0-form $C^{(0)}$ (we display the rank of a differential form as a
superscript in parentheses).  From the 3D point of view, the D7-brane is
a point particle and, as we go around it, the 3D scalar $\phi=C^{(0)}$
jumps as $\phi\to\phi+1$.  This discontinuous ``jump'' (or
multi-valuedness) of the scalar $\phi$ is allowed because it is a part
of the $SL(2,\bbZ)$ symmetry of Type IIB string theory, which is a
discrete gauge symmetry.
%
In 3D, this symmetry of shifting $\phi$ by 1 gets combined with other
dualities such as $T$-dualities to form the $U$-duality group
$G(\bbZ)=E_{8(8)}(\bbZ)$, and the scalar $\phi$ gets combined with other
scalars into a matrix $M$ parametrizing the moduli space $\CM =
SO(16)\backslash E_{8(8)}(\bbR) / E_{8(8)}(\bbZ)$.  Therefore, the
$\phi\to\phi+1$ monodromy around the D7-brane is only one of all
possible $U$-duality monodromies we can have in 3D, and we should
consider all possible 3D particles with $U$-duality monodromies
conjugate to the monodromy of the D7-brane.  In fact, we can even
consider 3D particles with $E_{8(8)}(\bbZ)$ monodromies which are
\emph{not} conjugate to the monodromy of the D7-brane.  Exotic states
are objects with such general $U$-duality monodromies.  Note that, being
general, exotic states are the rule, not the exception.
%

Now let us consider such a 3D particle with a general $U$-duality
monodromy and lift it to 10D/11D, where it becomes a codimension-2
object (namely, it becomes a 7-brane in 10D and a 8-brane in 11D)\@.
Because the 3D scalars lift to the internal components of
higher-dimensional fields, such as metric, $B$-field and RR potentials,
the 3D particle with a scalar monodromy around it lifts to a
codimension-2 object with a non-trivial monodromy for these
higher-dimensional fields.  In particular, this means that, as one goes
around the object, the metric does not generally come back to itself but
only to a $U$-dual version; the geometry is multi-valued.  Namely,
exotic states lift to \emph{exotic branes} which are \emph{non-geometric
backgrounds}, or ``$U$-folds'' \cite{Hull:2004in}.  We emphasize that
such non-geometric spacetime is the rule, not the exception, for
codimension-2 branes in string theory.

Being $U$-dual to ordinary branes, exotic branes are dynamical objects
which can move, wiggle, {\it etc.}, unlike orientifolds which are fixed
hyperplanes in spacetime.  It is also worth noting that, at this point,
it is only the higher-dimensional metric that has non-trivial monodromy
around an exotic brane whereas the Einstein metric in 3D is
single-valued; from the 3D point of view, it is only scalars that have
non-trivial monodromy.  We will discuss the possibility of having
non-geometric spacetime in lower dimensions later.

Note that there are two types of $U$-fold; in the first one, $U$-duality
is non-trivially fibered over a non-contractible circle in the internal
manifold (see, {\it e.g.}, \cite{Dabholkar:2002sy, Kachru:2002sk,
Flournoy:2004vn, Wecht:2007wu}) while, in the second one, $U$-duality is
non-trivially fibered over a contractible circle (see, {\it e.g.},
\cite{Hellerman:2002ax, McOrist:2010jw}). The exotic branes discussed in
the current paper are of the second type.  The relation between the two
types of $U$-fold is similar to that between a geometry with branes
wrapped on a non-trivial cycle and the geometry in which a geometric
transition has occurred and the branes have turned into fluxes.  It
would be interesting to study this similarity further.

\subsection{Codimension-2 objects in various dimensions}

As is clear from the above discussion, the non-geometric $U$-fold
structure is intrinsic to codimension-2 exotic branes and we do not
have to go to three dimensions to find them.  Here we discuss the
codimension-2 branes that appear when we compactify M-theory on $T^k$
or Type II string theory on $T^{k-1}$ down to $d=11-k=3,\dots,10$
dimensions.

As $k$ increases, the $U$-duality group (Cremmer--Julia groups) $G(\bbR)$
of the low $(d)$ dimensional theory becomes larger as listed in Table
\ref{table:CJgroups}.  In string theory, $G(\bbR)$ is believed to be
broken to the discrete subgroup $G(\bbZ)$.  The scalar moduli space is
$\CM=H(\bbR)\backslash G(\bbR)/G(\bbZ)$ where $H(\bbR)$ is the maximal
compact subgroup of $G(\bbR)$, which is also listed in Table \ref{table:CJgroups}.
The duality group $G(\bbR)$ is the isometry group of this scalar moduli space $\CM$.
\begin{table}[tbhp]
\begin{center}
  \begin{tabular}{|@{\,}c@{\,}||@{\,}c@{\,}|@{\,}c@{\,}|@{\,}c@{\,}|@{\,}c@{\,}|@{\,}c@{\,}|@{\,}c@{\,}|@{\,}c@{\,}|}
  \hline
  $d$& $G(\bbR)$ & $G(\bbZ)$ & $H$  & $\dim G$ & $\rank G$ & $\dim H$ & $\text{dim.\ of}\atop\text{orbit}$\\  \hline  \hline
  10A& $\bbR_+$                     & $\bf 1$                      & $\bf 1$             & 1   &1    & 0   &---\\ \hline
  10B& $SL(2,\bbR)$                 & $SL(2,\bbZ)$                 & $SO(2)$             & 3   &1    & 1   &2\\ \hline
  9  & $SL(2,\bbR)\times \bbR_+$    & $SL(2,\bbR)\times \bbZ_2$    & $SO(2)$             & 4   &2    & 1   &2 \\ \hline
  8  & $SL(3,\bbR)\times SL(2,\bbR)$& $SL(3,\bbZ)\times SL(2,\bbZ)$& $SO(3)\times SO(2)$ &$8+3$&$2+1$&$3+1$&$4+2$\\ \hline
  7  & $SL(5,\bbR)$                 & $SL(5,\bbZ)$                 & $SO(5)$             & 24  &4    & 10  &8\\ \hline
  6  & $SO(5,5,\bbR)$               & $SO(5,5,\bbZ)$               & $SO(5)\times SO(5)$ & 45  &5    & 20  &14\\ \hline
  5  & $E_{6(6)}$                   & $E_{6(6)}(\bbZ)$             & $USp(8)$            & 78  &6    & 36  &22\\ \hline
  4  & $E_{7(7)}$                   & $E_{7(7)}(\bbZ)$             & $SU(8)$             &133  &7    & 63  &34\\ \hline
  3  & $E_{8(8)}$                   & $E_{8(8)}(\bbZ)$             & $SO(16)$            &248  &8    & 120 &58 \\ \hline
 \end{tabular}
\end{center} 
 \vspace*{-3ex}
\caption{\sl The $U$-duality groups in various dimensions. $G(\bbR)$ is
the classical $U$-duality group while $G(\bbZ)$ is the quantum one. $H$
is the maximal compact subgroup of $G(\bbR)$. In the last column, we
listed the dimension of the conjugation orbit of half-supersymmetric
branes; see section 
\ref{ss:num_of_charges}.\label{table:CJgroups}}
\end{table}

\begin{table}
 \begin{center}
  \newcommand{\parboxM}[1]{\parbox{0.17\textwidth}{\raggedright #1}}
  \newcommand{\parboxII}[1]{\parbox{0.26\textwidth}{\raggedright #1}}
 \begin{tabular}{|@{~}c@{~}||l|l|l|}
 \hline
 $d$ & \multicolumn{1}{|c|}{M} & \multicolumn{1}{|c|}{IIA} & \multicolumn{1}{|c|}{IIB} \\ \hline\hline
 10 & --- &  ---
   & {\sl   2\/}: D7~({\bf 1}), $7_3$~({\bf 1})   \\ \hline
 9 & {\sl   2\/}: \parboxM {KKM~({\bf 2})}
   & {\sl   2\/}: \parboxII{D6~({\bf 1}), $6^1_3$~({\bf 1})}
   & {\sl   2\/}: \parboxII{D7~({\bf 1}), $7_3$~({\bf 1})}  \\ \hline
 8 & {\sl   6\/}: \parboxM {KKM~({\bf 6})} 
   & {\sl   6\/}: \parboxII{D6~({\bf 2}), KKM~({\bf 2}), $6^1_3$~({\bf 2})}
   & {\sl   6\/}: \parboxII{D7~({\bf 1}), D5~({\bf 1}), NS5~({\bf 1}), $7_3$~({\bf 1}), $5^2_2$~({\bf 1}), $5^2_3$~({\bf 1})}    \\ \cline{2-4}
   & {\sl   2\/}: \parboxM {M5~({\bf 1}), $5^3$~({\bf 1})}
   & {\sl   2\/}: \parboxII{NS5~({\bf 1}), $5^2_2$~({\bf 1}) }
   & {\sl   2\/}: \parboxII{KKM~({\bf 2})}    \\ \hline
 7 & {\sl  20\/}: \parboxM {M5~({\bf 4}), KKM~({\bf 12}), $5^3$~({\bf 4})} 
   & {\sl  20\/}: \parboxII{D4~({\bf 1}), D6~({\bf 3}), NS5~({\bf 3}), KKM~({\bf 6}), $6^1_3$~({\bf 3}), $4^3_3$~({\bf 1}), $5^2_2$~({\bf 3}) }
   & {\sl  20\/}: \parboxII{D5~({\bf 3}), NS5~({\bf 3}), D7~({\bf 1}), $7_3$~({\bf 1}), KKM~({\bf 6}), $5^2_2$~({\bf 3}), $5^2_3$~({\bf 3})} \\ \hline
 6 & {\sl  40\/}: \parboxM {M5~({\bf 10}), KKM~({\bf 20}), $5^3$~({\bf 10})}
   & {\sl  40\/}: \parboxII{D6~({\bf 4}), D4~({\bf 4}), $6^1_3$~({\bf 4}), $4^3_3$~({\bf 4}), KKM~({\bf 12}), NS5~({\bf 6}), $5^2_2$~({\bf 6})}
   & {\sl  40\/}: \parboxII{D7~({\bf 1}), $7_3$~({\bf 1}), D5~({\bf 6}), NS5~({\bf 6}), D3~({\bf 1}), $5^2_3$~({\bf 6}), $5^2_2$~({\bf 6}), $3^4_3$~({\bf 1}), KKM~({\bf 12})}  \\ \hline
 5 & {\sl  72\/}: \parboxM {M5~({\bf 20}), M2~({\bf 1}), KKM~({\bf 30}), $5^3$~({\bf 20}), $2^6$~({\bf 1})}
   & {\sl  72\/}: \parboxII{D6~({\bf 5}), D4~({\bf 10}), D2~({\bf 1}), $6^1_3$~({\bf 5}), $4^3_3$~({\bf 10}), $2^5_3$~({\bf 1}), KKM~({\bf 20}), NS5~({\bf 10}), $5^2_2$~({\bf 10})}
   & {\sl  72\/}: \parboxII{D7~({\bf 1}), $7_3$~({\bf 1}), D5~({\bf 10}), NS5~({\bf 10}), D3~({\bf 5}), $5^2_3$~({\bf 10}), $5^2_2$~({\bf 10}), $3^4_3$~({\bf 5}), KKM~({\bf 20})}  \\ \hline
 4 & {\sl 126\/}: \parboxM {M5~({\bf 35}), M2~({\bf 7}), KKM~({\bf 42}), $5^3$~({\bf 35}), $2^6$~({\bf 7})}
   & {\sl 126\/}: \parboxII{D6~({\bf 6}), D4~({\bf 20}), D2~({\bf 6}), $6^1_3$~({\bf 6}), $4^3_3$~({\bf 20}), $2^5_3$~({\bf 6}),  KKM~({\bf 30}), NS5~({\bf 15}), $5^2_2$~({\bf 15}), F1~({\bf 1}), $1^6_4$~({\bf 1})}
   & {\sl 126\/}: \parboxII{D7~({\bf 1}), $7_3$~({\bf 1}), D5~({\bf 15}), NS5~({\bf 15}), D3~({\bf 15}), D1~({\bf 1}), F1~({\bf 1}), $5^2_3$~({\bf 15}), $5^2_2$~({\bf 15}), $3^4_3$~({\bf 15}), $1^6_3$~({\bf 1}), $1^6_4$~({\bf 1}), KKM~({\bf 30})} \\ \hline
 \end{tabular}
 \end{center}
 \vspace*{-3ex}
 \caption{\sl $U$-duality multiplets of codimension-2 objects in
 toroidal compactifications and their brane interpretations in M-theory
 and string theory. The total multiplicities are written in slanted
 font in front in each cell.  The multiplicity for each brane is
 written in boldface numbers in the parentheses.  The latter
 multiplicities can be interpreted as those of $SL(8)$ representations for
 M-theory and those of $SL(7)$ representations for Type IIA/B\@.  For
 $D=3$, see Table
 \ref{table:exotic_states}. \label{table:codim2variousd}}
\end{table}

Just as we did in section \ref{ss:exotic_states_3D}, it is
straightforward to find the multiplet of codimension-2 objects, as
listed in Table \ref{table:codim2variousd}. The total multiplicity is
always given by $\dim G-\rank G$.  This is because one can associate
with each state in the multiplet a root vector of the Lie algebra $g$ of
the Lie group $G$, and $T$- and $S$-dualities are Weyl reflections of
the root lattice \cite{Obers:1998fb}.  Because the group $G$ for
toroidal compactifications always has roots of equal length ({\it i.e.},
it is simply-laced), any of the $\dim
G$ root vectors can be Weyl-reflected into each other, except for the
$\rank G$ zero vectors corresponding to the Cartan subalgebra.
Therefore, the number of states in the $U$-duality orbit of the ordinary
supersymmetric brane is $\dim G-\rank G$. It is perhaps worth emphasizing 
once more that this multiplet is what one obtains by acting with simple $S$-
and $T$-dualities only, and not by the most general $U$-duality. 
In \cite{Kleinschmidt:2011vu, Bergshoeff:2011se}, it was shown that it
is this multiplet of ($\dim G-\rank G$) branes that can couple to spacetime
potential fields by gauge-invariant and $U$-duality invariant Wess--Zumino
coupling in a supersymmetric manner.

The list of codimension-2 objects such as the one in Table
\ref{table:codim2variousd} has appeared in the literature in some way or
other \cite{Elitzur:1997zn, Blau:1997du, Hull:1997kb, Obers:1997kk, Obers:1998fb,
Kleinschmidt:2011vu, Bergshoeff:2011se}.  However, the content of the
multiplets in terms of various exotic branes was not explicitly written
down and we believe that it is useful to present it here.

\subsection{Exotic branes, F-theory, and U-branes}
\label{ss:exotic_branes,F-theory,U-branes}

In the above, we argued that exotic branes are nothing but codimension-2
objects with non-trivial $U$-duality monodromies around them.  This is
exactly the idea of F-theory \cite{Vafa:1996xn}, which is about
considering a configuration with non-trivial monodromies for the
axio-dilaton $\tau=C^{(0)}+ie^{-\Phi}$ under the $SL(2,\bbZ)$ duality group
of 10D Type IIB string, and interpreting the configuration as a
12-dimensional geometry obtained by fibering a $T^2$ with the modulus
$\tau$ over the original 10D space.  Indeed, the two states appearing in
the $d=10$ row of Table \ref{table:codim2variousd} are nothing but the
standard $(p,q)$ 7-branes of F-theory.

It was only natural to generalize this F-theory construction by
compactifying to lower dimensions where the $U$-duality gets enhanced to
a group $G(\bbZ)$ listed in Table \ref{table:CJgroups}.  This is what
was done in Ref.\ \cite{Kumar:1996zx}.  Specifically, one takes an
$SL(n,\bbZ)$ subgroup of $G(\bbZ)$ and interprets it as the large
diffeomorphism group of a torus $T^n$ whose moduli are embedded in the
scalar moduli space $\CM=H\backslash G(\bbR)/G(\bbZ)$.\footnote{Note
that this is geometrizing only a part of the full duality group
$G(\bbZ)$ and the full moduli space $\CM$.} Then, one considers a
non-trivial fibration of the $T^n$ over the non-compact directions.  For
example, Ref.\ \cite{Kumar:1996zx} studied the $d=8$ case with
$G=SL(3)\times SL(2)$, which they call ``S-theory'', and constructed
some $T^3$ fibrations whose total space turned out to be Calabi-Yau
3-folds.
More general configurations with non-trivial $U$-duality monodromies were
studied in \cite{Kumar:1996zx, Liu:1997mb, Curio:1998bv, Leung:1997tw,
Lu:1998sx}, and the $U$-duality generalizations of the original F-theory
7-brane are dubbed ``$U$-branes.''  

Therefore, (some of) the codimension-2 branes listed in Table
\ref{table:codim2variousd} have already been known in the context of
F-theory.  However, their relation to exotic states
\cite{Elitzur:1997zn, Blau:1997du, Hull:1997kb, Obers:1997kk, Obers:1998fb} as well as
their non-geometric interpretation was not appreciated until
\cite{deBoer:2010ud} and it is that connection that we are making here.
As we will see below, the identification of non-geometric $U$-folds (or
$U$-branes) as branes helps us understand the supertube effect involving
non-geometric monodromies and leads to interesting possible applications
in string theory.

\section{Aspects of exotic branes}
\label{sec:aspects}
%

\subsection{Charge as monodromy}
\label{ss:charge_as_monodromy}

As we discussed in the previous section, the charge of a codimension-2
brane is classified by the $U$-duality monodromy around it.  A
$U$-duality monodromy is an element in the discrete group $G(\bbZ)$.
This is a generalization of the notion of ordinary charge, which lives
in the lattice $\bbZ^n$ with some $n$.  Henceforth, we will often
use the words ``charge'' and ``monodromy'' interchangeably.

Let us make it more precise what we mean by charges of codimension-2
branes defined by the monodromies around them.  A brane with monodromy
$q$ means the following.  As we travel along a path $\gamma$
encircling the brane, the moduli matrix $M\in H\backslash
G(\bbR)/G(\bbZ)$ undergoes the monodromy transformation
\begin{align}
 M\to Mq, \qquad q\in G(\bbZ).\label{hknh19Mar12}
\end{align}
Actually, in order to define the monodromy of a configuration
unambiguously, one needs to fix a ``base point'' with the value of the
moduli value there, $M$, and always measure monodromies with respect to
that point.  Namely, the path $\gamma$ always starts and ends at the
base point.
Also note that the monodromy \eqref{hknh19Mar12} is when the value of
the moduli is $M$ at the base point and, if we instead start with a
different value of the moduli $\tilde{M}=MU$ at the base point, where
$U$ is some $G(\bbZ)$ matrix, then the monodromy along the same path
$\gamma$ is given instead by
\begin{align}
 \tilde M\to \tilde M \tilde q,
 \qquad \tilde q=U^{-1}qU.\label{niqk12Sep12}
\end{align}

\begin{figure}[tb]
\begin{quote}
 \begin{center}
 \begin{tabular}{ccc}
  \includegraphics[height=3cm]{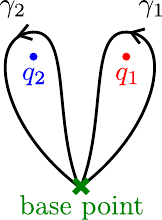} &\hspace*{1.5cm}&
  \includegraphics[height=2.5cm]{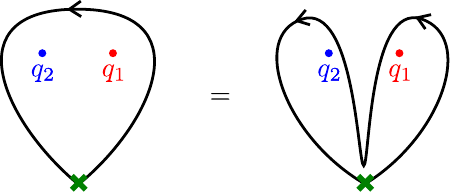}\\ (a) && (b)
 \end{tabular}
 \end{center}
 \caption{\sl To define the monodromy of a configuration unambiguously,
 one needs to fix a base point (shown as ``$\boldsymbol{\times}$\!'')
 with the value of the moduli there, $M$, and give monodromies of $M$
 with respect to the basis of paths which start and end on the base
 point.  (a): an example configuration with a base point and two
 paths $\gamma_1$ and $\gamma_2$ going around charges with
 monodromies $M\to M q_1$ and $M\to M q_2$, respectively.  (b): a path
 going around both charges corresponds to monodromy $Mq_1\to Mq_1\cdot
 q_1^{-1}q_2 q_1.$ See text for more detail. \label{fig:ref_pt1,2}}
\end{quote}
\end{figure}
If we know the monodromies for a basis of 1-cycles, it is possible to
determine the monodromy along any paths.  For example, in
Fig.~\ref{fig:ref_pt1,2}(a), we have a configuration with two charges
with individual monodromies $q_1$ and $q_2$ along paths $\gamma_1$ and
$\gamma_2$, respectively.  Let us consider going around both charges at
the same time as shown in Fig.~\ref{fig:ref_pt1,2}(b). By homotopically
deforming the path as shown, we see that this path is the composition of
$\gamma_1$ and $\gamma_2$, which we denote by $\gamma_2\gamma_1$.  As we
move along $\gamma_1$, we get the monodromy $M\to Mq_1$ by definition.
When we further go along $\gamma_2$, we are starting with the moduli
$\tilde{M}=Mq_1$.  Therefore, using \eqref{niqk12Sep12}, the moduli now
transform as $Mq_1\to Mq_1\cdot q_1^{-1}q_2 q_1$.  Namely, the monodromy
for going around both $q_1,q_2$ at the same time is $M\to Mq_1\cdot
q_1^{-1}q_2 q_1=Mq_1^{-1}(q_1q_2)q_1=Mq_2q_1$.

\begin{figure}[tb]
\begin{quote}
 \begin{center}
  \begin{tabular}{c@{\hspace{5ex}}c}
   \includegraphics[height=3cm]{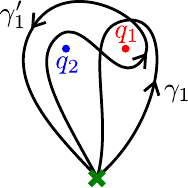}
   &
   \includegraphics[height=3cm]{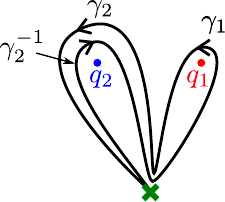}
   \\
   (a) \sl Two paths going around $q_1$.
   &
   (b) \sl Path $\gamma'_1$  is equal to $\gamma_2\gamma_1\gamma_2^{-1}$.
   \\[3ex]
   \multicolumn{2}{c}{\includegraphics[height=3cm]{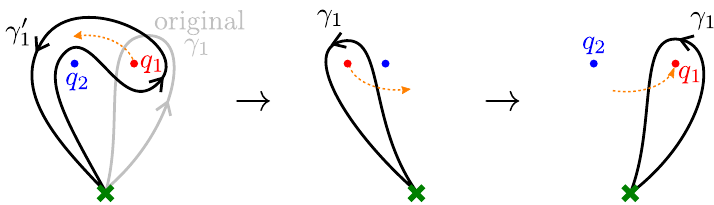}}
   \\
   \multicolumn{2}{c}{(c) \sl Interchanging charges will change paths.}
  \end{tabular}
 \end{center}
 \caption{\sl The monodromy charge depends on the choice of paths with
 respect to which to measure the monodromy.  See text for detail.
 \label{fig:ref_pt4}}
\end{quote}
\end{figure}
To define monodromy charge unambiguously, it is important to fix once
and for all the paths along which to measure monodromy, not just the
base point.  For example, in the present configuration, $\gamma_1$ is
not the only path that includes the base point and goes around $q_1$.
In Fig.~\ref{fig:ref_pt4}(a), we presented one other example denoted by
$\gamma_1'$.  The monodromy along $\gamma_1'$ can be easily computed to
be $M\to Mq_2q_1q_2^{-1}$, by noting that path $\gamma_1'$ is equal to
$\gamma_2\gamma_1\gamma_2^{-1}$ as shown in Fig.~\ref{fig:ref_pt4}(b).
In general, homotopically different paths that encircle the same brane
give different monodromy charges that are related to each other by
conjugation.

There is an related subtlety when we move one charge around another.
Let us consider moving charge $q_1$ around charge $q_2$
counterclockwise, as shown in Fig.~\ref{fig:ref_pt4}(c).  Before the
move, we have path $\gamma_1$ with monodromy $M\to Mq_1$ and path
$\gamma_1'$ with monodromy $M\to Mq_2q_1q_2^{-1}$ (see the left panel of
the Figure).  After the move, the system looks as if it went back to the
original configuration, with the original $\gamma_1'$ changed into a
``new $\gamma_1$'' (the right panel).  However, because monodromy living
in the discrete space $G(\bbZ)$ cannot change under continuous
deformation of the path, the monodromy associated with this new ``new
$\gamma_1$'' is given by $M\to Mq_2q_1q_2^{-1}$, not the original $M\to
Mq_1$.  So, every time we move a charge around another, it looks as if
the charge measured by path $\gamma_1$ jumps.  This is happening because
we are choosing different paths to measure the charge and, if we stick
to the original path by following its continuous deformation, the
monodromy charge remains the same.  This point is important in
understanding charge conservation in the presence of exotic branes, as
we will see in the next subsection.


\medskip
Let us ask a question of what
charges exist in a \emph{fixed} $U$-duality frame.  Let us start from
the situation given in \eqref{hknh19Mar12}, where a brane with monodromy
$q$ exists. Furthermore, let us assume that we can arrange codimension-2
branes so that the moduli $M$ at infinity tends to a constant value,
without a non-trivial monodromy. We can achieve this by having multiple
codimension-2 branes with canceling monodromies or curling up branes as
we will discuss in section \ref{sec:supertube_effect_and_exotic_branes}.
After dualization, one has a brane with monodromy $\tilde q$ as in
\eqref{niqk12Sep12} and, at the same time, the value of the moduli at
infinity have changed to $\tilde M$.  Now, let us change the moduli at
infinity adiabatically back to the original value $M$. If the brane
configuration is supersymmetric, we expect that the brane with monodromy
$\tilde q$ survives the adiabatic process, provided that there is no
wall of marginal stability.  So, a brane with charge $\tilde q$ should
exist even for the original value of the moduli $M$.  Namely, if a brane
with monodromy $q$ exists, branes with monodromies $\tilde q=U^{-1}qU$,
$U\in G(\bbZ)$ should also exist.  One caveat, however, is that this
does not mean that we can generate all charges that exist in the theory
by conjugation; there can be many conjugacy classes in the group
$G(\bbZ)$ and we cannot generate charges in different conjugacy classes.
Also, there can be non-supersymmetric configurations for which the above
argument of adiabatically changing moduli does not apply.

Note that the above argument is not a very strong one.  First, in a
situation where we cannot make the moduli to tend to a constant value at
infinity, the conjugated charges do not have to exist.  For example, if
there is a single charged particle $q$ in 3D, then the moduli has the
monodromy $M\to Mq$ even at infinity and the above argument does not
apply.  Also, if a wall of marginal stability exists, the above argument
can fail.

As a simple example, consider a D7-brane in Type IIB superstring.
Around it, there is a non-trivial monodromy of the $SL(2,\bbZ)$ duality
given by
\begin{align}
 T=\begin{pmatrix}1&1\\0&1\end{pmatrix}.
\end{align}
Let us
conjugate this with a general $SL(2,\bbZ)$ matrix
\begin{align}
 U=\begin{pmatrix}s&r\\ q&p\end{pmatrix},\qquad sp-rq=1.
\end{align}
The
conjugated charge is
\begin{align}
 \tilde T=U^{-1}TU=
 \begin{pmatrix}
  1+pq &p^2 \\
 -q^2 & 1-pq\end{pmatrix},
 \label{pq7-monodromy}
\end{align}
which is the monodromy of the standard $(p,q)$ 7-brane. So, if the
assumptions we made above are true, there should also exist $(p,q)$
7-branes with $p,q$ coprime. If we further assume the existence of a
bound state of $N$ 7-branes with monodromy
\begin{align}
 T^N=\begin{pmatrix}1&N\\0&1\end{pmatrix}.
\end{align}
for all $N\in\bbZ$,
then there should exist 7-branes the  monodromy
\begin{align}
 \tilde T^N=U^{-1}T^NU=
 \begin{pmatrix}
  1+Npq &Np^2 \\
 -Nq^2 & 1-Npq\end{pmatrix}.
 \label{pq7-monodromy2}
\end{align}

Note that the monodromy matrix thus obtained has always $\tr(\tilde
T^N)=2$.  Because trace (partly) characterizes the conjugacy classes of
$SL(2,\bbZ)$, we cannot reach objects whose monodromy matrices have
trace different from $2$ by starting from a $(p,q)$ 7-brane (for
classification of $SL(2,\bbZ)$ conjugacy classes, see
\cite{DeWolfe:1998eu, DeWolfe:1998pr}).  It is not clear whether in
string theory there exist 7-branes whose monodromy is not in the trace-2
conjugacy class.  Note that, although the orientifold 7-plane in Type
IIB superstring has a monodromy which is not in the trace-2 class, in
F-theory it is represented by a bound state of $(p,q)$ 7-branes, each of
which is in the trace-2 class \cite{Sen:1996vd, Dasgupta:1996ij}.
However, it is known that supersymmetric 7-brane solutions with general
conjugacy classes (``Q7-branes'') do exist at the level of classical
supergravity \cite{Bergshoeff:2006gs, Bergshoeff:2006jj,
Bergshoeff:2007aa}.  The meaning of such solutions in string theory is
not clear.

\subsection{Monodromies and charge conservation}
\label{sec:monodromies_and_charge_conservation}

In the presence of an exotic brane with a non-trivial $U$-duality
monodromy, moving a second brane around it will $U$-dualize the second
brane into a different brane.  Therefore, it appears that the associated
brane charge is not conserved in the presence of an exotic brane. Here
we demonstrate, based on explicit examples, that this is not the case
and charges are actually conserved even in such situations, if we use
the appropriate notion of charge.

More precisely, the question is the following.  An exotic brane is a
codimension-2 brane in $d$ dimensions and there is a non-trivial
monodromy of $d$ dimensional scalars around it.  Now, introduce a second
object which is charged under some gauge field in $d$ dimensions.  It
sometimes happens that moving the second object around the codimension-2
exotic brane apparently induces a new charge.  We would like to
understand how this phenomenon is consistent with charge
conservation.\footnote{A recent discussion on the apparent
non-conservation of brane charges in a configuration with exotic branes
can be found in \cite{Kikuchi:2012za}.}

\subsubsection{Charge conservation and Page charge}
\label{ss:chg_consv_and_Page_chg}

\begin{figure}[tb]
\begin{quote}
  \begin{center}
  \includegraphics[height=4cm]{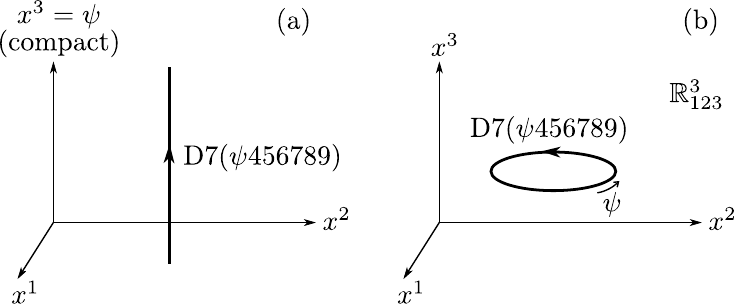}
 \end{center}
 \caption{\sl A D7-brane along $\psi 456789$ directions, with $\psi$
 being (a) a compactified direction and (b) a direction along a
 contractible circle.  \label{fig:d7}}
\end{quote}
\end{figure}

To start the discussion with, let us consider a D7-brane along $\psi
456789$ directions.  Here, we take $456789$ to be compact directions
with period $2\pi l_s$.  $\psi$ can be either a compactified direction
or the direction along a contractible circle in non-compact
$\bbR^{3}_{123}$.  The first case is simpler but the transverse
spacetime directions will not be asymptotically flat $\bbR^2$ once
backreaction is taken into account, as will be discussed in section
\ref{sec:sugra_description}.  In the second case, the transverse
spacetime remains asymptotically flat $\bbR^3$ even if backreacted, as
will be studied in section
\ref{sec:supertube_effect_and_exotic_branes}. See Figure \ref{fig:d7}
for a schematic description of the configurations.  In the second case,
the D7-brane can either be a static configuration supported by something
({\it e.g.}\ by a supertube effect, as will be studied in section
\ref{sec:supertube_effect_and_exotic_branes}) or just an instantaneous
configuration which will collapse eventually.
In either case, the D7-brane is a codimension-2 object already in $d=10$
dimensions and, around it, there is a non-trivial $SL(2,\bbZ)$ monodromy
described by the matrix
$q=(\begin{smallmatrix}1&1\\0&1\end{smallmatrix})$.  So, the scalar
$\tau=C^{(0)}+ie^{-\Phi}$ has the monodromy $\tau\to \tau+1$.

\begin{figure}[tb]
\begin{quote}
  \begin{center}
  \includegraphics[height=3cm]{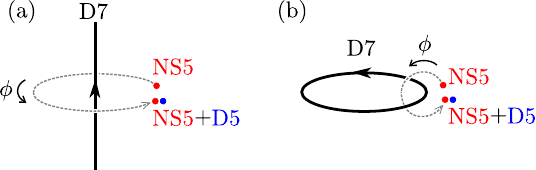}
 \end{center}
 \caption{\sl Moving an NS5-brane around a D7-brane appears to produce 
 D5-branes.  \label{fig:d7_ns5}}
\end{quote}
\end{figure}

Now, add to this configuration an NS5-brane along $56789$ directions and
consider moving it around the D7-brane.  The NS5-brane is magnetically
charged under the gauge field $B^{(2)}$.
Because
$A^T=(C^{(2)},B^{(2)})^T$ is a doublet under $SL(2,\bbZ)$ transforming
as $A\to q A$, $C^{(2)}$ changes as $C^{(2)}\to C^{(2)}+B^{(2)}$ as we
go around the D7.  Because the integral of $dC^{(2)}$ measures 
D5-brane charge, this means that moving an NS5 around a D7 produces
D5-branes; see Figure \ref{fig:d7_ns5}.  Let us write this as
\begin{align}
 \rm D7(\psi 456789):\quad  NS5(56789) \to  NS5(56789)+D5(56789).\label{foh7Aug12}
\end{align}
Another way to see that $C^{(0)}$ induces D5-brane charge on the
NS5-brane is as follows.  By the Wess--Zumino term in the D-brane
worldvolume action, non-vanishing $B^{(2)}_{89}\neq 0$ induces D3(567)
charge on the D5(56789) worldvolume.  The $S$-dual of this statement is
that $C^{(2)}_{89}$ induces D3(567) on NS5(56789).  Further by $T_{89}$,
we see that $C^{(0)}$ induces D5(56789) on NS5(56789).

By taking $S$, $T_{89}$, $S$, and then $T_{567}$ dualities of
\eqref{foh7Aug12}, we see that moving a D0 around a $5^2_2$ produces
D2-brane charge:
\begin{align}
\rm 5_2^2(\psi 4567;89): \quad D0\stackrel{?}{\to} D0+D2(89),\label{frq7Aug12}
\end{align}
where the D0 is smeared along 456789 directions and the D2(89) is
smeared along 4567 directions.  We have put a question mark in
\eqref{frq7Aug12} because we will later question this process.

Further by taking $T_{89}$ of \eqref{frq7Aug12}, we see that
moving a D2 around an NS5 produces D0 charge:
\begin{align}
\rm NS5(\psi 4567): \quad D2(89) \to D2(89)+D0,\label{fvg7Aug12}
\end{align}
where, again, branes are smeared along transverse directions within the
compact $456789$ directions.  The NS5($\psi$4567)-brane can be thought
of as a codimension-2 object in 8D if we compactify the 10D theory on
$T^2_{89}$.

Let us study the above processes \eqref{foh7Aug12}--\eqref{fvg7Aug12},
in which brane charges do not appear to be conserved, and examine in
what sense they can actually be conserved.

We begin with \eqref{fvg7Aug12} as the easiest situation to study,
although the NS5($\psi 4567$) is not an exotic brane.  Because the $89$
directions are compactified, from the viewpoint of the non-compact
directions ($123$ or $1234$, depending on the situations (a), (b) of
Fig.\ \ref{fig:d7}), the NS5($\psi 4567$) is a codimension-2 object.
The charge of the NS5($\psi 4567$)-brane is measured by
\begin{align}
 Q^{\text{NS5}}={1\over 2\pi(2\pi l_s)^2}\int_{\phi 89}H^{(3)}
 = {1\over 2\pi}[B_{89}^{(2)}]^{\phi=2\pi}_{\phi=0}
 \equiv {1\over 2\pi}\Delta B_{89}^{(2)},
\end{align}
where $\phi$ is an angular direction encircling the NS5-brane in the
non-compact space (see Figure \ref{fig:d7_ns5}).  This means that
$B_{89}^{(2)}$ increases by $\Delta B_{89}^{(2)}=2\pi Q^{\text{NS5}}$ as we go around
the NS5.  Now, the D2-brane Wess--Zumino coupling
\begin{align}
S^{\text{D2,WZ}} ={1\over (2\pi)^2l_s^{3}}\int
 (C^{(3)}-B^{(2)}\wedge  C^{(1)}) 
 \label{nclc7Aug12}
\end{align}
implies that moving
a D2(89) around the NS5 will induce  D0-charge
\begin{align}
 \Delta Q^{\text{D0,bs}}={1\over (2\pi l_s)^{2}}\int
 \Delta B^{(2)}=\Delta B_{89}^{(2)}=2\pi Q^{\text{NS5}}.\label{ncwl7Aug12}
\end{align}
The superscript ``bs'' will be explained below.
In \eqref{nclc7Aug12}, we assumed that the Chan--Paton (Born--Infeld)
gauge field strength $F^{(2)}$ in the D-brane worldvolume vanishes.
Therefore, it appears that D0 charge $Q^{\text{D0,bs}}$ is not conserved and
increases by \eqref{ncwl7Aug12} every time we move the D2 around the
NS5.

However, recall that, as discussed in \cite{Marolf:2000cb}, there are
multiple notions of charge and we should be careful about what charge we
are talking about.  \emph{Brane source charge} is gauge-invariant but
not conserved, whereas \emph{Page charge} is conserved and
gauge-invariant under small gauge transformation. Page charge changes
under global gauge transformation, but global gauge transformation
changes the state of the system and charge does not have to remain the
same under it in the first place.  Moreover, it is Page charge that is
quantized and appears in the asymptotic super-Poincar\'e algebra as
central charge \cite{Marolf:2000cb}.
So, if we want to discuss charge conservation, it is Page charge that we
should consider, not brane source charge.
We discuss brane source and Page charges in Appendix
\ref{app:page_chg}\@.  Here, we only use the results for the expression
of brane source and Page charges from there and refer the reader to the
appendix for details.

D-brane source current is obtained simply by varying the D-brane action
\eqref{nclc7Aug12} with respect to RR potential $C$, as discussed in
Appendix \ref{app:page_chg}\@.  Therefore, D-brane source charge
includes the D-brane charge induced by the spacetime $B^{(2)}$ field,
and $Q^{\text{D0,bs}}$ in \eqref{ncwl7Aug12} is brane source charge;
that is why the superscript ``bs''.  On the other hand, D-brane Page
charge is obtained from brane source charge precisely by subtracting the
charge induced by $B^{(2)}$.  Therefore, for Page charge,
\eqref{ncwl7Aug12} is modified to
\begin{align}
 \Delta Q^{\text{D0,Page}}=0.
\end{align}
Namely, even if we move D2 around NS5, no Page D0 charge is induced and
it is actually conserved.

Now let us turn to the process \eqref{foh7Aug12}.  Although the D7-brane
has codimension 2, it is not quite exotic in that the monodromy around
it does not involve metric; it is just an additive shift of $C^{(0)}$.
The analysis of this process is similar to that of \eqref{fvg7Aug12}.
Namely, although it appears that D5-brane charge is induced on the
NS5-brane by spacetime RR potentials $C$, as discussed in Appendix
\ref{app:page_chg}, D5-brane Page charge is defined by subtracting such
induced charge.  Therefore, in this case, there is no D5-brane charge
induced, {\it i.e.},
\begin{align}
 \Delta Q^{\text{D5,Page}}=0.
\end{align}
Even if we move NS5 around D7, no Page D5 charge is induced and
charge is conserved.

Finally, let us consider the process \eqref{frq7Aug12} in which the
exotic brane $5^2_2$ is involved.  First of all, we immediately notice
that D2-charge being induced on a D0 is strange, because the D0-brane
Wess--Zumino term
\begin{align}
S^{\text{D0,WZ}} 
 ={1\over l_s}\int C^{(1)}
\end{align}
does not involve $C^{(3)}$ and there is no way to induce D2 charge on a
D0.\footnote{Note that we are considering a single D0-brane and the
commutator couplings \cite{Myers:1999ps} which are present for multiple
branes and are responsible for Myers' effect do not exist here.}  The
mistake we made is the following.  The D0-brane charge on the right hand
side of \eqref{fvg7Aug12}, which we confirmed is induced, is brane
source charge.  Actually, brane source charge do not transform
covariantly under duality transformations, and therefore the existence
of induced D0-brane source charge in the duality frame \eqref{fvg7Aug12}
does \emph{not} imply that D2-brane source charge is induced in the
duality frame \eqref{foh7Aug12} as we naively presumed.  Therefore, in
order to study the charge in \eqref{foh7Aug12}, we need instead a notion
of charge that transforms covariantly under duality.

It turns out to that the charge that transformscovariantly under
duality is Page charge.  This can be seen as follows.  Let us focus on
the $T$-duality transformation that we performed in going between
\eqref{frq7Aug12} and \eqref{fvg7Aug12}.
If we compactify the 10D theory on $T^2_{89}$ down to 8D, we have
$T$-duality group $SO(2,2,\bbZ)=SL(2,\bbZ)_\tau\times SL(2,\bbZ)_\rho$.
Here, $\tau,\rho$ are 8D moduli scalars defined by
\begin{align}
 \rho&=B^{(2)}_{89}+i\det\nolimits^{1/2}G_{ab},\qquad
 G_{ab}={\rho_2\over \tau_2}\begin{pmatrix}1&\tau_1 \\ \tau_1&|\tau|^2\end{pmatrix},\qquad a,b=8,9\label{hcjy26Sep12}
\end{align}
and transform in the standard way under respective $SL(2,\bbZ)$ factors; namely,
\begin{align}
 \rho\to \rho'={a\rho+b\over c\rho+d},\qquad
 \begin{pmatrix}  a&b\\ c&d \end{pmatrix}
 \in SL(2,\bbZ)_\rho~,
\end{align}
and likewise for $\tau$ and $SL(2,\bbZ)_\tau$.  As we will study in
detail in section \ref{sec:sugra_description}, NS5($\psi 4567$) and
$5^2_2(\psi 4567,89)$ can be both thought of as codimension-2 branes in
8D with non-trivial monodromies for the scalar $\rho$, but no such
details are necessary for the current discussion.
The $T$-duality transformation that we performed in going between
\eqref{frq7Aug12} and \eqref{fvg7Aug12} belongs to $SL(2,\bbZ)_\rho$
\cite{Polchinski:1998rq}.
Upon reducing to 8D, the D0 and D2(89)-branes both become 0-branes, and
the 10D RR potentials $C_\mu$ and $C_{89\mu}$ ($\mu=0,\dots,7$) that
they couple to reduce to 8D 1-forms, $\CC_{ \alpha\mu}$ ($\alpha=1,2$),
which form a doublet under $SL(2,\bbZ)_\rho$ \cite{Obers:1998fb,
Fukuma:1999jt}.  Ref.\ \cite{Fukuma:1999jt} showed that the covariant
field $\CC_{\alpha\mu}$ in 8D is related to $C_\mu$ and $C_{89\mu}$ by
\begin{align}
 \CC_{1,\mu}=C_\mu,\qquad
 \CC_{2,\mu}=C_{89\mu}-B_{89}C_\mu.
\end{align}
If we define charge currents covariant under the $T$-duality group by
variation of the action with respect to $\CC_{\alpha\mu}$, then, using
the relation
\begin{align}
 {\delta\over \delta \CC_{1,\mu} } =
 {\delta\over \delta C_{\mu} }+B_{89}{\delta\over \delta C_{89\mu} },
 \qquad 
 {\delta\over \delta \CC_{2\mu} } =
 {\delta\over \delta C_{89\mu} },
\end{align}
we can show that the charges associated with $\CC_{\mu\alpha}$ are
related to the brane source charges as
\begin{align}
 \CQ^{\alpha=1}=Q^{\rm D0,bs}+B_{89}Q^{\rm D2(89),bs},\qquad
 \CQ^{\alpha=2}=Q^{\rm D2(89),bs},
\end{align}
where the volume of $T^2_{89}$ is $(2\pi l_s)^2$.  $\CQ^{\alpha=1}$ is
D0-brane source charge minus the one induced by the $B$-field on the
D2-brane worldvolume.  This is nothing but Page charge for the D0-brane.

So, Page charge covariantly transforms under $T$-duality transformation.
In the duality frame \eqref{fvg7Aug12}, we have shown that D0 Page
charge is conserved.  By $T$-dualizing to the present duality frame
\eqref{frq7Aug12}, this automatically implies that D2 Page charge is
conserved.  The apparent non-conservation of D2-brane charge in
\eqref{fvg7Aug12} was simply because we were looking at a wrong notion
of charge inappropriate to discuss duality transformation and charge
conservation.

This argument is valid for any systems which are related to the above
configuration by $T$-duality.  Namely, as long as one measures Page
charge, charges are always conserved.  Brane source charge, on the other
hand, is not conserved, but that does not contradict with charge
conservation.  We expect that this holds true generally, not just in the
examples we considered.  Namely, even in the presence of exotic scalar
monodromies,  brane Page charges are always 
unambiguously defined and conserved.

It would be interesting to show this in full generality in a more
systematic way.  In particular, it would be desirable to generalize the
result of Ref.~\cite{Fukuma:1999jt} to $U$-duality transformations.
Because Page charge is the central charge that appears in the asymptotic
algebra and transform covariantly under $U$-duality, this would amount
to finding explicit expressions for Page charges that transform
covariantly under $U$-duality.

\subsubsection{Monodromies and Page charge}
\label{ss:monodromies_page_chg}

We thus showed that, as long as we use Page charge to define charge,
there is no induced charge even when one moves a charge around an exotic
brane, around which there is non-trivial $U$-monodromy.  We also argued
that, because Page charges transform covariantly under $U$-duality, the
conservation of Page charge must hold in any frames.

But there is an apparent tension here.  On one hand, we said that Page charge
remain the same even when we go around an exotic brane around which
there is $U$-duality monodromy.  On the other hand, we said that Page
charges transform covariantly under $U$-duality.  How can the two
statements be consistent with each other?

The resolution is closely related to the subtlety in defining monodromy
charges that we discussed in section \ref{ss:charge_as_monodromy}.  As
we discussed there, unambiguously defining monodromies requires that we
fix a base point and the value of moduli at that point, once and for
all, and that we measure monodromy with respect to them.  A similar
consideration is needed when defining Page charge. Page charge involves
an integral of form fields around certain cycles that enclose the object
in question. However, in the presence of exotic branes, those form
fields themselves are also not globally well-defined. To define all
quantities consistently, we need to choose a base point $P$ plus a
choice of moduli and form fields at $P$.  One can then follow the moduli
and form fields continuously along any path that does not intersect the
exotic brane.  Namely, Page charge should properly be defined as the
integral over a cycle that contains the point $P$, where the form fields
and moduli take the given values $M$ and $(B,C)$ respectively, and such that
moduli and form fields are continuous everywhere along the cycle. The
cycle is not allowed to intersect the exotic brane.

\begin{figure}[tb]
\begin{quote}
  \begin{center}
  \includegraphics[height=5cm]{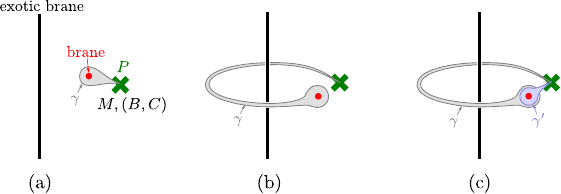}
 \end{center}
 \caption{\sl Defining Page charges in the presence of exotic branes
 with non-trivial monodromies requires that we fix a base point $P$
 (denoted ``$\times$'') along with the values of moduli $M$ and form
 fields $(B,C)$ there, and measure the charge with respect to
 them. Continuously deforming the cycle to measure charge does not
 change the charge.  However, there can be multiple, non-equivalent
 cycles \label{fig:page_monodromy}}
\end{quote}
\end{figure}

In Fig.~\ref{fig:page_monodromy}(a), we described a situation where we
measure Page charge of a brane sitting near the base point $P$ by
integrating a form field through a cycle $\gamma$ that encloses the
brane and goes through $P$.   If we move the brane and at the same
time continuously deform $\gamma$ so that it always goes through $P$ and
does not intersect the exotic brane, Page charge does not change as we
have explicitly demonstrated in section \ref{ss:monodromies_page_chg}
above.  We described this process in Fig.~\ref{fig:page_monodromy}(b).

With this definition of Page charge, it is still not unique, as there
can be different topological types of cycles which one can use to
compute it. Because cycles are not allowed to intersect the exotic
brane, not all cycles can be continuously deformed. But cycles that can
be continuously deformed into one another will give rise to the same
Page charge. This should be the case since Page charge is quantized.  
If we take a path that encircles the exotic brane and comes back to the
original base point $P$, the fields will undergo a monodromy and be
mapped into a $U$-dual version of themselves.  Thus, different
topological types of charges will give rise to different Page charges
that are related to each other through $U$-dualities.

In Fig.~\ref{fig:page_monodromy}(c), we described a situation where we
have moved the brane once around the exotic brane and brought it back
near the base point $P$.  Besides cycle $\gamma$ that now has a ``tail''
going around the exotic brane, there is another cycle $\gamma'$ that
goes through $P$ and encloses the brane but does not go around the
exotic brane.  If we use $\gamma$ to measure Page charge, we get the
same answer as the one measured in Fig.~\ref{fig:page_monodromy}(a).  On
the other hand, if we use $\gamma'$ to measure Page charge, we get the
$U$-dual version.  There is no contradiction here because the two cycles
$\gamma$ and $\gamma'$ cannot be deformed into each other without
intersecting the exotic brane; they measure physically different
charges.

Now the puzzle raised at the beginning of this section
\ref{ss:monodromies_page_chg}  is resolved.  When we said in
section \ref{ss:chg_consv_and_Page_chg} that Page charge is
conserved, we meant that the Page charge measured by cycle $\gamma$ is
unchanged even if we move the brane around the exotic brane along with
the cycle that encloses it.  This is what we must do physically, because
when we talk about charge conservation we should adopt one notion of
charge and stick to it as we continuously change the configuration.
On the other hand, the fact that Page charges transform covariantly
under $U$-duality is not related to such continuous deformation, and
there is no contradiction here.\footnote{This $U$-duality transformation
is in some sense similar to going between charges defined by $\gamma$
and $\gamma'$, but not quite.  Here the difference between $\gamma$ and
$\gamma'$ is related to the monodromy charge of the exotic brane present
in the configuration, but the $U$-duality transformation in section
\ref{ss:chg_consv_and_Page_chg} has nothing to do with the exotic brane
present in the configuration.}

\subsection{Number of charges}
\label{ss:num_of_charges}

As we discussed above, the charge of a codimension-2 brane is
characterized by the monodromy around it.  Since the monodromy matrix
$q$ is an element of the discrete non-Abelian ``lattice'' $G(\bbZ)$, it
does not really make sense to ask how many different charges there are,
in contrast to the case of an ordinary charge lattice $\bbZ^n$ where one
can say that there are $n$ different charges.  However, to get a
qualitative idea, we can replace $G(\bbZ)$ by the continuous group
$G(\bbR)$ and study the dimension of the (now continuous) space of
possible charges.  This is expected to be the dimension of the space of
charges that we see in the classical limit where the charges are large.

There are multiple notions that one can mean by the number of charges.  In
toroidal compactifications, we have a scalar moduli space of the form
$\CM=H(\bbR)\backslash G(\bbR)/G(\bbZ)$ whose isometry group is
$G(\bbR)$.  Since $G(\bbR)$ has
\begin{align}
 m\equiv \dim G
\end{align}
generators, there are $m$ associated conserved Noether currents in the
theory.  In this sense, the number of charges that are in principle
possible to occur in the theory is $m$.  In the 3D theory,
we have $G=E_{8(8)}$ and $m=248$.

If one could introduce a dual $(d-2)$-form gauge field for each of these
$m$ currents, it would seem like there are $m$ different
codimension-2 branes with different charges.  However, as is manifest in
Table \ref{table:exotic_states} for $d=3$ and in Tables
\ref{table:CJgroups} and \ref{table:codim2variousd} for $d>3$,
there are only
\begin{align}
 h\equiv \dim G-\rank G<m
\end{align}
branes that can be obtained by $U$-dualizing standard
half-supersymmetric branes such as D7-branes.  For example, in the $d=3$
case, $h=240$ ($<m=248$), and these are the point particle states listed
in Table \ref{table:exotic_states}.
This discrepancy between $m$ and $h$ is understood as follows
\cite{Bergshoeff:2010xc, Bergshoeff:2011zk, Bergshoeff:2012ex}.
Although there are $m$ gauge fields, only $h$ of them couple to 1/2-BPS
branes.  More precisely, if one tries to construct a $U$-duality and
gauge invariant Wess--Zumino coupling of a possible brane to the gauge
fields, it is possible to do so preserving half of supersymmetry only
for $h$ gauge fields out of $m$, and these $h$ branes are in the
$U$-duality orbit of the standard 1/2-BPS branes.  In this sense, $h$ is
the number of fundamental 1/2-BPS codimension-2 branes allowed in
string theory.
At the time of writing, it is not understood whether there exist states in
string theory that couple to the remaining $m-h$ gauge fields.
For counting of 1/2-BPS branes based on an $E_{11}$ group
theoretical argument, see \cite{Kleinschmidt:2011vu}.

Yet another other notion that one may associate with the number of
charges is the dimension of the $U$-duality orbit of a brane.  A general
analysis on the dimension of the orbits of BPS configurations in string
theory was done in \cite{Lu:1997bg} for codimension $>2$ branes.  For
codimension-2 branes, such an analysis was done \cite{deBoer:2010ud} and
later by \cite{Bergshoeff:2012ex} along the line of \cite{Lu:1997bg}.
Here, we repeat the analysis of \cite{deBoer:2010ud}, including some
details omitted there.
Let us start from a given charge $q\in G(\bbR)$.  Using $\CQ\in
g(\bbR)$, where $g(\bbR)$ is the Lie algebra of $G(\bbR)$, we can write
$q$ as $q=e^{\CQ}$.  In particular, if we consider 1/2 BPS objects such
as D7-branes, the matrix is nilpotent, $\CQ^2=0$, and hence $q=1+\CQ$\@.
Now, generate a new charge by conjugation by $U=e^{\epsilon t}\approx
1+\epsilon t$ where $t\in g(\bbR)$ and $\epsilon$ is infinitesimal. The
new charge matrix is $ \tilde q =U^{-1}qU\approx
q+\epsilon\,[\CQ,t]$. So, the number of different charges is given by $\dim g$ minus
the dimension of the stabilizer subspace $\{[\CQ,t]=0\,|\, t\in
g(\bbR)\}$.\footnote{This is the same procedure followed in
\cite{Lu:1997bg, Bergshoeff:2012ex}} This is given as follows.  First,
we find an $sl(2)$ subalgebra in which $\CQ$ is the raising operator.
Then, decompose the adjoint representation of $g(\bbR)$ into $sl(2)$
representations as
\begin{align}
 {\rm adj}=\bigoplus_{k=1}^d ({\bf 2j_k+1})
\end{align}
where $d$ is the number of representations appearing in the
decomposition.  Since $\CQ$ acts effectively on all states in each
$sl(2)$ representation except for the highest spin state, we find that the
dimension of the orbit, $b$, is
\begin{align}
 b=
 \mathop{\rm dim}g
 -
 d.
\end{align}
In the 3D case where $G=E_{8(8)}$, the adjoint representation decomposes as
${\bf 248}={\bf 3}+ 56\cdot {\bf 2}+ 133\cdot{\bf 1}$, and therefore 
$b=248-(1+56+133)=58$.  This number agrees
with the one obtained in \cite{Bergshoeff:2012ex} by a slightly
different argument.
In the last column of Table \ref{table:CJgroups}, we listed 
the values of $b$ in various dimensions.

\section{Supergravity description of exotic states}
\label{sec:sugra_description}

In this section, we study supergravity solutions corresponding to exotic
branes.  In higher dimensions, they correspond to infinitely long
defects in spacetime, around which there are non-geometric monodromies.
These solutions are not new, but our focus will be on their exotic
non-geometric aspects.

\subsection{An example: the supergravity solution for $5^2_2$}

To demonstrate that exotic branes are non-geometric objects, let us
compute the supergravity solutions for them and analyze the structure.
Using the duality rules shown in Tables
\ref{table:T-duality_IIA/B}--\ref{table:M2IIA}, it is straightforward to
start with any known standard brane backgrounds, act by supergravity
duality transformation on them, and obtain the background for exotic
branes.  Here, as an example, let us compute the metric for $5^2_2$ by
$T$-dualizing the KK monopole metric transverse to its worldvolume (cf.\
\eqref{fxxw20Mar12}).  A simplified version of the following analysis
was given in \cite{deBoer:2010ud}.

The metric for KK monopoles wrapped on compact $345679$
directions, with $x^9$ being the special circle (namely, they are
$5^1_2(34567,9)$), placed at ${\bf x}={\bf x}_p$ in the transverse space
$\bbR^3_{128}$, is
\begin{align}
\begin{split}
  ds^2&=dx_{034567}^2+H dx_{128}^2+H^{-1}(dx^9+\omega)^2,\qquad
 e^{2\Phi}=1,\\
 H&=1+\sum_p H_p,\qquad 
 H_p={R_9\over 2|{\bf x}-{\bf x}_p|},
\end{split}
\label{KKM_dilaton_omega}
\end{align}
where the 1-from $\omega$ satisfies
\begin{align}
  d\omega=*_3 dH
\end{align}
and $R_9$ is the radius of the $x^9$ direction.  Also,
$dx_{034567}^2=-(dx^0)^2+(dx^3)^2+\dots+ (dx^7)^2$ and
$dx_{128}^2=(dx^1)^2+(dx^2)^2+(dx^8)^2$.  The labeling of the
coordinates is slightly perverse for later convenience.
This solution preserves half of
supersymmetry.
In order to be able to $T$-dualize along a transverse direction, let us
compactify $x^8$, which is the same as arraying centers at intervals of
$2\pi \tilde R_8$ along $x^8$.  So,
\begin{align}
H&=1+ \sum_{n\in\bbZ}{R_9\over 2\sqrt{r^2+(x^3-2\pi \tilde R_8 n)^2}}
 \approx 1+\sigma
 \log{\Lambda+\sqrt{r^2+\Lambda^2}\over r},\qquad
\sigma\equiv{R_9\over 2\pi \tilde R_8},
  \label{KKM_harmonic}
\end{align}
where we took a cylindrical
coordinate system
\begin{align}
 ds_{128}^2=dr^2+r^2d\theta^2+(dx^8)^2.\label{hkjo20Mar12}
\end{align} 
We approximated the sum in \eqref{hkjo20Mar12} by an integral and
introduced a cutoff $\Lambda$ to make it convergent.  The approximation
is valid for $r\gg \tilde R_8$.  Such computations of arraying centers were
done in \cite{Sen:1994wr, Blau:1997du}.  We could have done the
summation exactly \cite{Ooguri:1996me} but the above approximation is
sufficient for our purposes.
$H$ in \eqref{KKM_harmonic} diverges as we send $\Lambda\to \infty$, but
this can be formally shifted away by introducing a ``renormalization
scale'' $\mu$ and writing
\begin{align}
 H(r)=h_0+\sigma\log{\mu\over r}
\label{H=h0+log}
\end{align}
where $h_0$ is a ``bare'' quantity which diverges in the
$\Lambda\to\infty$ limit.  When $H$ is given by \eqref{H=h0+log}, eq.\
\eqref{KKM_dilaton_omega} gives $\omega = -\sigma\theta\, dx^8$.  The
log divergence of $H$ implies that such an infinitely long
codimension-2 object is ill-defined as a stand-alone object.  In
physically sensible configurations, this must be regularized either by
taking a suitable superposition of codimension-2 objects
\cite{Greene:1989ya} or, as we will do later, by considering instead a
configuration which is of higher codimension at long distance.  So, the
present analysis should be regarded as for illustration purposes only.

Now let us take $T$-duality along $x^8$.  By the standard Buscher rule,
we obtain the metric and other fields for $5^2_2(34567,89)$:
\begin{align}
   ds^2_{\text{10,str}}&=H \,\, (dr^2+r^2d\theta^2) +HK^{-1}dx_{89}^2+dx_{034567}^2,
 \notag\\
 e^{2\Phi}&=HK^{-1},\qquad
 B^{(2)}=-K^{-1}{\theta \sigma}dx^8\wedge dx^9,\qquad
 B^{(6)}=-H^{-1}K dx^0\wedge dx^3\wedge \cdots dx^7,\notag\\
 K&\equiv H^2+\sigma^2\theta^2.
\label{metric_522}
\end{align}
In terms of the radii in this frame,
\begin{align}
 \sigma={R_8R_9\over 2\pi\ap}.
\end{align}
Such metric of exotic branes has been written down in the literature in
various papers; some early work includes \cite{Blau:1997du,
Meessen:1998qm} and  more recent papers include
\cite{LozanoTellechea:2000mc, Kleinschmidt:2011vu, Bergshoeff:2011se}.
However, it does not appear to have been discussed in the context of
$U$-folds.  Although we arrived at the $5^2_2$ solution by arraying
(smearing) KK monopoles and $T$-dualizing it, which one might find
uncomfortable with, we could have obtained it without arraying by taking
a different route, {\it e.g.}, by starting with a D7-brane metric and
dualizing it \cite{LozanoTellechea:2000mc, Kikuchi:2012za}.  We will
derive the same solution (actually its generalizations) again in section
\ref{ss:susy} more directly in supergravity without using duality.

As can be seen from \eqref{metric_522}, as we go around the
$5^2_2$-brane at $r=0$ by changing $\theta=0$ to $2\pi$, the size of the
8-9 torus does not come back to itself:
\begin{align}
\begin{split}
 \theta=0   &:\quad G_{88}=G_{99}=H^{-1},\\
 \theta=2\pi&:\quad G_{88}=G_{99}={H\over H^2+(2\pi\sigma)^2}.
\end{split}
 \label{T-monodromy_522}
\end{align}
Therefore, indeed, the exotic $5^2_2$-brane has a non-geometric
spacetime around it.  This non-geometric spacetime can be understood as
a $T$-fold as follows.  If we package the 8-9 part of the metric and
$B$-field in a $4\times 4$ matrix \cite{Maharana:1992my}
\begin{align}
 M=\begin{pmatrix}G^{-1} & G^{-1}B \\ -BG^{-1} & G-BG^{-1}B \end{pmatrix}
\end{align}
then the $SO(2,2)$ $T$-duality transformation matrix $\Omega$
satisfying
\begin{align}
 \Omega^t \eta \Omega=\eta,\qquad
	    \eta=\begin{pmatrix} 0 & {\bf 1}_2 \\ {\bf 1}_2 & 0 \end{pmatrix},
\end{align}
acts on $M$ as
\begin{align}
 M\to M'=\Omega^t M\Omega.
\end{align}
It is easy to see that the matrix
\begin{align}
 \Omega=
 \left(\begin{matrix} {\bf 1}_2 &  0 \\ 2\pi \sigma & {\bf 1}_2 \end{matrix}\right)
 \label{522_monodromy_mat}
\end{align}
relates the $\theta=0,2\pi$ configurations in \eqref{T-monodromy_522}.
Namely, $5^2_2$ is a non-geometric $T$-fold with the monodromy $\Omega$.

Another way to represent the monodromy is in terms of the
$SO(2,2,\bbZ)=SL(2,\bbZ)_\tau\times SL(2,\bbZ)_\rho$ mentioned in
section \ref{ss:chg_consv_and_Page_chg}. Let us introduce
moduli $\tau,\rho$  defined in \eqref{hcjy26Sep12}
and take a complex coordinate $z=re^{i\theta}$.  From
\eqref{metric_522}, we can read off
\begin{align}
 \rho={iH-\sigma\theta\over K}={i\over h_0+\sigma\ln(\mu/z)},\qquad
 \tau=i.
\end{align}
So, as we go around $5^2_2$ ($z\to ze^{2\pi i}$), it is not $\rho$ but $\rho'=-1/\rho$ that
undergoes a simple shift:
\begin{align}
 \rho'\to \rho'+2\pi \sigma.\label{nepy12Jun12}
\end{align}
(If $\rho$ underwent a shift $\rho\to\rho+1$ as $z\to ze^{2\pi
i}$ instead, the configuration would be simply an NS5(56789)-brane.)
In terms of the original $\rho$, we have
\begin{align}
 \rho\to {a\rho+b\over c\rho+d},\qquad
 \begin{pmatrix}  a&b\\c&d \end{pmatrix}
 =\begin{pmatrix}  1&0\\-2\pi \sigma&1 \end{pmatrix}.\label{jjbz25Sep12}
\end{align}

Let us study the behavior of the $5^2_2$ metric in the string frame,
\eqref{metric_522}, near $r=0$.  Near $r=0$, the functions in \eqref{metric_522}
behave as
\begin{align}
 H\sim \sigma\ln({\mu/ r}),\qquad K\sim [\sigma\ln({\mu/ r})]^2,
\end{align}
where we absorbed the constant $h_0$ into $\mu$.  Therefore, the $r\to
0$ behavior of the metric is
\begin{align}
   ds^2_{\text{10,str}}&\sim [\sigma \ln ({\mu/r})] (dr^2+r^2d\theta^2)
 +[\sigma \ln ({\mu/r})]^{-1}dx_{89}^2+dx_{034567}^2.\label{meua12Jul12}
\end{align}
Let us introduce a new coordinate $\rho$ by
\begin{align}
 d\rho=\sqrt{\sigma \ln ({\mu/r})}\,dr,
\qquad
 \rho={\sqrt{\pi \sigma}\mu\over 2}\left[1-{\rm erf}\left(\sqrt{\log(\mu/r)}\right)\right]
 +\sqrt{\sigma\log(\mu/r)}\,r,\label{jixu31Jul12}
\end{align}
where
\begin{align}
 {\rm erf}(x)={2\over \sqrt{\pi}}\int_0^x e^{-t^2}dt.
\end{align}
For $r\sim 0$, $\rho\sim 0$, the relation between $r$ and $\rho$ is
simply
\begin{align}
 \rho\sim \sqrt{\sigma\log(\mu/r)}\,r
\end{align}
and therefore the metric \eqref{meua12Jul12} becomes
\begin{align}
   ds^2_{\text{10,str}}&\sim d\rho^2+\rho^2d\theta^2
 +{r^2\over\rho^2}dx_{89}^2+dx_{034567}^2.\label{meua11Jul12}
\end{align}
One sees that the linearly wrapped directions $x^{3,\dots,7}$ remain
finite while the quadratically wrapped directions $x^{8,9}$ shrink at
the position of the brane, $\rho=0$ ($r=0$). On the other hand, the
metric along the transverse directions $(\rho,\theta)$ are actually flat
near the brane.

Similarly, it is easy to show that the Einstein metric in 3D,
\begin{align}
 ds_{\text{3,Ein}}^2=-dt^2+H dx^2_{12},\label{exoticbrane_metric_3Ein}
\end{align}
is also flat at $r=0$ and there is no conical deficit there.  This means
that the mass of the brane is not localized at $r=0$ but is spread over
the space.  We can compute the mass of this configuration
\eqref{exoticbrane_metric_3Ein} by the following {\it ad hoc} procedure,
even though the mass of a codimension-2 object is not strictly
well-defined.  Let $\gamma_{ij}$ be the spatial metric for constant $t$
slices and $G_{\mu\nu}$ the Einstein tensor.  We find that
$\sqrt{\gamma}\,G_0^0 ={1\over 2}\partial_i^2\log H$.  So, the energy is
\begin{align}
 M&
 =-{1\over 8\pi G_3}\int d^2x\sqrt{\gamma}\,G_0^0
 =-{1\over 16\pi G_3}\int dS\cdot \nabla\log H.
\end{align}
If we use \eqref{H=h0+log} and assume that
$H(r=\infty)=1$, then
\begin{align}
 M
 ={1\over 16\pi G_3}\left.\left[ {{2\pi \sigma}\over H(r)}\right]\right|_{r\to \infty}
 ={R_3\cdots R_7 (R_8 R_9)^2 \over g_s l_s^9},\label{svf21Mar12}
\end{align}
as expected of a $5^2_2(34567,89)$.  Here, we used $16\pi
G_3={g_s^2l_s^8/ R_3\cdots R_{9}}$. Although the $5^2_2$ changes the
asymptotics, setting $H(r=\infty)=1$ effectively puts it in an
asymptotically flat space and allows us to compute its mass.

Because the $5^2_2$ background \eqref{metric_522} has non-vanishing NSNS
$B$-field, it is natural to ask if it carries F1 and/or NS5 charges.
First, it does not carry F1 charge, because $H^{(7)}=dB^{(6)}$ has no
purely spatial component.  On the other hand, since $H^{(3)}=dB^{(2)}$
has non-vanishing spatial components, it appears that there is
non-vanishing NS5(34567)-brane charge in this solution.  However, as one
can easily derive from \eqref{metric_522}, $H^{(3)}$ is not single
valued as $\theta\to \theta+2\pi$. Therefore, it does not make sense to
integrate the flux $H^{(3)}$ around the $5^2_2$ to measure the NS5
charge; the integral is not well-defined and its value changes as one
goes around the $5^2_2$.  This state of matter can be understood by
noting that the pair of charges (NS5(34567),\,$5_2^2$(34567,89)) can be
$U$-dualized to (D7(3456789),\,$7_3$(3456789)) in Type IIB\@.  Because
$7_3$ is $S$-dual of D7, the axio-dilaton $\tau=C^{(0)}+ie^{-\Phi}$ behaves
around $7_3$ as $\tau(r,\theta)=-{2\pi i/ \log z}$ where
$z=re^{i\theta}$.  Therefore, around a $7_3$-brane, the RR 0-form is
given by $C^{(0)}=-2\pi \theta/[(\log r)^2+\theta^2]$.  One can define the
1-from flux $G^{(1)}=dC^{(0)}$ and try to define the D7-brane charge by the
integral $\oint G^{(1)}=[C^{(0)}]^{\theta+2\pi}_{\theta}$, but this is
nonsensical because this integral depends on $\theta$.  The only
sensible way to define the charge of $(p,q)$ 7-brane is via the
monodromy matrix \eqref{pq7-monodromy}.
Similarly, the only sensible way to define the (NS5,$5^2_2$) charge is
by the monodromy matrix, which in the current situation is
\eqref{522_monodromy_mat} and implies that there is no NS5 charge. We
will discuss the explicit monodromy matrices of NS5 and $5^2_2$ in
section \ref{ss:susy}.

Because only NSNS fields are excited in the solution \eqref{metric_522},
the $5^2_2$-brane exists in all string theories, including Type I and
heterotic strings. Moreover, the tension of the $5^2_2$-brane is
proportional to $g_s^{-2}$, just like that of the NS5-brane and KK
monopole.  Therefore, the solution \eqref{metric_522} must represent a
legitimate configuration of string theory and give an approximate
description of the physics, much as the supergravity solutions of the
NS5 and KK monopole do (again, with the caveat that it cannot exist as a
stand-alone object). What is interesting about this $5^2_2$ background
is that, because the involved duality monodromy is the perturbative
$T$-duality, it should allow a string theory description in terms of a
worldsheet sigma model.  It would very interesting to find such a sigma
model description.  In particular, to derive the $5^2_2$ metric by
$T$-duality, we used the Buscher rule, which is a valid prescription
only at the supergravity level.  In string theory, the $T$-duality rule
can be corrected by stringy effects and it would be interesting to
examine such effects for $5^2_2$ using sigma model.  In the case of
$T$-duality between NS5 and KKM, it was a non-trivial matter how the
position of the NS5-brane along the direction of $T$-duality is encoded
in the $T$-dual KKM background by worldsheet instanton effects
\cite{Tong:2002rq, Harvey:2005ab, Jensen:2011jn} (see also
\cite{Sen:1997zb, Sen:1997js, Marsano:2007fe}). Exactly the same issue
arises in the $5^2_2$ background also and it would be interesting to
understand it better.

The $5^2_2$-brane is the only exotic brane with mass proportional to
$g_s^{-2}$ and all other exotic branes in \eqref{IIAexotics} and
\eqref{IIBexotics} have mass proportional to $g_s^{-3}$ or $g_s^{-4}$.
For example, formal applications of duality transformations on
\eqref{metric_522} give the following $4^3_3(3456,789)$-brane metric:
\begin{align}
 ds_{\rm str}^2=H^{1/2}K^{1/2}(dr^2+r^2d\theta^2)+H^{-1/2}K^{1/2}dx_{03456}^2+
 H^{1/2}K^{-1/2}dx_{789}^2.\label{metric_433}
\end{align}
$H$ and $K$ are the same ones as given in \eqref{H=h0+log} and
\eqref{metric_522}, except that now
\begin{align}
 \sigma={NR_7R_8R_9\over 2\pi g_s l_s^3}.
\end{align}
Just as we did in \eqref{svf21Mar12}, we can formally show that mass of
this object is
\begin{align}
 M={NR_3\cdots R_6(R_7R_8R_9)^2\over g_s^3 l_s^{11}}.
\end{align}
One may think that such exotic branes with mass $\sim g_s^{-3},g_s^{-4}$
have too large backreaction for supergravity solutions to give a meaningful
description.  However, that is too quick.  One can show
that the $4^3_3$ solution \eqref{metric_433} has no conical deficit at
$r=0$, just as for $5^2_2$.  This means that there is no localized
energy at $r=0$ but, as is clear from the computation leading to
\eqref{svf21Mar12}, the energy of an exotic brane is delocalized and
spread in the surrounding space over a large distance.  What we mean by
the mass of the exotic brane being proportional to $g_s^{-3}$ or
$g_s^{-4}$ is that, if we took metrics such as \eqref{metric_522} and
\eqref{metric_433} at face value and integrated the energy density
distributed over long distances up to $r=\infty$, then the total would
be proportional to $g_s^{-3}$ or $g_s^{-4}$, enough to destroy the
spacetime picture.  This means that the metric of a stand-alone exotic
brane such as \eqref{metric_522} and \eqref{metric_433} should be
thought of as an approximation near $r=0$ and must be replaced at some
large $r$ by some other solution so that the total energy stored in
space is at most $\sim g_s^{-2}$.  This is precisely what one does in
F-theory \cite{Vafa:1996xn}, where one considers a configuration of 24
$(p,q)$ 7-branes to stop the space from extending to
$r=\infty$. Instead, the transverse space terminates at a finite
distance and becomes a compact $S^2$.  Note that a $(p,q)$ 7-brane is
nothing but a bound state of D7-branes and $7_3$-branes with mass $\sim
g_s^{-3}$.  Nevertheless, the configuration has finite energy because
the space is now finite; actually, the size of the transverse $S^2$ is
a modulus and can be arbitrarily large.
This clearly shows that it is too quick to regard exotic branes with
nominal tension $\sim g_s^{-3},g_s^{-4}$ as physically irrelevant.
To examine the physics of such heavy exotic branes, one possibility is
to extend the framework of the original F-theory to geometrize the
moduli space of the $U$-duality group and study the geometry of the
extended spacetime, {\it i.e.}\ the moduli space fibered over the
physical spacetime.  As mentioned in section
\ref{ss:exotic_branes,F-theory,U-branes}, this direction has been
already undertaken in \cite{Kumar:1996zx} followed by a spur of activity
\cite{Liu:1997mb, Curio:1998bv, Leung:1997tw, Lu:1998sx,
Vegh:2008jn}. It would be desirable to revisit this with the improved
understanding of exotic branes provided in the current paper.

Even if the problem of the superficial mass being proportional to
$g_s^{-3}$, $g_s^{-4}$ can be evaded, it should be noted that, as
exemplified by the $(p,q)$ 7-branes of F-theory, the monodromy of exotic
branes involve $U$-duality and therefore the string coupling cannot
generally be made small at all points in spacetime.  In such cases, we
should regard the supergravity solution as a qualitative guide for the
physics at best. However, for protected BPS quantities, they should
probably give precise predictions.

%
%

\subsection{Supersymmetry analysis of $5^2_2$ solution}
\label{ss:susy}

Being dual to the KKM solution which preserves half of supersymmetry,
the exotic $5^2_2$ solution \eqref{metric_522} should also preserve half
of supersymmetry.  It is an instructive exercise to see how this works.
Because of the non-trivial duality monodromy, the Killing spinor is not
single-valued around an exotic brane.  Supersymmetry analyses of
non-geometric solutions have already appeared in the literature; our
purpose here is to only illustrate how supersymmetry is compatible with
exotic $U$-duality monodromies.  For example, an essentially identical
but somewhat less general analysis of the supersymmetry of the
$5^2_2$-brane solution was done in \cite{LozanoTellechea:2000mc}.  Ref.\
\cite{Hellerman:2002ax} provides a supersymmetry analysis of this system
in a more general setup but from a different perspective from ours.

In Type IIA/B supergravity with purely NS background fields, the
supersymmetry transformation for dilatino $\lambda$ and gravitino
$\Psi_M$ is, respectively \cite{Grana:2002tu, Bena:2002kq},
\begin{align}
 \delta \lambda&=\left[{1\over 2}\slash{\partial}\Phi-{1\over 4}\slash{H}\sigma^3
 \right]\epsilon,\qquad
 \delta \Psi_M=
 \left[
 \nabla_M 
-{1\over 8 }\Gamma^{NP}H_{MNP}\sigma^3\right]\epsilon.\label{dmp9Jun12}
\end{align}
Here, the supersymmetry transformation parameter
$\epsilon=\left(\begin{smallmatrix}\epsilon_1\\
\epsilon_2\end{smallmatrix}\right)$ is a doublet of Majorana--Weyl
spinors $\epsilon_{1,2}$ with appropriate chirality (see Appendix
\ref{app:10D-8D})\@. The Pauli matrices such as $\sigma^3$ in
\eqref{dmp9Jun12} acts on the doublet index. For our convention, see
Appendix \ref{app:conv}\@.

Although the $5^2_2$ solution we obtained by dualizing a known solution
is \eqref{metric_522}, it is instructive to study the supersymmetry of
the following more general configuration:
\begin{align}
\begin{split}
  ds_{10}^2&=f^2\eta_{\mu\nu}dx^\mu dx^\nu
 +g^2 \delta_{ij}dx^i dx^j+h^2\delta_{ab} dx^adx^b,\qquad
 H^{(3)}=H_i dx^i\wedge dx^8\wedge dx^9,
\\
 \mu,\nu&=0,3,4,5,6,7,\qquad
 i,j=1,2,\qquad a,b=8,9.
\end{split}
 \label{jhaw9Jun12}
\end{align}
We assume that $f,g,h,H_i$ and $\Phi$ are functions of $x^i$.    If we take the vielbein to
be
\begin{align}
 e^{\hat{\mu}}_\nu =f \delta^\mu_\nu,\qquad
 e^{\hat{\imath}}_j =g \delta^i_j,\qquad
 e^{\hat{a}}_b =h \delta^a_b,
\end{align}
then the non-vanishing components of the spin connection are
\begin{align}
 \omega_{\mu\hat \nu \hat{\imath}}=
 g^{-1}\eta_{\mu\nu}\partial_i f,\qquad
 \omega_{a\hat b \hat{\imath}}=
 g^{-1}\delta_{ab}\partial_i h,\qquad
 \omega_{i\hat{\jmath} \hat k}=
 g^{-1}(\delta_{ij}\partial_k g-\delta_{ik}\partial_j g).
\end{align}
For this configuration, the supersymmetry variation \eqref{dmp9Jun12} 
becomes
\begin{align}
 \delta \lambda&=
 \left[{1\over 2}\Gamma^i\partial_i \Phi-{1\over 4}H_i \Gamma^{i89}\sigma^3\right]\epsilon,
 &\delta \Psi_\mu &=\left[{1\over 2}\eta_{\mu\nu}f\partial_i f \Gamma^{\nu i}\right]\epsilon,\notag\\
\delta \Psi_a&=\left[
 {1\over 2}h\partial_i h \Gamma^{ai}
 -{1\over 4}\epsilon_{ab}H_i\Gamma^{bi}\sigma^3
 \right]\epsilon,&
\delta \Psi_i&=\left[
 \partial_i+{1\over 2}\epsilon_{ij}g\partial_j g\, \Gamma^{12}
 -{1\over 4}H_i \Gamma^{89}\sigma^3
 \right]\epsilon.
\label{gwue8Jun12}
\end{align}
We assumed that $\epsilon$ depends only on $x^i$.  $\epsilon_{ij}$ and
$\epsilon_{ab}$ are antisymmetric symbols with
$\epsilon_{12}=\epsilon_{34}=1$.  For the configuration
\eqref{jhaw9Jun12} to be supersymmetric, there should exist $\epsilon$
for which all of \eqref{gwue8Jun12} vanish.  First, in order that
$\delta \Psi_\mu=0$, we see that $\partial_if=0$ for all $i$ and
therefore we can take $f=1$ by an appropriate rescaling of $x^\mu$.

Next, let us look at the condition $\delta \lambda=0$, which can be
written as
\begin{align}
 \Gamma^{\hat{\imath}}[2h^2\partial_i \Phi-H_i \Gamma^{\hat{8}\hat{9}}\sigma^3]\epsilon=0.\label{feui9Jun12}
\end{align}
It is not difficult to see that this can be rewritten as  a projection condition
\begin{align}
 (1+P)\epsilon=0, \qquad
P^2=1,\label{feuw9Jun12}
\end{align}
if $\Phi$ and $h$ are related by
\begin{align}
 H_i=\pm
 2\epsilon_{ij}h^2\partial_j \Phi.\label{gyyh8Jun12}
\end{align}
The matrix $P$ is given explicitly as
\begin{align}
 P&=\pm
 \Gamma^{\hat{1}\hat{2}\hat{8}\hat{9}}\sigma^3.
\end{align}
Since $\tr P=1$, the condition \eqref{feuw9Jun12} annihilates exactly one
half of the components of $\epsilon$.  

Because we want a 1/2-BPS configuration, the remaining conditions
$\delta\Psi_a=0$, $\delta\Psi_i=0$ must give no additional constraint on
the spinor $\epsilon$.  The $\delta\Psi_a=0$ condition can be easily
seen to reduce to the same condition \eqref{feuw9Jun12} if we set
\begin{align}
 h^{-1}\partial_i h=\partial_i\Phi,\qquad
 \text{therefore}\qquad e^\Phi=h.\label{jdls3Sep13}
\end{align}
Finally, for the $\delta \Psi_i$ condition \eqref{gwue8Jun12} to give no
additional constraint on $\epsilon$, we must set
\begin{align}
 \epsilon=\exp\left[t(x^i)\,\Gamma^{\hat{1}\hat{2}}\right]\epsilon_0,
\end{align}
where $\epsilon_0$ is a constant spinor satisfying
\begin{align}
  (1+P)\epsilon_0=0,
\end{align}
so that \eqref{gwue8Jun12}  becomes
\begin{align}
\delta \Psi_i&=\Gamma^{\hat{1}\hat{2}}\left[
 \left(\partial_i t+{1\over 2}\epsilon_{ij}\partial_j \log g\right)
 +{1\over 4}h^{-2}H_i \Gamma^{\hat{1}\hat{2}\hat{8}\hat{9}}\sigma^3
 \right]\epsilon.
\end{align}
For this to give the same condition as \eqref{feuw9Jun12}, it should be that
\begin{align}
 \partial_i t+{1\over 2}\epsilon_{ij}\partial_j \log g
 =\pm
 {1\over 4}h^{-2}H_i
 ={1\over 4}\epsilon_{ij}\partial_j \log h,
\end{align}
where in the last equality we used \eqref{gyyh8Jun12},
\eqref{jdls3Sep13}. Therefore,
\begin{align}
 \partial_i t
 ={1\over 2}\epsilon_{ij}\partial_j \log{h\over g}.\label{gzpf8Jun12}
\end{align}
Let us introduce complex coordinates by
\begin{gather}
 z=x^1 \pm ix^2,
 \qquad \zb=x^1 \mp ix^2,
\end{gather}
where the signs are chosen to make the later results simple.  In terms
of $z,\zb$,  \eqref{gzpf8Jun12} can be written as
\begin{align}
 \bar\partial(t\pm ir)=0,\qquad r\equiv{1\over 2}\log{h\over g}.
\end{align}
The solution to this is
\begin{align}
 t\pm ir=\mp{i\over 2}\varphi(z),\qquad
 \text{namely}\quad t=\pm{\varphi_2\over 2},\quad r=-{\varphi_1\over 2},
 \label{hkqi30Jul12} 
\end{align}
where $\varphi=\varphi_1+i\varphi_2$ is a holomorphic function of $z$.
The factor $\mp i/2$ on the right hand side was inserted to make
the later results simple.

The above solution satisfies all field equations provided that $h$
satisfies
\begin{align}
 \p\pb h^2(z,\zb)=0.
\end{align}
Namely, $h^2$ is a real harmonic function and can be written as
\begin{align}
 h^2(z,\zb)={\rho(z)-\rhob(\zb)\over 2i}=\rho_2,
\end{align}
where $\rho(z)$ is a holomorphic function of $z$ and
$\rho=\rho_1+i\rho_2$.  From \eqref{hkqi30Jul12}, this means that
$g^2=e^{2\varphi_1}\rho_2$.

Substituting the above results into \eqref{jhaw9Jun12}, the
configuration that locally preserves half of supersymmetry is
\begin{align}
\begin{split}
 ds^2_{10}&=\eta_{\mu\nu}dx^\mu dx^\nu +\rho_2 e^{2\varphi_1} dzd\zb+ \rho_2(dx^a)^2,\qquad
 e^{2\Phi}={\rho_2},\\
 B^{(2)}&=
 \rho_1\,dx^8\wedge dx^9,\qquad
 B^{(6)}=
 {1\over \rho_2}\,dt\wedge dx^3\wedge\cdots\wedge dx^7,\\
 \epsilon&
 =\exp\left(
 \pm
 {\varphi_2\over 2}\,\Gamma^{\hat{8}\hat{9}}\sigma^3\right)\epsilon_0,
 \qquad (1+P)\epsilon_0=0,\\
\end{split}
\label{mwfz27Jul12}
\end{align}
where $\rho=\rho(z),\varphi=\varphi(z)$ are holomorphic functions, and
we used \eqref{feuw9Jun12} for the expression for $\epsilon$.  However,
in order for this configuration to be globally well-defined and
supersymmetric, we must impose further conditions. To see this, it is
convenient to compactify the 10D theory on $T^2_{89}$ to 8D $\CN=2$
supergravity \cite{Andrianopoli:1996ve}.  The 8D metric in the Einstein
frame is
\begin{align}
 ds^2_{\text{8,Ein}}&=\eta_{\mu\nu}dx^\mu dx^\nu
 +{\rho_2 e^{2\varphi_1}} dz d\zb.\label{cvc29Jul12}
\end{align}
8D $\CN=2$ supergravity has $U$-duality group $SL(3,\bbZ)\times
SL(2,\bbZ)$, which contains the $T$-duality subgroup
$SO(2,2,\bbZ)=SL(2,\bbZ)_\tau\times SL(2,\bbZ)_\rho$.  Associated with
this subgroup are moduli parametrizing $(SO(2)\times SO(2))\backslash
SO(2,2,\bbR)/SO(2,2,\bbZ)=\CM_\tau\times \CM_\rho$, where
$\CM=SO(2)\backslash SL(2,\bbR)/SL(2,\bbZ)$.  The first factor
$\CM_\tau$ corresponds to the complex structure $\tau$ of the torus
$T^2_{89}$, which has been defined in \eqref{hcjy26Sep12} and is fixed
to $\tau=i$ in \eqref{mwfz27Jul12} in the present case.  The second
factor $\CM_\rho$ corresponds to $B_{89}$ and the volume of $T^2_{89}$,
and is parametrized by the complex field $\rho$ defined in
\eqref{hcjy26Sep12}.  This is the same as the $\rho$ introduced above.
The 10D supersymmetry transformation parameter
$\epsilon=\left(\begin{smallmatrix}\epsilon_1\\
\epsilon_2\end{smallmatrix}\right)$ reduces to a pair of 8D Weyl spinors
$\eta_A$, $A=1,2$.  Under a duality transformation, $\eta_A$ will also
transform as we will discuss below.

If there is a codimension-2 exotic brane at $z=z_0$ on the $z$-plane
then, as we move around it on the $z$-plane, there is a non-trivial
duality monodromy $q\in G(\bbZ)$.  Here we are focusing on the
$SL(2,\bbZ)_\rho$ subgroup and hence $q\in SL(2,\bbZ)_\rho$.  Let us
consider a brane at $z=z_0$ with the following
monodromy
\begin{align}
 \label{mhac29Jul12}
 q=\begin{pmatrix} a&b\\c &d\end{pmatrix}
 \in SL(2,\bbZ)_\rho,
 \qquad a,b,c,d\in\bbZ,\qquad ad-bc=1.
\end{align}
As we go around $z=z_0$, all fields must jump according to this
transformation \eqref{mhac29Jul12}.  First, the modulus $\rho(z)$ should
have the monodromy
\begin{align}
 \rho&\to {a\rho+b\over c\rho+d}.\label{mqlg29Jul12}
\end{align}
The 8D spinors $\eta_A$ must have a monodromy corresponding to the same
\eqref{mhac29Jul12}.  Depending on whether we have compactified 10D type
IIA or type IIB, the transformation rule of the 8D spinors $\eta_A$
under \eqref{mhac29Jul12} is different \cite{Andrianopoli:1996ve} and
given by
\begin{align}
 \label{ndez29Jul12}
 \eta_A\to 
 \begin{cases}
  e^{{i\over 2}\arg(c\rho+d)}\eta_A                   &\qquad \text{(type IIA)},\\[1ex]
  [e^{{i\over 2}\arg(c\rho+d)\sigma^3}]_A{}^B\,\eta_B &\qquad \text{(type IIB)}.
 \end{cases}
\end{align}
By examining how the 10D spinor $\epsilon$ reduces to the 8D spinor
$\eta_A$, one can show that this corresponds to the following monodromy
for the 10D spinor $\epsilon$,
\begin{align}
\label{msqq29Jul12}
 \epsilon
 \to
 \exp\left[\pm{1\over 2}\Gamma^{\hat{8}\hat{9}}\sigma^3 \arg(c\rho+d)\right]\epsilon
 ~,
\end{align}
where the $\pm$ signs correspond to type IIA/IIB\@.  For more detail
about \eqref{ndez29Jul12} and \eqref{msqq29Jul12}, see Appendix
\ref{app:10D-8D}\@.  By comparing \eqref{msqq29Jul12} with
\eqref{mwfz27Jul12}, we see that $\varphi_2$ must have the following
monodromy:
\begin{align}
 \varphi_2\to  \varphi_2 +\arg(c\rho+d).\label{gpl29Jul12}
\end{align}
Note that the $\pm$ signs in \eqref{mwfz27Jul12} are now understood to apply for type IIA/IIB\@.

There is another condition: the 8D Einstein frame metric
\eqref{cvc29Jul12} must be invariant under the duality
\eqref{mhac29Jul12}.  This means that $\varphi_1$ must have the
following monodromy:
\begin{align}
 \varphi_1&\to \varphi_1+\log|c\rho+d|.\label{gpj29Jul12}
\end{align}
Combining \eqref{gpl29Jul12} and \eqref{gpj29Jul12}, we see that
$\varphi$ must have the following monodromy:
\begin{align}
 \varphi\to\varphi + \log(c\rho+d),\qquad
 \text{or}\quad
 e^\varphi\to(c\rho+d)e^\varphi.
 \label{mqlj29Jul12}
\end{align}

To summarize, for the solution \eqref{mwfz27Jul12} to be globally
well-defined and supersymmetric, the holomorphic functions
$\rho(z),\varphi(z)$ must satisfy the monodromy conditions
\eqref{mqlg29Jul12}, \eqref{mqlj29Jul12} around a brane with charge
\eqref{mhac29Jul12}.

The $5^2_2$ solution \eqref{metric_522} corresponds to the following
particular choice
\begin{align}
 \rho(z)={i\over \sigma\log{\mu\over z}},
 \qquad e^{\varphi(z)}=\sigma\log{\mu\over z}\label{ndtn12Jun12}
\end{align}
where we absorbed the constant $h_0$ into $\mu$.  At $z=0$, there is the following monodromy:
\begin{align}
 q^{}_{5^2_2}=\begin{pmatrix}1&0\\ -2\pi \sigma & 1\end{pmatrix},\label{jdmc31Jul12}
\end{align}
which already appeared in \eqref{jjbz25Sep12}.
It is easy to show that $\rho,\varphi$ in \eqref{ndtn12Jun12} do have
the monodromy \eqref{mqlg29Jul12}, \eqref{mqlj29Jul12} as $z\to ze^{2\pi
i}$.  On the other hand, the NS5(34567)-brane solution smeared along
$x^{8,9}$ is
\begin{align}
\begin{split}
  ds^2_{\text{10,str}}&=\eta_{\mu\nu}dx^\mu dx^\nu +  H dzd\zb+ H (dx^a)^2,\qquad
 e^{2\Phi}=H,\\
 B^{(2)}&=\sigma' \theta\, dx^8\wedge dx^9,\qquad
 H=\sigma' \log{\mu'\over r},\qquad
 \sigma'={N' \ap\over 2\pi R_8 R_9},
\end{split}
\end{align}
where $\mu'$ is a constant and $N'$ is the number of NS5-branes.  From
this, we read off
\begin{align}
 \rho(z)=i\sigma'\log{\mu'\over z},\qquad e^{\varphi(z)}=1.
\end{align}
This corresponds to the monodromy
\begin{align}
 q_{\text{NS5}}^{}=\begin{pmatrix}1&2\pi\sigma' \\ 0 & 1\end{pmatrix}.\label{jdmf31Jul12}
\end{align}
Comparing the monodromy matrices \eqref{jdmc31Jul12} and
\eqref{jdmf31Jul12} with that of $(p,q)$ 7-branes
\eqref{pq7-monodromy2}, we see that the monodromy matrix $q^{}_{5^2_2}$
is the same as that of $(1,0)$- or D7-brane while $q_{\text{NS5}}^{}$
is the same as that of $(0,1)$- or $7_3$-branes, although
$q^{}_{5^2_2},q^{}_{\text{NS5}}$ are about the $SL(2,\bbZ)_\rho$
$T$-duality while \eqref{pq7-monodromy2} is about the $SL(2,\bbZ)$
$S$-duality of type IIB superstring.  In fact, by a chain of dualities
($S$, $T_{89}$ and then $S$), $5^2_2$ and NS5 are mapped into $7_3$ and
D7, respectively.  So, just as one can consider configurations of
various $(p,q)$ 7-branes in type IIB, we can consider configurations of
branes with general $SL(2,\bbZ)_\rho$ monodromies.

In more general configurations with multiple branes on the $z$-plane,
the holomorphic function $\rho(z)$ is determined by the monodromies
(charges) of the branes.  On the other hand, to determine $\varphi(z)$,
the monodromy condition \eqref{mqlj29Jul12} is not enough and we need to
specify the boundary condition at infinity, which should be chosen
based on the physical situation under consideration.  This is always the
case for codimension-2 branes, which is not well-defined as a
stand-alone object.  For example, the same undetermined function
appears in the context of F-theory \cite{Vafa:1996xn} (see also
\cite{Greene:1989ya}) and one determines it requiring that the
transverse space should close smoothly to $S^2$.  For explicit examples
of and a detailed discussion on how to determine $\varphi$ in the context
of ($p,q$) 7-branes in type IIB, see \cite{Bergshoeff:2006jj}.  We will
see later another example where this freedom is fixed by the boundary
condition at infinity.

An essentially identical analysis of the supersymmetry of the
$5^2_2$-brane solution was done in \cite{LozanoTellechea:2000mc},
although they did not make $\varphi(z)$ arbitrary.\footnote{The solution
\eqref{mwfz27Jul12} reduces to the one in \cite{LozanoTellechea:2000mc}
if we set
\begin{align}
 \rho=-\CH^{-1},\qquad e^{\varphi}=\CH.
\end{align}
in their notation.  The $5^2_2$-brane here is called the $S5_2$-brane
there.}  They also discussed supersymmetry of other exotic
branes,\footnote{The relation between their notation and ours is:
$Dp_{7-p}^{}=p_3^{7-p}$, $S5_2=5^2_2$, $F1_6=1^6_4$, $W_6=0_4^{(1,6)}$;
$M2_6=2^6$, $M5_3=5^3$, $WM_7=0^{(1,7)}$.  } which are all related to
$U$-duality to $5^2_2$, and have explicitly written down the
supersymmetry projector for each of them.  The monodromy condition on
the Killing spinors for these solutions must work just the same way as
for the (NS5,$5^2_2$) solution above, although we do not try to check it
here.
Ref.\ \cite{Hellerman:2002ax} gave a more general supersymmetry analysis
of the system allowing both $\tau$ and $\rho$ to vary but our discussion
above is more focused on the monodromic structure of the solution.

\subsection{Metrics for other exotic branes}

It is straightforward to derive the metric for other exotic branes
appearing in Table \ref{table:exotic_states}\@.  As discussed above,
they must give approximate descriptions of exotic branes near its core.
In the previous subsection, we discussed the metric of $5^2_2$, a unique
exotic brane in string theory with tension $\sim g_s^{-2}$.  Here, as
examples, let us discuss exotic branes in M-theory in some details.
Again, such exotic metrics have been written down \cite{Blau:1997du,
LozanoTellechea:2000mc, Kleinschmidt:2011vu, Bergshoeff:2011se}, but we
discuss them from a different perspective.

We represent the $x^{10}$ direction by ``\ten''.

\subsubsection*{$\boldsymbol{5^3(34567,89\ten)}$}\vspace*{-1.5ex}
The metric and form fields in 11D are given by
\begin{align}
\label{kmnw18Feb10}
\begin{split}
  ds_{11}^2&=
 H^{2/3}K^{1/3}(dr^2+r^2d\theta^2)
 +H^{-1/3}K^{1/3}dx_{034567}^2 +H^{2/3}K^{-2/3}dx_{89\ten}^2\\
 A^{(3)}&=-K^{-1}{\theta\sigma}\,dx^8\wedge dx^9\wedge dx^{\ten},\qquad
 \sigma={N R_8 R_9 R_{\ten}\over 2\pi l_{11}^3},\\
 A^{(6)}&=-H^{-1}K\,dt\wedge dx^3\wedge \dots\wedge dx^7\\
 F^{(4)}&=dA^{(3)}=-\sigma K^{-2}\left[{2\sigma \theta H}{dr\over r}+(H^2-\sigma^2\theta^2)d\theta\right]
 \wedge dx^8\wedge dx^9\wedge dx^{\ten}\\
 F^{(7)}&=*F^{(4)}=\sigma H^{-2}
 \left[{2\sigma \theta H}d\theta-(H^2-\sigma^2\theta^2){dr\over r}\right]
 \wedge dt\wedge dx^{3}\wedge\cdots\wedge dx^{7}
\end{split}
\end{align}
Here, $H$ and $K$ are the same ones as given in \eqref{H=h0+log} and
\eqref{metric_522}, except that $\sigma$ is now as given above.

Just as we discussed before in the case of $5^2_2$, one cannot measure
the M5(34567)-charge based on the integral of $F^{(4)}$.  The pair of
charges (M5(34567),\,$5^3$(34567,89\ten)) is $U$-dual to
(D7(3456789),\,$7_3$(3456789)), for which one cannot use the integral of
form fields to define charge.  Again, one should instead look at the
monodromy to define charges.

The behavior of the metric \eqref{kmnw18Feb10} for $r\to 0$ is
\begin{align}
 ds_{11}^2\sim [\sigma\ln(1/r)]^{4/3}(dr^2+r^2d\theta^2)+[\sigma\ln(1/r)]^{1/3}dx_{034567}^2
 +[\sigma\ln(1/r)]^{-1/3}dx_{89\ten}.
\end{align}
As $r\to 0$, the quadratically wrapped directions $x^{8,9,\ten}$ shrink
to zero while the linearly wrapped directions $x^{0,3,\dots,7}$ blow up.
This is in contrast with ordinary branes (M2- and M5-branes) which
shrink the wrapped directions.  One can show that the $(r,\theta)$ part
of the metric is flat at $r=0$, just as we did around
\eqref{jixu31Jul12}--\eqref{meua11Jul12}.

The Ricci scalar is
\begin{align}
 \CR={1\over 6r^2H^{8/3}K^{1/3}}\sim {1\over \sigma^{4/3}r^2[\log(\mu/ r)]^{10/3}}
 \qquad (r\sim 0),
\end{align}
which blows up as $r\to 0$.  The value of $r$ where the Ricci scalar
becomes of the Planck scale is estimated as
\begin{align}
 \CR\sim l_{11}^{-2}\qquad \Longrightarrow \qquad 
 r\sim \sigma^{-2/3}l_{11}\sim N^{-2/3}l_{11},
\end{align}
where we assumed that $R_8,R_9,R_\ten\sim l_{11}$.  Therefore, by making
$N$ large, we can make the supergravity description valid down to very
small value of $r$.

\subsubsection*{$\boldsymbol{2^6(34,56789\ten)}$}\vspace*{-1.5ex}
The metric and form fields are
\begin{align}
 ds^2_{11}&=
 H^{1/3}K^{2/3}(dr^2+r^2d\theta^2)
+H^{-2/3}K^{2/3}dx_{034}^2
 +H^{1/3}K^{-1/3}dx_{56789\ten}^2\\
 A^{(3)}&=-H^{-1}K\, dt\wedge dx^3\wedge dx^{4},\qquad
 \sigma={N R_5\cdots R_{\ten}\over 2\pi l_{11}^6}
\end{align}
For this solution, one cannot measure the M2(34) charge based on the
integral of $F^{(7)}$.  As $r\to 0$, the linearly wrapped directions
$x^{3,4}$ blow up, while the quadratically wrapped directions
$x^{5,6,\dots,\ten}$ shrink.  The $(r,\theta)$ part of the metric is
flat at $r=0$. The behavior of the Ricci scalar is qualitatively similar
to that for $5^3$ and, if $N$ is large, supergravity description is good
down to small $r$.

\subsubsection*{$\boldsymbol{0^{(1,7)}(\,,3456789,\ten)}$}\vspace*{-1.5ex}
The metric and form fields are
\begin{align}
 ds^2_{11}&=
 K(dr^2+r^2d\theta^2)-H^{-1}K dt^2
 +dx_{3456789}^2
 +HK^{-1}(dx^{\ten}-KH^{-1}dt)^2\\
 A^{(3)}&=0,\qquad
 \sigma={N R_3\cdots R_9 R_{\ten}^2\over 2\pi l_{11}^9}
\end{align}
For this solution, one cannot measure the $\rm P(\ten)$ charge (momentum
along $x^{\ten}$) based on $g_{\ten\mu}$.  As $r\to 0$, the linearly
wrapped direction $t=x^0$ blows up, quadratically wrapped directions
$x^{3,4,\dots,9}$ remain finite, while the cubically wrapped direction
$x^{\ten}$ shrinks.  The $(r,\theta)$ part of the metric is flat at
$r=0$.  This solution is purely metrical and the Ricci flat.

\bigskip
Although we presented supergravity solutions with one stack of exotic
branes in the above, it is straightforward to work out exotic solutions
with more than one stack by dualizing known solutions.  For example, if
we start from the D1(5)+D5(56789) system, take $S$, $T_{34}$-dualities
and lift it to 11 dimensions, one can obtain the solution for
M2(34)+$5^3(56789,34\ten)$.  In the next section, we will consider more
complicated solutions involving exotic and standard charges at the same
time.

\section{Supertube effect and exotic branes}
\label{sec:supertube_effect_and_exotic_branes}

%
%

\subsection{Exotic supertube effects}

As we discussed in the introduction, the supertube effect
\cite{Mateos:2001qs} is a spontaneous polarization phenomenon that
occurs when a particular combination of brane charges are put
together. For example, as we saw in \eqref{supertube_D0+F1}, if D0s and
F1(1)s are put together, they polarize, or ``puff up,'' into a
D2(1$\psi$)-brane along a closed but arbitrary curve $\psi$.  It is
important that this D2-brane represents a genuine bound state of the
system, not just a non-interacting superposition of D0s and F1s.
Although the D2-brane did not exist in the original configuration, it
does not violate charge conservation because D2 is only a dipole and
there is no net D2 charge.

By taking duals of the original supertube effect
\eqref{supertube_D0+F1}, one can derive other possible polarization
phenomena.  For example,
\begin{align}
 \rm F1(1)+P(1)     &\to \rm f1(\psi)+p(\psi),\label{hkns21Mar12}\\
 \rm D1(1)+D5(12345)&\to \rm kkm(2345\psi,1)+p(\psi),\label{hlni21Mar12}\\
 \rm M2(12)+M2(34)  &\to \rm m5(1234\psi)+p(\psi).\label{hlvd21Mar12}
\end{align}
The first one \eqref{hkns21Mar12} is the so-called F-P system or the
Dabholkar--Harvey system \cite{Dabholkar:1989jt}.  This is perhaps the
duality frame in which it is easiest to understand why the spontaneous
polarization occurs in the first place.  If one takes an F1 string along
$x^1$ and add momentum along the same direction, then the F1 should
oscillate in the transverse direction, because the F1 worldvolume does
not have longitudinal oscillation modes.  That is why the system puffs
up in the transverse directions.  The second one \eqref{hlni21Mar12} is
the so-called D1-D5 system and the puffed-up configuration of the KKM is
nothing but the Lunin--Mathur geometries \cite{Lunin:2001jy,
Lunin:2002iz} that played an essential role in Mathur's conjecture
\cite{Mathur:2005zp,
Bena:2007kg, Skenderis:2008qn, Balasubramanian:2008da, Chowdhury:2010ct}
The last one \eqref{hlvd21Mar12} was the basic process for the
construction of supersymmetric black rings \cite{Bena:2004wv,
Elvang:2004rt, Gauntlett:2004wh, Bena:2004de}.

In the above, we considered polarization processes involving only
ordinary branes.  However, it is easy to find ones with exotic branes.
For example, by $T$-dualizing \eqref{hlni21Mar12} along $236$ directions
and relabeling coordinates, we obtain
\begin{align}
 \rm  D4(6789)+D4(4589)\to 5^2_2(4567\psi,89)+p(\psi).\label{hoqz21Mar12}
\end{align}
The configuration on the left can be thought of as a pointlike
configuration in asymptotically flat 4D spacetime, which puffs up into
an extended configuration of an exotic dipole charge along a curve
$\psi$ in $\bbR^3_{123}$ on the right hand side.  Such exotic dipole
charges do not change the asymptotics of spacetime. Note that the
original configuration of D4-branes is part of the standard D0-D4
configuration used for the black hole microstate counting in 4D
\cite{Maldacena:1997de}. So, to understand the physics of such black
holes, it is unavoidable to consider exotic charges.

If we want more exotic mass by the supertube effect, we can for example
apply $T_6$ and $S$ dualities to \eqref{hoqz21Mar12} to get an object
with mass $\sim g_s^{-3}$:
\begin{align}
  \rm D3(789)+NS5(45689)\to 5^2_3(4567\psi,89)+p(\psi).
\end{align}
If we further perform $T_{4567}$ and $S$, we get a $g_s^{-4}$ object:
\begin{align}
  \rm D5(45689)+KKM(45689,7)\to 1^6_4(\psi,456789)+p(\psi).
\end{align}

As one can guess from the above examples, the general rule for the
tension of the product brane can be written schematically as
\begin{align}
 g_s^{-a} +  g_s^{-b} \to g_s^{-(a+b)}.\label{ickt17Jun12}
\end{align}
Also,  the radius of the puffed up configuration depends on $g_s$ as
\begin{align}
 R\sim g_s^{a+b\over 2}.\label{ickz17Jun12}
\end{align}
The relations \eqref{ickt17Jun12}, \eqref{ickz17Jun12} can be derived as
follows.  For example, take the FP system \eqref{hkns21Mar12}, where the
two original charges have $M_1=M_{F1(1)}\sim {R_1/\alpha'}$,
$M_2=M_{P(1)}\sim {1/ R_1}$, where $R_1$ is the radius of the $x^1$
direction.  The tension of the product brane, $F1(\psi)$, is $T\sim
{1/\alpha'}$, while the size of the puffed up configuration is
$R\sim\ap{}^{1/2}$ \cite{Lunin:2001fv}.  If we go to a duality frame
where the same charges are expressed as $M_1\sim \tilde g_s^{-a}$,
$M_2\sim \tilde g_s^{-b}$, with $\tilde g_s$ the string coupling in the
new frame, then the original $\ap,R_1$ are expressed as $\ap\sim \tilde
g_s^{a+b},R_1\sim \tilde g_s^{b}$.  So, the tension in the new frame in
terms of $\tilde g_s$ is $T\sim 1/\ap\sim \tilde g_s^{-a-b}$ and the
size of the system is $R\sim\ap{}^{1/2}\sim\tilde g_s^{(a+b)/ 2}$.

The cautious reader should have noticed that the mass of an object with
tension $T\sim g_s^{-a-b}$ extending along a distance $R\sim
g_s^{(a+b)/2}$ does not reproduce the mass of the original object,
$M_1+M_2\sim g_s^{-a}+g_s^{-b}$.  This is because, in the puffed-up
configuration, it is not precisely an object with tension $T\sim
g_s^{-a-b}$ that is extending along $\psi$.  Instead, it is the
combination of the puffed-up charge and the original charges before
puffing up.  For example, in the F-P frame, it is the fundamental string
that wraps the $x^1$ and $\psi$ directions simultaneously and is moving
in the transverse direction (see {\it e.g.}\ the appendix of
\cite{Bena:2011uw} for a more detailed explanation).  One can show that,
in the general duality frame, the tension of this combined object is
schematically $T'\sim g_s^{-3a/2-b/2}+g_s^{-a/2-3b/2}$, which reproduces
the original mass $T'R\sim g_s^{-a}+g_s^{-b}$.

In \eqref{supertube_D0+F1}, the $\psi$ direction of the puffed-up
D2-brane can be an arbitrary curve in the eight transverse directions
$x^2,\dots, x^9$.  Generically, the $\psi$ curve is non-trivial in all
eight directions.  If one dualizes such generic configurations just as
we did above, the puffed-up branes will have extra dipole charges
dissolved in the worldvolume. For example, the more generic puffing-up
of the D1-D5 system \eqref{hlni21Mar12} can be derived as follows.
Consider compactifying the 12345 directions.  From
\eqref{supertube_D0+F1}, following puffing-up is possible:
\begin{align}
\rm D0+F1(1)
 \quad \to\quad
 \begin{array}{l@{\,}c@{\,}l}
 \rm d2(1\psi)& + & \rm p(\psi)   \\
 \rm d2(12)   & + & \rm p(2)     \\
 \rm d2(13)   & + & \rm p(3)\\
 \rm d2(14)   & + & \rm p(4)  \\
 \rm d2(15)   & + & \rm p(5)  \\
 \end{array}.
\label{eq:D0F1d2_more_general}
\end{align}
This diagram is understood as follows.  The first line on the right
means that, after puffing up, we have a dipole D2-brane which extends
along $x^1$ as well as an arbitrary closed curve $\psi$ in the
non-compact 6789 directions.  As we move along $\psi$, the D2-brane can
also move in the internal $x^2$ direction, which amounts to having local
d2(12) charge; this is the meaning of the second line.  Similarly, the
remaining lines mean that the D2-brane can move in the 345 directions.
As one circumnavigates the closed $\psi$, the D2-brane must not have a
net winding number along any of the 2345 directions, because the
original configuration did not have any D2 charge.  All this is a
complicated way to say that the D0-F1 system puffs up into an arbitrary
curve in the transverse 23456789 directions, without having net winding
number along the compact 12345 directions.  From the viewpoint of the
non-compact 06789 directions, we have a 1-brane along $\psi$, with four
different kinds of charge density varying along $\psi$.

The advantage of the above way of writing the puffing-up pattern is that
it is easier to take duality transformations.  If we $T$-dualize
\eqref{eq:D0F1d2_more_general} in the 234 directions, we obtain
\begin{align}
\rm D3(234)+F1(1)
 \quad \to\quad
 \begin{array}{l@{\,}c@{\,}l}
 \rm d5(1234\psi)& + & \rm p(\psi)   \\
 \rm d3(134)     & + & \rm f1(2)     \\
 \rm d3(124)     & + & \rm f1(3)\\
 \rm d3(123)     & + & \rm f1(4)  \\
 \rm d5(12345)   & + & \rm p(5)  \\
 \end{array}.
\end{align}
By further applying $S$ and then $T_{15}$, we obtain the general puffing-up
pattern of the D1-D5 system as follows:
\begin{align}
\rm D5(12345)+D1(5)
 \quad \to\quad
 \begin{array}{l@{\,}c@{\,}l}
 \rm kkm(1234\psi,5)& + & \rm p(\psi)   \\
 \rm d3(345)        & + & \rm d3(125)\\
 \rm d3(245)        & + & \rm d3(135)  \\
 \rm d3(235)        & + & \rm d3(145)  \\
 \rm ns5(12345)     & + & \rm f1(5)     \\
 \end{array}.
\end{align}
This diagram is understood as follows: the general configuration is a
KKM along the $\psi$ curve as displayed in the first line on the right,
with four other dipole charges listed in the subsequent four lines
dissolved in the worldvolume of the KKM as fluxes.  These fluxes are in
addition to the ones that induce the original charges on the left.  Note
that, in order to account for the microscopic degeneracy of the 2-charge
D1-D5 system, it is crucial to have 8 arbitrary functions worth of
possible configurations.  Therefore, not only the 4 functions associated
with the $\psi$ curve but also 4 dissolved dipole charges are important
for reproducing the correct microscopic entropy, including the numerical
factor.\footnote{This is as far as bosonic degrees of freedom are
concerned; in order to reproduce the degeneracy including fermionic
ones, one should consider fermionic excitations in addition to these
bosonic ones. See \cite{Taylor:2005db} for an related attempt.}  In the
original Lunin--Mathur geometries \cite{Lunin:2001jy, Lunin:2002iz},
these four extra dipole charges are turned off, which were later
included in \cite{Kanitscheider:2007wq}.

Similarly, the more general puffing-up of the M2-M2 system \eqref{hlvd21Mar12} is
\begin{align}
\rm M2(12)+M2(34)
 \quad \to\quad
 \begin{array}{l@{\,}c@{\,}l}
 \rm m5(1234\psi)& + & \rm p(\psi)\\
 \rm m2(13)      & + & \rm m2(24)   \\
 \rm m2(14)      & + & \rm m2(23)     \\
 \end{array},
\end{align}
where the $1234$ directions are compact and $\psi$ is an arbitrary curve
in the non-compact $56789\ten$ directions.  The general configuration is
an M5 along the $\psi$ curve in the first line on the right, with two
other dipole charges dissolved in its worldvolume as written in the last
two lines.

Finally, the exotic puffing-up \eqref{hoqz21Mar12} should more generally be
\begin{align}
\label{D4D4-522_general}
\rm D4(6789)+D4(4589)
 \quad \to\quad
 \begin{array}{l@{\,}c@{\,}l}
 \rm 5^2_2(4567\psi,89) & + & \rm p(\psi)   \\
 \rm kkm(45678,9)       & + & \rm f1(8)     \\
 \rm kkm(45679,8)       & + & \rm f1(9)     \\
 \rm d2(89)             & + & \rm d6(456789)\\
 \rm d4(4789)           & + & \rm d4(5689)  \\
 \rm d4(5789)           & + & \rm d4(4689)  \\
 \end{array},
\end{align}
where the $456789$ directions are compact and $\psi$ is an arbitrary curve in the
non-compact 123 directions.  The general configuration is a $5^2_2$
brane along the $\psi$ curve in the first line on the right, with five
other dipole charges dissolved in its worldvolume as written in the
subsequent five lines.

It would be interesting to study a description of such supertubes in
terms of the worldvolume action of the highest dimensional brane.  Some
analysis of the worldvolume action of codimension-2 branes can be found
in \cite{Eyras:1999at, Bergshoeff:2012ex}.

\subsection{Supergravity solution for an exotic supertube}

To demonstrate the idea of exotic branes spontaneously generated via the
supertube effect out of standard branes, let us study the supergravity
solution corresponding to the puff-up \eqref{hoqz21Mar12} where the
$5^2_2$-brane dipole charge is produced from two stacks of D4-branes (we
do not consider the more general case \eqref{D4D4-522_general}).  A
simplified version of the following discussion has appeared in
\cite{deBoer:2010ud}.

As we saw above, the desired exotic supertube \eqref{hoqz21Mar12} can be
obtained by dualizing  standard (non-exotic) supertubes.  As the initial
configuration, let us take the F-P system \eqref{hkns21Mar12}.  More
precisely,  Consider Type IIB superstring in $\bbR_t\times
\bbR^{3}\times S^1\times T^{5}$, and denote the coordinates of
$\bbR_t,\bbR^{3},S^1$, and $T^{5}$ by $t$, ${\bf x}=x^i=(x^1,x^2,
x^{3})$, $x^4$, and ${\bf z}=(x^{5},\dots,x^9)$, respectively.  Let the
radius of $S^1$ be $R_4$. In this setup,  wind an F1 string $N_1$
times along $S^1$ and put $N_2$ units of momentum along $S^1$.  This
system undergoes the supertube transition
\begin{align}
 {\rm F1}(4)+{\rm P}(4)\to {\rm f1}(\psi)+ {\rm p}(\psi),\label{iwfq11Jun12}
\end{align}
where $\psi$ is a curve in the non-compact $\bbR^3$.

After such puffing up, the corresponding backreacted solution in
supergravity is \cite{Gauntlett:2002nw, Lunin:2001fv}
\begin{align}
\begin{split}
  ds_{\text{str}}^2&=f_1^{-1}\left[-(dt-A)^2+(dx^4-A)^2+(f_2-1)(dt-dx^4)^2\right]
 +d{\bf x}^2+d{\bf z}^2,\\
 e^{2\Phi}&=f_1^{-1},\qquad B^{(2)}=-(f_1^{-1}-1)dt\wedge dx^4+ f_1^{-1}(dt-dx^4)\wedge A,
\end{split}
\label{FPsys}
\end{align}
where $A=\sum_{i=1}^3 A_i dx^i$.  The functions $f_1,f_2$ and 1-form $A$
are functions of the transverse coordinates ${\bf x}$ defined by
\begin{align}
\label{kjkv25Mar10}
\begin{split}
  f_1&=1+{Q_1\over L}\int_0^L {dv\over |{\bf x}-{\bf F}(v)|},\qquad
 f_2=1+{Q_1\over L}\int_0^L {|\dot{\bf F}(v)|^2 \over |{\bf x}-{\bf F}(v)|}dv,
 \\
 A_i&=-{Q_1\over L}\int_0^L {\dot F_i(v)dv\over |{\bf x}-{\bf F}(v)|},
\end{split} 
\end{align}
where $L\equiv 2\pi w R_4$ and $\dot{\,}=\partial_v$.  The function
${\bf F}(v)=(F_1(v),F_2(v),F_3(v))$ is an arbitrary function
parametrizing the curve along which the puffed-up F1 is extended (the
right hand side of \eqref{iwfq11Jun12}); we call ${\bf F}(v)$ the
profile function. Because the curve is closed, it is periodic: ${\bf
F}(0)={\bf F}(L)$. We also define
\begin{align}
  Q_2={Q_1\over L}\int_0^L {|\dot{\bf F}(v)|^2 }dv.
\end{align}

The functions \eqref{kjkv25Mar10} would logarithmically diverge if the
profile extended over an infinite distance.  However, because the
profile is a finite closed curve, these functions are finite, except on
the profile, ${\bf x}={\bf F}(v)$.
Near the profile, these functions diverge as
\begin{align}
 f_1\sim {2Q_1\over L|\dot{\bf F}|}\log\left({L_1\over \rho}\right),\quad
 f_2\sim {2Q_1|\dot{\bf F}|\over L}\log\left({L_2\over \rho}\right),\quad
 A\sim -{2Q_1 \dot{F}_\xi\over L|\dot{\bf F}|}\log\left({L_\ten\over \rho}\right)\,d\xi,\label{jjnv12Jun12}
\end{align}
where $\rho\to 0$ is the distance from the profile, $\xi$ is the
coordinate along the profile, and $|\dot{\bf F}|,\dot F_\xi$ are
evaluated at the point that we are zooming in onto.  $L_1,L_2,L_\ten$
are some distance scales of the order of the size of the profile, whose
precise values depend on the detail of the profile.

The functions defined in \eqref{kjkv25Mar10} satisfy $\Box f_{1,2}=\Box
A_i=\sum_{i=1}^3 \partial_i A_i=0$ away from the profile ${\bf x}={\bf
F}(v)$, where $\Box=\sum_{i=1}^3\partial_i^2$.  This means that
\begin{align}
 d{*_3 df_I}=d{*_3 dA}=0,\qquad I=1,2,
\end{align}
where $*_3$ is the Hodge star for the flat $\bbR^3$.  So, we can define
a scalar $\gamma$ and 1-forms $\beta_{1,2}$ by
\begin{align}
 d\gamma=*_3 dA,\qquad  d\beta_I=*_3 df_I\label{jtti16Feb10},
\end{align}
which will be used below.

By the following duality chain,
\begin{align}
 \begin{bmatrix}F1(4)\\P(4)\\f1(\psi)\\p(\psi)\end{bmatrix}
 &\xrightarrow{S}
 \begin{bmatrix}D1(4)\\P(4)\\d1(\psi)\\p(\psi)\end{bmatrix}
 \xrightarrow{T4567}
 \begin{bmatrix}D3(567)\\F1(4)\\d5(\psi 4567)\\p(\psi)\end{bmatrix}
 \xrightarrow{S}
 \begin{bmatrix}D3(567)\\D1(4)\\ns5(\psi 4567)\\p(\psi)\end{bmatrix}\notag\\
 &
 \xrightarrow{T5}
 \begin{bmatrix}D2(67)\\D2(45)\\ ns5(\psi 4567)\\p(\psi)\end{bmatrix}
 \xrightarrow{T8}
 \begin{bmatrix}D3(678)\\D3(458)\\ kk(\psi 4567;8)\\p(\psi)\end{bmatrix}
 \xrightarrow{T9}
 \begin{bmatrix}D4(6789)\\D4(4589)\\ 5^2_2(\psi 4567;89)\\p(\psi)\end{bmatrix},
 \label{dualitychain}
\end{align}
we can dualize the above F-P solution to the desired exotic supertube,
\eqref{hoqz21Mar12}.  After a long but straightforward computation, we
arrive at the following field configuration describing the $\rm
D4(6789)+D4(4589)\to 5^2_2(4567\psi,89)+p(\psi)$ supertube in Type IIA
superstring:\footnote{The duality rules we used are summarized in
Appendix \ref{app:duality_rules}.  }
\begin{align}
 \label{D4D4522_metric}
\begin{split}
  ds^2&=
 -{1\over \sqrt{f_1f_2}}\dtt^2
 +\sqrt{f_1f_2\,}\,dx_{123}^2
 +\sqrt{f_1\over f_2}\,dx_{45}^2
 +\sqrt{f_2\over f_1}\,dx_{67}^2
 +{\sqrt{f_1f_2}\over f_1 f_2+\gamma^2}dx_{89}^2\\
 B^{(2)}&={\gamma\over f_1 f_2+\gamma^2}dx^8\wedge dx^9,\qquad
 e^{2\Phi}={\sqrt{f_1f_2}\over f_1 f_2+\gamma^2}\\
 C^{(3)}&=-\gamma \rho +\sigma,\qquad
 C^{(5)}={f_1 f_2\rho+\gamma \sigma\over f_1 f_2+\gamma^2}\wedge dx^8\wedge dx^9,\qquad
 C^{(1)}=C^{(7)}=0,
\end{split}
\end{align}
where we defined
\begin{align}
\begin{split}
 \rho&= (f_2^{-1}\dtt-dt)\wedge dx^4\wedge dx^5
 + (f_1^{-1}\dtt-dt)\wedge dx^6\wedge dx^7\\
 \sigma&= (\beta_1-\gamma\,dt)\wedge dx^4\wedge dx^5
       +(\beta_2-\gamma\,dt)\wedge dx^6\wedge dx^7,\\
 \dtt&= dt-A.
\end{split}
\end{align}
Gauge-invariant field strengths are
\begin{align}
 G^{(4)}&\equiv 
 dC^{(3)}-H^{(3)}\wedge C^{(1)} = -d\gamma\wedge \rho-\gamma\, d\rho+d\sigma\notag\\
 &=\left[\left(-f_2^{-1}d\gamma+f_2^{-2}\gamma\, df_2\right)\wedge \dtt
 +f_2^{-1}\gamma\, dA+d\beta_1\right]\wedge dx^4\wedge dx^5
 +(1\leftrightarrow2,45\leftrightarrow 67)\\
 G^{(6)}&=dC^{(5)}-H^{(3)}\wedge C^{(3)} \notag\\
 &={1\over f_1f_2+\gamma^2}\left[-f_1^{-1}f_2df_1\wedge \dtt
 -f_2dA+\gamma d\beta_2-\gamma f_1^{-1}d\gamma\wedge \dtt
 \right]\wedge dx^6\dots dx^9\notag\\
 &\qquad\qquad +(1\leftrightarrow2,45\leftrightarrow 67),
\end{align}
which satisfy $*_{10}G_4=G_6$.

It is easy to see that this solution indeed has a non-geometric $T$-fold
structure, as should be the case for a solution with the exotic $5^2_2$
dipole charge, as follows.  From the definition \eqref{kjkv25Mar10} of
$\gamma,\beta_I$, we can derive
\begin{align}
 \int_{c}d\gamma&=\int_{c}*_3dA = {4\pi n Q_1\over L},\label{ickd12Jun12}\\
 \int_{\Sigma}d\beta_I &=\int_{\Sigma}*_3df_I=  -4\pi Q_I,\qquad I=1,2,\label{icrg12Jun12}
\end{align}
where $c$ is a closed curve which links with the curve with linking
number $n$, while $\Sigma$ is a 2-dimensional surface that encloses the
entire curve (see Figure \ref{fig:profile_link}). See Appendix
\ref{app:derive_int_db,int_dgamma}
for derivation.
\begin{figure}[tb]
\begin{quote}
  \begin{center}
  \includegraphics[height=4cm]{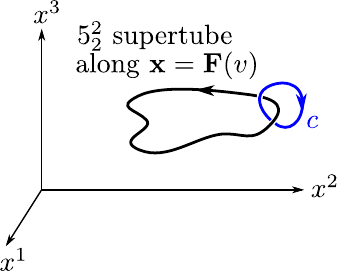}
 \end{center}
 \caption{\sl The $5^2_2$ supertube along the curve ${\bf x}={\bf
 F}(v)$, and curve $c$ linked with it. \label{fig:profile_link}}
\end{quote}
\end{figure}
Eq.\ \eqref{ickd12Jun12} means that, as we go
once along $c$, $\gamma$ undergoes an additive shift,
\begin{align}
 \gamma\to \gamma+{4\pi n Q_1\over L}.
\end{align}
As is clear from the $x^8,x^9$ part of the metric
\eqref{D4D4522_metric}, this means that, as we travel along $c$, the
radii of the internal torus $T_{89}^2$ do not come back to the original
value.  This is exactly what we observed in the straight $5^2_2$
solution \eqref{metric_522}.  Of course, this solution is more
complicated with non-vanishing RR potentials $C^{(3)},C^{(5)}$ because
$5^2_2$ also carries D4 charges dissolved in its worldvolume.  The
$T$-duality monodromy acts on these RR fields as well.

If we move around in the spacetime without going through the curve, then
$\gamma,\beta_I$ are single-valued and we do not see the non-geometric
structure.  Also, as is clear from the definition \eqref{kjkv25Mar10},
the harmonic functions behave as $f_I\to 1,A\to 0$ as $|{\bf x}|\to
\infty$. Namely, the spacetime is asymptotically flat.  Therefore, this
solution \eqref{D4D4522_metric} has all the advertised features; it
describes a puffed-up configuration of D4-branes with non-geometric
exotic dipole charge which, being along a closed curve, does not
destroy the asymptotic structure of spacetime.

The solution \eqref{D4D4522_metric} has non-vanishing RR fields
$C^{(3)},C^{(5)}$.  Let us check that it has the expected charges.  As
we discussed in section \ref{sec:monodromies_and_charge_conservation}
and Appendix \ref{app:page_chg}, there are multiple notions of charge
and we must use Page charge when we want to discuss charge conservation.
First, let us compute the D4 charges that we started with in the
polarization process \eqref{hoqz21Mar12}.  If $S^2$ is a 2-sphere that
encloses the entire curve in $\bbR^3$, the D4(4589) Page charge is
measured by
\begin{align}
 Q_{\rm D4(4589)}^{\text{Page}}
 =\int_{S^2\times T^2_{67}} (G^{(4)}-B^{(2)}\wedge G^{(2)})
 =\int_{S^2\times T^2_{67}} dC^{(3)}
 =\int_{S^2} d(\gamma f_1^{-1}A+\beta_2).\label{mvpu11Jun12}
\end{align}
Actually, for D4, Page charge or brane source charge does not
make a difference because $G^{(2)}=0$.  The first term in the last
expression of \eqref{mvpu11Jun12} vanishes, since $S^2$ encloses the
entire curve without going through it and therefore $\gamma,f_1,A$ are
all single-valued.  The second term gives $Q_2$ by \eqref{icrg12Jun12}.
One can similarly compute the D4(6789) charge to obtain $Q_1$.  So, the
D4 charges are given by $Q_I$, as expected.

Because $C^{(5)}$ is non-vanishing, one might think that this solution
\eqref{D4D4522_metric} carries non-vanishing D2 charges as well.
However, if we look at D2 Page charges, we can show that they vanish as
they should.  For example, D2(45) Page charge is
\begin{align}
 Q_{\rm D2(45)}^{\text{Page}}
 &=\int_{S^2\times T^2_{6789}} (G_6-B_2\wedge G_4)
 =\int_{S^2\times T^2_{6789}} d(C_5-B_2\wedge C_3)\notag\\
 &=\int_{S^2\times T^2_{6789}} d(-f_1^{-1}A \wedge dx^6\dots dx^9)
 =\int_{S^2} d(-f_1^{-1}A)=0,\label{csq9Sep11}
\end{align}
because $f_1,A$ are single valued on $S^2$ enclosing the entire curve.
Therefore, there is no D2 charge, as expected.
One can similarly see that the only monopole charges that the solution
\eqref{D4D4522_metric} carries are the D4-brane charges on the left of
\eqref{hoqz21Mar12}.

It is instructive to see what happens to these charges if we ``go
through'' the curve.  As discussed in section
\ref{ss:monodromies_page_chg}, this process corresponds to $U$-dualizing
the notion of charge according to the $U$-duality monodromy of the
exotic brane.  In this process, the quantities $\gamma,\beta_I$ undergo
additive shifts as described by \eqref{ickd12Jun12},
\eqref{icrg12Jun12}.  However, we can see that D4 and D2 charges
computed in \eqref{mvpu11Jun12} and \eqref{csq9Sep11} remain unchanged,
because $\gamma,f_{1,2},A$ are still single-valued on $S^2$ and because
\eqref{icrg12Jun12} is still true.
This is understood in the language we used in section
\ref{sec:monodromies_and_charge_conservation} as follows.  If we
compactify the 10D theory on $T^2_{89}$ to 8D, we have $T$-duality group
$SO(2,2,\bbZ)=SL(2,\bbZ)_\tau\times SL(2,\bbZ)_\rho$, with moduli $\tau,\rho$
defined in \eqref{hcjy26Sep12} transforming in the respective
$SL(2,\bbZ)$ factors.  The $5^2_2$-brane has the $T$-duality monodromy
\eqref{jjbz25Sep12} acting on $\rho$.  In 8D, D2-branes not wrapping 89
and D4-branes wrapping 89 both become 2-branes, and the 10D RR
potentials $C_{\mu\nu\rho}$ and $C_{89\mu\nu\rho}$
($\mu,\nu,\rho=0,\dots,7$) that they couple to, respectively, reduce to
8D 3-forms $\cC_{\alpha\,\mu\nu\rho}$
$(\alpha=1,2)$ which transform covariantly under
$SL(2,\bbZ)_\rho$ as a doublet.  The precise relation is
\cite{Fukuma:1999jt}
\begin{align}
 \cC_{1,\mu\nu\rho}=C_{\mu\nu\rho},\qquad
 \cC_{2,\mu\nu\rho}=C_{89\mu\nu\rho}-B_{89}C_{\mu\nu\rho}.
\end{align}
Just as we discussed in section
\ref{sec:monodromies_and_charge_conservation}, this means that 
$\cC_{\alpha\,\mu\nu\rho}$ couple to
\begin{align}
 \cQ^{\alpha=1}
 =Q^{\text{D2,bs}} +B_{89}Q^{\text{D4,bs}},\qquad
 \cQ^{\alpha=2}
 =Q^{\text{D4,bs}},
\end{align}
where $Q^{\text{D2,bs}}$ and $Q^{\text{D4,bs}}$ mean the brane source
charge for D2($ij$) and D4($ij89$), respectively, with $i,j=1,\dots,7$.
The covariant charges $\cQ^{\alpha}$ are nothing but Page charges.
Under the $SL(2,\bbZ)_\rho$ duality monodromy around the $5^2_2$ given
in \eqref{jjbz25Sep12}, $\cQ^{\alpha}$ transform as
\begin{align}
 \begin{pmatrix}  \cQ^{1}\\ \cQ^{2} \end{pmatrix}
 \to
 \begin{pmatrix}  \cQ'^{1}\\ \cQ'^{2} \end{pmatrix}
 =
 \begin{pmatrix}  1&0\\ -2\pi \sigma & 1 \end{pmatrix}
 \begin{pmatrix}  \cQ^{1}\\ \cQ^{2} \end{pmatrix}
 =
 \begin{pmatrix}  \cQ^{1}\\ -2\pi\sigma\cQ^1+\cQ^{2} \end{pmatrix}.\label{jtdl25Sep12}
\end{align}
Because we originally had non-vanishing D4 charges but vanishing D2
charges, $\cQ^{1}=0$ and $\cQ^{2}\neq 0$.  So, even after the duality
transformation \eqref{jtdl25Sep12}, we have $\cQ'^{1}=0,\cQ'^{2}=\cQ^2$.
Namely, D2 and D4 Page charges remain unchanged even if we go through
the $5^2_2$ ring.

Because $S^1\times T^5$ is compact, the solution \eqref{D4D4522_metric}
can be thought of as a solution of 4D supergravity, where D4-branes are
point particles which have puffed up into a one-dimensional object,
$5^2_2$.  The 4D metric in the Einstein frame is
\begin{align}
 ds_4^2&=
 -{1\over \sqrt{f_1f_2}}\dtt^2
 +\sqrt{f_1f_2\,}\,dx_{123}^2. 
\end{align}
The 4D Einstein metric is single-valued.  In the 4D viewpoint, the
monodromy of the $5^2_2$-brane appears as the monodromy of 
scalar moduli.

\subsection{Non-geometric microstates}

The 2-charge system, which is nothing but the system of two stacks of
branes that appear on the left side of the puffing-up relation
\eqref{supertube_D0+F1}, \eqref{hkns21Mar12}--\eqref{hoqz21Mar12}, is
known to have large microscopic degeneracy.  Using weak coupling
descriptions, the microscopic entropy of the system can be computed as
\begin{align}
 S_{\text{micro}}=2\sqrt{2}\pi\sqrt{N_1N_2},\label{Smicro_2chg}
\end{align}
where $N_1,N_2$ are the numbers of the two branes, in the large
$N_{1,2}$ limit.  The strong coupling, {\it i.e.}\ gravity, description
of the system is given by backreacted solutions of supergravity that
represent the right hand side of the puffing-up relation
\eqref{supertube_D0+F1}, \eqref{hkns21Mar12}--\eqref{hoqz21Mar12}.  
In different duality frames, the supergravity solutions come in
different guises.  In particular, in the D1-D5 duality frame
\eqref{hlni21Mar12}, the puffed-up supergravity solutions, the so-called
Lunin--Mathur geometries, are configurations of Kaluza--Klein monopole
with flux and are completely regular \cite{Lunin:2001jy, Lunin:2002iz}.
By quantizing these solutions, one can reproduce the correct order of
the microscopic entropy \eqref{Smicro_2chg} \cite{Rychkov:2005ji}.
Therefore, the Lunin--Mathur geometries give a genuine description of the
microstates in supergravity, and are now called microstate geometries or
geometric microstates.

In the D4-D4 system \eqref{hoqz21Mar12}, on the other hand, we have
shown that the puffed-up configurations are non-geometric solutions
\eqref{D4D4522_metric}.  They are parametrized by the profile function
${\bf F}(v)$ just as the Lunin--Mathur geometries are, and represent
string theory configurations produced by back-reaction of the
$5^2_2$-brane.  Therefore, they are on complete equal footing with the
Lunin--Mathur geometries and can accordingly equally well be viewed as a
strongly-coupled description of
the microstates.  Namely, they are \emph{non-geometric microstates} or
\emph{microstate non-geometries}.  Note that $5^2_2$ has tension $\sim
g_s^{-2}$, just like KKM, unlike other exotic branes with tension $\sim
g_s^{-3}$ which makes a gravity description more questionable.  Although
the $5^2_2$ has singularity at the core, it must be regarded as an
acceptable singularity in string theory, as the singularities of
D-branes are.  It is nothing more than an accident that the microstates
are regular in the D1-D5 frame.

In \cite{Sen:2009bm}, Sen claimed that, in a fixed duality frame, in the
classical limit ({\it i.e.}, $g_s\to 0$ with fixed $\ap$), only either
one of the following two possibilities must be true: (a) there only
exists a small black hole solution representing the whole ensemble, or
(b) there are regular gravity solutions representing individual
microstates.
Whether this claim is true or not remains highly controversial at the
time of writing \cite{FuzzballFAQ}.  Note however that, according to
Sen, the D4-D4 system is a duality frame in which there is no small
black hole solution. In \cite{Sen:2009bm}, based on a scaling argument,
he argued that in the D1-D5 system there should be no classical small
black hole solution.  Since the D1-D5 system is related to the D4-D4
system by $T$-duality, which is perturbative and does not modify the
classical limit, there must be no small black hole solution in the
latter frame either.  Therefore, if the claim of \cite{Sen:2009bm} is
true, the D4-D4 system belongs to case (b) and there should be gravity
microstates.  The fact that we found the non-geometric microstates
\eqref{D4D4522_metric} seems to be in accord with this claim.  However,
note that Ref.\ \cite{Sen:2009bm} actually makes a stronger claim that
there the gravity solution in case (b) must be smooth.  The fact the
metric of \eqref{D4D4522_metric} is not smooth, may mean that the claim
of Ref.\ \cite{Sen:2009bm} must be reconsidered in the case of
non-geometric solutions; the non-geometric microstates
\eqref{D4D4522_metric} offer an important touchstone for the validity of
the claim of \cite{Sen:2009bm}.

\subsection{Circular case}
\label{ss:circular_ring}

When the profile function represents a circular ring, the D4+D4$\to
5^2_2$ supertube solution \eqref{D4D4522_metric} can be written down
more explicitly.  Let us take the profile to be
\begin{align}
 F_1(v)+iF_2(v)=Re^{i\omega v},\qquad F_3(v)=0,\qquad 
 \omega={2\pi n\over L},
\end{align}
where $n\in\bbZ$  corresponds to the number of times
the $5^2_2$ worldvolume winds around the circle of radius $R$.
Let us introduce the $(y,\psi,x)$ coordinate system,
\begin{align}
 dx_{123}^2&=(dx^1)^2+(dx^2)^2+(dx^3)^2={R^2\over (x-y)^2}
 \left[{dy^2\over y^2-1}+(y^2-1)d\psi^2+{dx^2\over 1-x^2}\right],\\
&-\infty<y\le -1,\qquad 0\le\psi<2\pi, \qquad -1\le x\le 1 ,
\end{align}
which is a 3D version of the coordinate system used for black rings
\cite{Elvang:2004rt}.  $y$ is roughly a radial coordinate which goes
to $-\infty$ near the ring and $-1$ near infinity (and on the axis of
the ring; see Figure 1 of \cite{Elvang:2004ds}). $x$ is an angular
variable around the ring while $\psi$ is an angular variable along the
ring.  The explicit relations between $(x^1,x^2,x^3)$ and $(y,\psi,x)$ are
\begin{align}
 x^1&={\sqrt{y^2-1}\over x-y}R\cos\psi,\qquad x^2={\sqrt{y^2-1}\over x-y}R\sin\psi,\qquad
 x^3=\pm{\sqrt{1-x^2}\over x-y}R.\label{niir10Jan10}
\end{align}
As the $\pm$ sign in \eqref{niir10Jan10} indicates, we need two
patches of $(y,\psi,x)$ spaces to cover $\bbR^3$.  Or, we can use the
$(y,\psi,\phi)$ coordinates with
\begin{align}
 x^1&={\sqrt{y^2-1}\over \cos\phi-y}R\cos\psi,\qquad x^2={\sqrt{y^2-1}\over \cos\phi-y}R\sin\psi,\qquad
 x^3={\sin\phi\over \cos\phi-y}R.
\end{align}
and
\begin{align}
 x=\cos\phi,\qquad 0\le\phi\le 2\pi.\label{hcrb19Feb10}
\end{align}
With this range \eqref{hcrb19Feb10}, the $(y,\psi,\phi)$ coordinates
cover the entire $\bbR^3$.  The inverse relation of
\eqref{niir10Jan10} is
\begin{align}
 x=-{{\bf x}^2-R^2\over \Sigma},\qquad
 y=-{{\bf x}^2+R^2\over \Sigma},\qquad
 \Sigma^2=({\bf x}^2+R^2)^2-4R^2x_3^2.
\end{align}

In this $(y,\psi,x)$ coordinate system, the functions $f_I,A$ can
be computed explicitly as
\begin{align}
\begin{split}
  f_{I}&=1+{Q_{I}\over R}\sqrt{x-y\over -2y}\,\,_{2}F_{1}\!\left({1\over 4},{3\over 4};1;\zeta^2\right)
 =1+{2Q_{I}\over \pi R}\sqrt{{x-y\over -2y}}\,{{\bf K}({2\zeta\over \zeta-1})\over \sqrt{1-\zeta}},\label{nijj10Jan10}\\
 A&
 =-{qR\over 2}{y^2-1\over (x-y)^{1/2}(-2y)^{3/2}}\,\,_2F_{1}\!\left({3\over 4},{5\over 4};2;\zeta^2\right)d\psi\\
 &=-{qR\over 8}\zeta^2\sqrt{-2y\over x-y}\,\,_{2}F_{1}\!\left({3\over 4},{5\over 4};2;\zeta^2\right)d\psi,
\end{split}
\end{align}
where $I=1,2$; $_2F_1(\alpha,\beta;\gamma;z)$ is the
hypergeometric function and
\begin{align}
 Q_2=Q_1R^2\omega^2,\qquad
 q\equiv Q_1\omega={2\pi n Q_1 \over L},\qquad
 \zeta\equiv \sqrt{1-{1\over y^2}}.
 \label{ibnq12Jun12}
\end{align}
Furthermore, the equations \eqref{jtti16Feb10} can be solved to give
\begin{align}
 \gamma
 &=-{q\sqrt{u+1}\over 4\sqrt{2}\,u^{3/2}}\biggl\{
 (1-u)\,
 \EllipticF{\phi\over 2}{2\over u+1}\,
 {}_{2}F_1\!\left({3\over 4},{5\over 4};2;\zeta^2\right)\notag\\
 &\qquad\qquad
 +u\,\EllipticE{\phi\over 2}{2\over u+1} \left[
 3\,{}_{2}F_1\!\left({3\over 4},{1\over 4};2;\zeta^2\right)
 +\,{}_{2}F_1\!\left({3\over 4},{5\over 4};2;\zeta^2\right)
 \right]
 \biggr\}
 \\
 &=-{q\cosh{\nu\over 2}\over 4\cosh^{3/2}\nu}\biggl\{
 (1-u)\,\EllipticF{\phi\over 2}{1\over \sinh^2{\nu\over 2}} 
 \,{}_{2}F_1\!\left({3\over 4},{5\over 4};2;\tanh^2\nu\right)\notag\\
 &\qquad\qquad
 +u\,\EllipticE{\phi\over 2}{1\over \sinh^2{\nu\over 2}} \left[
    3\,{}_{2}F_1\!\left({3\over 4},{1\over 4};2;\tanh^2\nu\right)
 +\,{}_{2}F_1\!\left({3\over 4},{5\over 4};2;\tanh^2\nu\right)
 \right]
 \biggr\},
 \\
 (\beta_I)_u&=(\beta_I)_\phi=0,\\
(\beta_I)_\psi&={Q_I\over\pi\cosh{\nu\over 2}}\biggl\{
 -e^{\nu/2}\,{\bf E}(2e^{-\nu}\sinh\nu)\,\EllipticF{\phi\over 2}{1\over\cosh^2{\nu\over 2}}
 \notag\\
 &\qquad
 +e^{-\nu/2}{\bf K}(2e^{-\nu}\sinh\nu)
 \Bigl[
 -(\cosh\nu+1)\,\EllipticE{\phi\over2}{1\over\cosh^2{\nu\over 2}}
 +\cosh\nu\,  \,\EllipticF{\phi\over2}{1\over\cosh^2{\nu\over 2}}
 \Bigr]
 \biggr\}
 \notag\\
&\quad
 +
 {\sqrt{2}\,Q_I\,e^{-\nu/2}\sin\phi\over\pi\sqrt{\cosh\nu+\cos\phi}}
 {\bf K}(2e^{-\nu}\sinh\nu)
\end{align}
where
\begin{align}
 u=-y=\cosh\nu,\qquad \zeta=\sqrt{1-u^{-2}}=\tanh\nu.
\end{align}
In the above formulae, ${\bf F}(z|m)$ and ${\bf E}(z|m)$ are the elliptic
integral of the first and second kind, respectively, and
${\bf K}(m)$ and ${\bf E}(m)$
are the complete elliptic
integral of the first and second kind, respectively, defined by
\begin{align}
 {\bf E}(z|m)&=\int_0^z{dt\over\sqrt{1-m\sin^2t}},\qquad
 {\bf F}(z|m)=\int_0^z \sqrt{1-m\sin^2t\,}\,dt
 \\
 {\bf E}(m)&={\bf E}\left({\pi\over 2}\Big|m\right),\qquad
 {\bf K}(m)={\bf F}\left({\pi\over 2}\Big|m\right).
\end{align}

\begin{figure}[tb]
\begin{quote}
  \begin{center}
  \includegraphics[height=2.5cm]{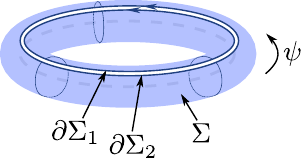}
 \end{center}
 \caption{\sl 2-surface $\Sigma$ with toric topology, enclosing the
 circular $5^2_2$ ring.  Its boundaries $\partial \Sigma_{1,2}$ are both
 directed along $\psi$. \label{fig:profile_sigma_torus}}
\end{quote}
\end{figure}

By using the property of the elliptic integrals,
\begin{align}
 {\bf E}(z+\pi|m)&=2{\bf E}(m)+{\bf E}(z|m),\qquad
 {\bf F}(z+\pi|m)=2{\bf K}(m)+{\bf F}(z|m).
\end{align}
One can show that, as one goes through the ring, {\it i.e.}\ as
$\phi\to\phi+2\pi$, the quantities $\gamma, \beta_I$ undergo the following additive
shifts:
\begin{align}
 \gamma\to\gamma-2q,\qquad
 (\beta_I)_\psi\to (\beta_I)_\psi-2Q_I.\label{idtx12Jun12}
\end{align}
We can easily see that the first relation is consistent\footnote{Up to a
sign that depends on the convention.} with \eqref{ickd12Jun12}, by
looking at \eqref{ibnq12Jun12}. On the other hand, the second relation
is consistent with \eqref{icrg12Jun12}, as follows.  As the surface
$\Sigma$, take a torus that encloses the ring.  Although the torus has
no boundaries, make ones by cutting it along its length (namely, along
$\psi$). This way, one creates two boundaries $\partial \Sigma_1$ and
$\partial \Sigma_2$ such that $\partial \Sigma_1-\partial \Sigma_2=0$,
both going in the $\psi$ direction; see Figure
\ref{fig:profile_sigma_torus}.
Then \eqref{icrg12Jun12} can be understood as saying
\begin{align}
 \int_{\partial\Sigma_1} \beta_I
 -\int_{\partial\Sigma_2} \beta_I
= \int d\psi \, \Delta (\beta_I)_\psi
=  4\pi Q_I,
\end{align}
which is consistent with \eqref{idtx12Jun12}.
In the near-ring region $y\to-\infty$ (or equivalently, $\nu\to\infty$
or $u\to\infty$), one can show
\begin{align}
 \gamma\approx -{q\phi\over\pi}
,\qquad
 \beta_\psi\approx -{Q\phi\over\pi},
\end{align}
from which the additive shifts \eqref{idtx12Jun12} are easy to see.

\subsection{Near-ring analysis}

Let us study the near-ring region of the circular exotic supertube
above.  For a general profile, \eqref{jjnv12Jun12} was all we could say
about the near-supertube behavior, but in this circular case we can be
more explicit.

Let us define $(r,\theta,\xi)$ coordinates by
\begin{align}
 (x^1)^2+(x^2)^2=(R+r\cos\theta)^2-r^2,\qquad x^3=r\sin\theta,\qquad \psi={\xi\over R}.
\end{align}
In the near-ring region $R\gg r,|x^3|$, these coordinates give a
cylindrical coordinate system; namely, the $\bbR^3_{123}$ part of the
metric is
\begin{align}
 ds_{123}^2\approx dr^2+r^2d\theta^2+d\xi^2.\label{jtqk16Feb10}
\end{align}
In the near-ring region, the harmonic functions \eqref{nijj10Jan10}
reduce to
\begin{align}
 f_{I}&\approx 1+{Q_{I}\over \sqrt{2}\,R}\,\,_{2}F_1\!\left({1\over 4},{3\over 4};1;1-{r^2\over R^2}\right),
 \quad
 A\approx -{qR\over 4\sqrt{2}}\,\,_{2}F_1\!\left({3\over 4},{5\over 4};2;1-{r^2\over R^2}\right)\,d\psi.
\end{align}
Expanding the hypergeometric functions, we get
\begin{align}
 f_{I}&\approx 1+{Q_{I}\over \pi R}\log\left({8R\over r}\right),\quad
 A\approx -{q\over \pi}\log\left({8e^2R\over r}\right)\,d\xi,
 \label{jueb16Feb10}
\end{align}
from which we can explicitly see the asymptotic behavior
\eqref{jjnv12Jun12}.  Also, in these coordinates,
\begin{align}
 \gamma&=-{q\over \pi}\theta,\qquad
 \beta_I=-{Q_I\over \pi R}\theta \, d\xi.\label{hmhr13Jun12}
\end{align}

In this near-ring region, the supertube can be regarded as straight.  The metric becomes
\begin{align}
 ds^2\approx -\dtt^2+\sqrt{f_1f_2}\,dz d\zb+{\sqrt{f_1f_2}\over f_1f_2+\gamma^2}[(dx^8)^2+(dx^9)^2]
 +\dots,\qquad z=re^{i\theta}.\label{gifv13Jun12}
\end{align}
Let us compare\footnote{In \eqref{gifv13Jun12}, we have $\dtt=dt-A$
instead of $dt$ in \eqref{gigs13Jun12}. This is because the former
represents a supertube solution with momentum flowing along the
worldvolume, which does not exist in the latter.} this with the straight
$5^2_2$ metric given by \eqref{mwfz27Jul12}:
\begin{align}
 ds^2&=-dt^2+{\rho_2 e^{2\varphi_1}}dz d\zb+\rho_2[(dx^3)^2+(dx^4)^2]+\dots,\qquad\label{gigs13Jun12}
\end{align}
By matching the corresponding metric components as
\begin{align}
 \sqrt{f_1f_2}\leftrightarrow {\rho_2 e^{2\varphi_1}}, 
 \qquad {\sqrt{f_1f_2}\over f_1f_2+\gamma^2} \leftrightarrow \rho_2,
\end{align}
using the asymptotic behavior \eqref{jueb16Feb10}, \eqref{hmhr13Jun12},
we see that
\begin{align}
 \rho(z)\sim {i\over {\sqrt{Q_1Q_2}\over \pi R}\log{L\over z}},\qquad
 e^{\varphi(z)}\sim {\sqrt{Q_1Q_2}\over \pi R}\log{L\over z},\label{gyzr13Jun12}
\end{align}
where $L=\CO(R)$.  As we discussed in section \ref{ss:susy}, the function
$e^{\varphi(z)}$ represents the freedom associated with the boundary
condition at infinity, which makes the spacetime well-defined there.  In
the present case, we can say that the functional form of
$e^{\varphi(z)}$ in \eqref{gyzr13Jun12} was determined so that the
geometry is asymptotically flat.  We see that \eqref{gyzr13Jun12} has
the same form as for the naive, infinite stand-alone straight $5^2_2$
solution \eqref{ndtn12Jun12}.
This functional form is also the same as the F-theory case
\cite{Greene:1989ya, Vafa:1996xn, Bergshoeff:2006jj}, in which one takes
$e^{\varphi(z)}\propto \eta(\rho)^{-2}z^{-N/12}$ so that the transverse
space has conical deficit angle $\pi N/ 12$.  Here, $\eta(\rho)$ is the
Dedekind eta function and $N$ is the number of 7-branes at $z=0$.
We do not have a full understanding of why they all agree, but it is
probably related to the fact that the boundary condition in all the
cases is such that the energy of the codimension-2 brane does not spread
all the way to infinity.

\subsection{Exotic solutions versus ``all supersymmetric solutions''}

Based on \cite{Gutowski:2004yv}, it was shown in \cite{Gauntlett:2004qy}
that, all supersymmetric solutions in $d=5,\CN=1$ supergravity with
vector multiplets in the timelike class can be written as\footnote{In
\cite{Gauntlett:2004qy}, they also assume that the vector-multiplet
scalars live in a symmetric space, which is applicable to the current
situation. This condition was relaxed in \cite{Gutowski:2005id}.}
\begin{align}
 ds_5^2=-f^2(dt+\omega)^2+f^{-1}ds^2(M_4),\label{mrtl13Jun12}
\end{align}
where $M_4$ is an arbitrary four-dimensional hyper-K\"ahler manifold
and, and $f$ and $\omega$ are a scalar and a 1-form on $M_4$,
respectively.  The timelike class means that the Killing vector constructed
from the bilinear of the Killing spinor is timelike.

It is interesting to see whether our exotic supertube solution fits in
this framework or not.  The solution \eqref{D4D4522_metric} describing
the exotic $\rm D4(4589)+D4(4567)\to 5^2_2(\psi 6789,45)$ supertube is
not appropriate for this for the following reason.  In order to put this
in the framework of $d=5$ supergravity, we first lift this to $\rm
M5(4589\ten)+M5(4567\ten)\to 5^3(\psi 6789,45\ten)$, where $\ten$
denotes the 11th direction, and compactify it on $456789$. However, this
solution belongs to the null class, not the timelike one, because the
naive M5-M5 solution is in the null class and the Killing spinors of the
two configurations must agree asymptotically.\footnote{It is interesting
to see whether or not this solution fits within the null class
supersymmetric solutions classified in \cite{Gutowski:2005id}.}

So, we should go to a different duality frame where the solution is
timelike and involves exotic charge.  By $T$-dualizing
\eqref{D4D4522_metric} along $67$, we obtain $\rm D6(456789)+D2(89)\to
5^2_2(\psi 4567,89)$, which is an exotic supertube.  By further lifting
it to 11 dimensions, we get an exotic supertube in M-theory, $\rm
KKM(456789,\ten)+M2(89)\to 5^3(\psi 4567,89\ten)$.  This must belong to
the timelike class, because the naive KKM+M2 solution does. The 11D
metric and 3-form for this exotic supertube solution are
\begin{align}
\begin{split}
 ds_{11}^2
 &=
 -f_1^{-2/3}F^{-2/3}\left[dt+f_2^{-1}\gamma (dx^\ten+\beta_2)-F A\right]^2+f_1^{1/3}f_2 F^{1/3}dx_{123}^2\\
 &\qquad
 +f_1^{1/3}f_2^{-1}F^{1/3}(dx^\ten-f_1^{-1}\gamma A+\beta_2)^2
 +f_1^{1/3}F^{1/3}dx_{4567}^2+f_1^{-2/3}F^{-2/3}dx_{89}^2,\\
 A_3
 &=
 \left[{f_2\dtt+\gamma (dx^\ten+\beta_2)\over f_1f_2+\gamma^2}-dt\right]
 \wedge dx^8\wedge dx^9,
\end{split}
\end{align}
where
\begin{align}
 F\equiv 1+{\gamma^2\over f_1f_2}.
\end{align}
To put this in the form of $d=5$ supergravity, we compactify it on
$456789$ and keep the $0123\ten$ part.  The 5D metric is
\begin{align}
 ds_5^2&=-f^2\left[dt+f_2^{-1}\gamma (dx^\ten+\beta_2)-F A\right]^2
 +f^{-1}ds_4^2,\label{msyh13Jun12}
\end{align}
where
\begin{align}
 ds_4^2&= f_2^{-1}(dx^\ten-f_1^{-1}\gamma A+\beta_2)^2+f_2 dx_{123}^2,\qquad
 f=f_1^{-1/3}F^{-1/3}.\label{mryo13Jun12}
\end{align}
The metric \eqref{msyh13Jun12} appears to be of the general form
\eqref{mrtl13Jun12}.  However, it is easy to see that the 4D base metric
\eqref{mryo13Jun12} is not hyper-K\"ahler; the Ricci tensor does not
vanish due to the $f_1^{-1}\gamma A$ term.

This suggests that, because exotic supertubes describe non-geometric
spacetimes in which fields have non-trivial monodromies, the existing
classification of ``all supersymmetric solutions'' does not apply to
them.  It would be very interesting to nail down what assumptions in the
existing analysis should be relaxed and generalize the classification of
supersymmetric configurations to incorporate exotic solutions.

\section{Toward ``truly non-geometric'' configurations}
\label{sec:toward_truly_non-geometric_config}

One could say that the exotic solutions we have discussed so far are not
truly non-geometric, in the sense that there are duality frames where
they are geometric, and that the Einstein metric in the lower
(non-compact) dimensions is single-valued.  In this section, we discuss
the existence of exotic solutions for which
\begin{itemize}
 \item [(i)] spacetime is non-geometric, {\it i.e.}, the metric is multi-valued, in any duality frames,
 \item [(ii)] even the Einstein metric in the lower (non-compact)
       dimensions is multi-valued.
\end{itemize}
The non-geometric solutions satisfying condition (i) have been known
(see, {\it e.g.}, \cite{Hellerman:2002ax, Narain:1986qm,
Hellerman:2006tx, Schulgin:2008fv}).  Here we discuss it using the model
related to the one in previous sections.  Furthermore, using the same
model, we argue for the existence of solutions satisfying the stronger
condition (ii).

For this purpose, let us consider the following simple model.  We
compactify 10D string on $T^2\times T^5$, where the $T^5$ metric is flat
and there is no form fields along it.  So, the 10D string frame
metric can be written as
\begin{align}
 ds_{\text{10,str}}^2&=ds_5^2+\sum_{a=5}^9 (dx^a)^2,\qquad
 ds_5^2=e^{2\Phi}G^{-1}ds_{\text{3,Ein}}^2+ds^2_{T^2},\label{iwgl17Jun12}
\end{align}
where $\Phi$ is the 10D dilaton.  Moreover, $ds_{\text{3,Ein}}^2$ is the Einstein
metric in non-compact 3D and $ds_{T^2}^2$ is the metric on $T^2$, 
given by
\begin{align}
\label{jhus23Jan12}
\begin{split}
 ds_{\text{3,Ein}}^2&=g_{\mu\nu}dx^\mu dx^\nu ,\qquad \mu,\nu=0,1,2,\\
 ds_{T^2}^2&=G_{ij}(dx^i+A^i_\mu dx^\mu)(dx^j+A^j_\nu dx^\nu),\qquad
 G=\det G_{ij},\qquad i,j=3,4,
\end{split}
\end{align}
where $g_{\mu\nu},G_{ij},A^i_\mu$ depend only on $x^\mu$.  If we write
the internal $T^2$ fields as
\begin{align}
\label{gpnc14Jun12}
\begin{split}
  ds_{T^2}^2&={\rho_2\over \tau_2}|dx^3+\tau dx^4|^2,\qquad
 G_{ij}={\rho_2\over \tau_2}\begin{pmatrix} 1 & \tau_1 \\ \tau_1 & |\tau|^2 \end{pmatrix},\qquad
 \tau=\tau_1+i\tau_2,\\
 \rho&=B_{34}+iG^{1/2}=\rho_1+i\rho_2,
\end{split}
\end{align}
then the 3D action is
\begin{align}
 S_{3D}=\int d^3x\sqrt{-g}\left(R_g-4(\nabla\phi)^2-{|\nabla \tau|^2\over 2\tau_2^2}
 -{|\nabla \rho|^2\over 2\rho_2^2}\right),\qquad
 \phi\equiv \Phi-{1\over 4}\log G.\label{iwiw17Jun12}
\end{align}
The equations of motion for $\phi$ and $\tau$ are
\begin{align}
 \nabla^2\phi=0,
 \qquad\qquad
 \nabla_\mu\left(
 {\nabla^\mu \tau_1 \over \tau_2^2}
 \right)=0,\qquad
 \nabla_\mu\left(
 {\nabla^\mu \tau_2 \over \tau_2^2}
 \right)+{|\nabla \tau|^2\over \tau_2^3}=0.\label{iviu23Jan12}
\end{align}
The $\rho$ equations of motion are obtained from the $\tau$ equations by
setting $\tau\to\rho$. In addition, $g_{\mu\nu}$ should satisfy the 3D
Einstein equation
\begin{align}
\label{hsym14Jun12}
\begin{split}
 & R_{\mu\nu}-{1\over 2}g_{\mu\nu}R=T_{\mu\nu},\\
 & T_{\mu\nu}=
 4\Bigl(\partial_\mu \phi\partial_\nu \phi
 -{1\over 2}{g_{\mu\nu}}|\partial \phi|^2\Bigr)
 +{1\over 4\tau_2^2}
 \left(\partial_\mu \taub\partial_\nu \tau+
 \partial_\nu \taub\partial_\mu \tau-{g_{\mu\nu}}|\partial \tau|^2\right)
 +(\tau\to\rho).
\end{split}
\end{align}

This theory describes 3D gravity coupled to two scalars $\tau,\rho$, and
has duality group $SL(2,\bbZ)\times SL(2,\bbZ)\times \bbZ_2^2$.  The
$SL(2,\bbZ)\times SL(2,\bbZ)$ factor acts on $\tau,\rho$ in an obvious
way.  The first $\bbZ_2$ exchanges $\tau\leftrightarrow \rho$ while the
second $\bbZ_2$ sends $(\tau,\rho)\to(-\taub,-\rhob)$
\cite{Polchinski:1998rq}.  This is the theory studied in
\cite{Greene:1989ya} in relation to cosmic strings and in
\cite{Hellerman:2002ax} further in the context of non-geometric
compactifications.  A particular solution of this is given by
\begin{align}
 ds^2_{\text{3,Ein}}&=-dt^2+e^U dz d\zb,\qquad z=x^1+ix^2,\\
 \phi&=0,\qquad \tau=\tau(z),\qquad \rho=\rho(z),\qquad
 e^U=\tau_2\rho_2 |f(z)|^2,
\end{align}
where $\tau(z),\rho(z),f(z)$ are arbitrary holomorphic functions subject to the condition that $e^U$ is single valued.  If
$\tau(z)$ has an $SL(2,\bbZ)$ monodromy around a point on the $z$-plane,
say $z=\alpha$, there is a codimension-2 brane (which we call a
$\tau$-brane) at $z=\alpha$ and the internal $T^2$ is non-trivially
fibered around it.  However, the total 5D spacetime is still geometric
because $T^2$ is glued with the same $T^2$ via a basis change.
On the other hand, if $\rho(z)$ has an $SL(2,\bbZ)$ monodromy around
$z=\alpha$ (which we call a $\rho$-brane), this generally means that the
5D spacetime is non-geometric, because the metric components and
$B$-field are mixed into each other as we go around $z=\alpha$.

Not all $\rho$-brane branes are non-geometric; the $\rho$-brane around
which $\rho$ simply shifts as $\rho\to \rho +1$ is geometric.
Therefore, if there is a single non-geometric $\rho$-brane, we can
always do an $SL(2,\bbZ)$ duality transformation to make it geometric.
However, this is not possible if there are multiple $\rho$-branes with
different monodromies at different points, because we cannot dualize all
non-geometric $\rho$-branes into geometric ones at the same time.  One
could use the $\bbZ_2$ duality $\tau\leftrightarrow \rho$ to transform
non-geometric $\rho$-branes into $\tau$-branes which are always
geometric, but if there exist multiple $\tau$-branes and $\rho$-branes
at the same time,\footnote{ For examples of solutions both with
$\tau$-branes and $\rho$-branes, see \cite{Hellerman:2002ax}} such
$\bbZ_2$ duality transformation cannot reduce the solution geometric.
The other $\bbZ_2$ symmetry does not help either.
Therefore, such solutions are concrete examples that are non-geometric
in the sense of (i).

Now let us move on to the second issue (ii), whether we can have a
configuration that is non-geometric even in lower dimensions.  Here, we
would like to use the same model \eqref{jhus23Jan12} to argue that there
are configurations which are (initially) asymptotically flat
$\bbR^{1,3}$ and whose \emph{four}-dimensional metric has non-trivial
monodromies.

%
%
%
To begin with, let us take flat $\bbR^{1,3}\times S^1$ with 
metric
\begin{align}
 ds_5^2&=-dt^2+dx^2+d\xi^2+d\eta^2+dy^2,\qquad t,x,\xi,\eta\in\bbR,\qquad y\cong y+2\pi.
\label{mcdd17Jun12}
\end{align}
This is a 4D spacetime (times a compact $S^1$), but let us do an
\emph{angular} compactification of $\bbR^{1,3}$ by going to polar
coordinates $(r,\theta)$ for the $(\xi,\eta)$ plane as
\begin{align}
\label{hkii2Apr12}
\begin{split}
  \xi&=r\cos\theta,\quad \eta=r\sin\theta;\qquad
  r\ge 0,\quad \theta\cong\theta+2\pi,\\
 ds_5^2&=-dt^2+dx^2+dr^2+r^2d\theta^2+dy^2,
\end{split}
\end{align}
and compactifying on the angle $\theta$.  Namely, we regard $y,\theta$
as the $T^2$ coordinates of the 5D$\to$3D compactification
\eqref{iwgl17Jun12}, \eqref{jhus23Jan12}, by taking $x^3=y,x^4=\theta$.  Then the ``3D''
fields \eqref{gpnc14Jun12}, \eqref{iwiw17Jun12} are,
\begin{align}
 ds_{\text{3,Ein}}^2&=r^2(-dt^2+dx^2+dr^2),\qquad
 \tau=\rho=ir,\qquad
 \phi=-{1\over 2}\log r,
 \label{jgoa23Jan12}
\end{align}
where we set the 10D dilaton $\Phi=0$ so that the flat spacetime
\eqref{mcdd17Jun12} is trivially a solution. This is a perverse way to
write non-compact $\bbR^{1,3}$ times compact $S^1$ as a $T^2$ fibration
over $\bbR^{1,2}$, although there is nothing wrong with it. The 4D
Einstein frame metric is simply $\bbR^{1,3}$:
\begin{align}
 ds^2_{\text{4,Ein}}&={e^{2\phi}\over \sqrt{\tau_2\rho_2}}ds_{\text{3,Ein}}^2
 +{\sqrt{\tau_2\rho_2}\over e^{2\phi}}(dx^4)^2
 =
 -dt^2+dx^2+dr^2+r^2d\theta^2.\label{jetx17Jun12}
\end{align}

\begin{figure}[tb]
\begin{quote}
  \begin{center}
  \includegraphics[height=3cm]{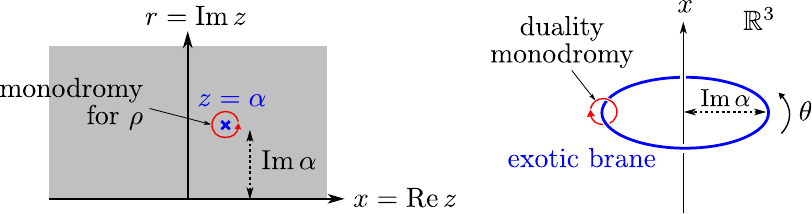}
 \end{center}
 \caption{\sl A point on the upper half $z$-plane around which there is
 non-trivial monodromy for $\rho$ (left) corresponds to a circular
 exotic brane extended in the $\theta$ direction in $\bbR^3$
 (right). \label{fig:noncompact_nongeometry}}
\end{quote}
\end{figure}

In this setup, a configuration that is non-geometric in the non-compact
4D spacetime is one that has non-trivial monodromy of $\rho$ over the 3D
base and asymptotes to \eqref{jgoa23Jan12} as $r\to\infty$.
Furthermore, the fields should behave as \eqref{jgoa23Jan12} also as
$r\to 0$, in order for the point $r=0$ to represent a smooth point in
spacetime.  See Figure \ref{fig:noncompact_nongeometry} for a schematic
explanation.

It is difficult to find an actual \emph{solution} that satisfies the
equations of motion \eqref{iviu23Jan12}, \eqref{hsym14Jun12} and has
such non-geometric monodromy for $\rho$.  Here, let us content ourselves
by writing down a static \emph{configuration} which has such
non-geometric monodromy but does \emph{not} satisfy the equations of
motion.  One can certainly use such a configuration as an initial
condition to find a time-dependent solution of the equations of motion.
Therefore, the existence of such a configuration is evidence for a
non-geometric solution in low dimensions.

To find such a non-geometric configuration, let us fix the 3D
metric to be simply \eqref{jgoa23Jan12}, which we write as
\begin{align}
 ds_{\text{3,Ein}}^2=(\Im z)^2(-dt^2+dzd\zb),\qquad
 z\equiv x+ir,\qquad \Im z\ge 0.
\end{align} 
On this fixed base metric, let us consider 3D fields
$\tau(z,\zb),\rho(z,\zb)$ which do not necessarily satisfy the
equations of motion.
Assume that, at $z=\alpha$ ($\Im \alpha>0$), the field $\rho$
undergoes an $SL(2,\bbZ)$ monodromy
\begin{align}
 \rho
 ~\to~
 \rho'={a\rho+b\over c\rho+d},\qquad
 a,b,c,d\in\bbZ,\qquad ad-bc=1.
\end{align}
This corresponds to having a circular exotic brane in the spatial
$\bbR^3$ at $z=\alpha$; see Figure \ref{fig:noncompact_nongeometry}.
%
Because $\Im z=r=0$ is a special point (the origin of the polar
coordinates), the fields $\rho,\tau$ should behave in a particular way
near $r=0$ for the 5D geometry to be regular there.
However, in the presence of a $\rho$-monodromy, even if we start from a
metric for which $r=0$ is a regular point, after we go around the exotic
brane, we may end up with a metric for which $r=0$ is  singular.
Because this is physically unacceptable, let us require that the 5D
metric be regular even if we go around the brane and come back near
$r=0$.  Then, we ask what constraint this requirement imposes on the
possible monodromy
$M=\left(\begin{smallmatrix}a&b\\c&d\end{smallmatrix}\right)$.

Let us assume that there is a brane at $z=\alpha$ with $\Im z>0$, and
that the behavior of $\rho$ near $r=0$ is still given by
\eqref{jgoa23Jan12}.  Namely,
\begin{align}
 \rho_1=\CO(r),\qquad \rho_2= r+\CO(r^2)\qquad\qquad
 (r\to 0).
 \label{jitt23Jan12}
\end{align}
We take this $\rho$ because we know that this gives a regular 5D
geometry.  Now consider starting from the small $r$ region where
\eqref{jitt23Jan12} is valid, moving away from $r=0$, going around
$z=\alpha$, and coming back to the small $r$ region again. After this
process, the $\rho$ gets transformed into
\begin{align}
 \rho'_1&={ac|\rho|^2+(ad+bc)\rho_1+bd\over |c\rho+d|^2}
 ={bd+acr^2\over d^2+c^2r^2}
 ,\\
 \rho'_2&={\rho_2\over |c\rho+d|^2}
 ={r\over d^2+c^2r^2}.
\end{align}
For simplicity, we assume that $\phi,\tau$ are unchanged and has no
monodromy: $\phi=-(1/2)\log r$, $\tau'=ir$.  As $r\to 0$, we want the volume of the torus
$\rho_2'$ to vanish just as $\rho_2$ did.  So, let us set $d^2=1$.  If
we require also that the $B$-field $B_{34}'=\rho_2'$ vanish at $r=0$, we
need $b=0$.  Then the condition $ad-bc=1$ says that
\begin{align}
 M&=\begin{pmatrix}1 & 0\label{jhqk23Jan12} \\ c & 1\end{pmatrix},\qquad
 c\in\bbZ,
\end{align}
where we took $d=+1$ since the overall sign of $M$ does not matter.  For
this monodromy matrix \eqref{jhqk23Jan12}, the transformed $\rho$ is
\begin{align}
 \rho'_1&={cr^2\over 1+c^2r^2},\qquad
 \rho'_2={r\over 1+c^2r^2}.
\end{align}
The transformed 5D fields are, from \eqref{jhus23Jan12},
\begin{align}
 ds_5'^2=-dt^2+dx^2+dr^2
 +{r^2d\theta^2+dy^2\over 1+c^2r^2},\quad
 B'_{y\theta}={acr^2\over 1+c^2r^2}
 \qquad (r\to 0).
\label{kkxt23Jan12}
\end{align}
The 4D Einstein metric is, using the middle expression in
\eqref{jetx17Jun12},
\begin{align}
 ds_{\text{4,Ein}}'^2=\sqrt{1+c^2r^2}(-dt^2+dx^2+dr^2)+{r^2\over\sqrt{1+c^2r^2}}d\theta^2.\label{jeyl17Jun12}
\end{align}
This geometry is indeed regular at $r=0$, both in 4D and 5D\@.
So, in this configuration, we started from a regular, asymptotically
flat 4D geometry (times a compact $S^1$) and, even after going through
an exotic monodromy (which mixes the internal $S^1$, the angular
direction $\theta$, and the $B$-field through these two directions), we
still have a regular geometry.  Note that, the transformed
Einstein metric \eqref{jeyl17Jun12} is not just different from the
original form \eqref{jetx17Jun12} but has an asymptotics totally
different from the one we started from.  By going through an exotic
brane, one ends up with a totally different universe!

We still have to find an explicit field configuration $\rho(z,\zb)$
which does have the monodromy \eqref{jhqk23Jan12} at $z=\alpha$.  A
naive first try is
\begin{align}
 \rho(z,\zb)&={1\over {2\over z-\zb}+{c\over 2\pi i}\ln (z-\alpha)},\label{fxhe2Apr12}
\end{align}
which indeed behaves near $r=\Im z\to 0$ as
\begin{align}
 \rho(z,\zb)&\sim ir\qquad (r\sim 0)
\end{align}
and has the desired monodromy around $z=\alpha$:
\begin{align}
 \rho& \to \rho'={\rho\over 1+c\rho}.
\end{align}
However, at large $r$, this $\rho$ goes like
\begin{align}
 \rho(z,\zb)&\sim {1\over {c\over 2\pi i}\ln (z-\alpha)} \qquad (r\to\infty )
\end{align}
instead of $\rho(z,\zb)\to ir$, which is necessary to have the correct
original asymptotics (even before going through the exotic brane). So,
\eqref{fxhe2Apr12} is too simple.

This problem can be circumvented by considering a configuration with
branes and anti-branes at the same time.  For example, take
\begin{align}
 \rho(z,\zb)&=
 {1\over {2\over z-\zb}+{c\over 2\pi i}\ln {(z-\alpha)(z-\beta)\over (z-\gamma)(z-\delta)}},\qquad \alpha+\beta-\gamma-\delta=0.
\end{align}
Namely, we placed branes at $z=\alpha,\gamma$ and anti-branes at
$z=\beta,\delta$.  As long as we keep away from the centers, this has
the desired asymptotics,
\begin{align}
 \rho(z,\zb)=
 \begin{cases}
  ir
  +\CO(1)
  &\qquad r\to \infty,\\
  ir
  +\CO(r^2)
  &\qquad r \to 0,
\end{cases}
\end{align}
so that the original geometry is regular at $r=0$ and is asymptotically
flat $\bbR^{1,3}\times S^1$, as desired.

If we go around one of the branes (but not all of them), $\rho$ changes
into
\begin{align}
 \rho\to \rho'&= {1\over {2\over z-\zb}+c+{c\over 2\pi i}\ln {(z-\alpha)(z-\beta)\over (z-\gamma)(z-\delta)}}.
\end{align}
The $r\to 0,\infty$ behavior is now
\begin{align}
 \rho(z,\zb)=
 \begin{cases}
  {1\over c}+{i\over c^2r}+\CO(r^{-2})&\qquad r\to \infty,\\[1ex]
  ir+\CO(r^2) &\qquad r\to 0.
\end{cases}
\end{align}
This is exactly the same as \eqref{kkxt23Jan12}, as it should be.

Note that, the position of the branes and anti-branes,
$\alpha,\beta,\gamma$ and $\delta$, does not have to be in the $r>0$ region.
So, for example, we can have only one real brane at $z=\alpha$ with
$\Im\alpha>0$, and all others can be some kind of ``image'' branes with
$\Im\beta,\Im\gamma,\Im\delta<0$ which are necessary to make the
asymptotics right.  This is just like the method of image charges in
electromagnetism.

So, there certainly is a configuration of exotic branes leading to
non-geometric structure in the Einstein metric for the lower, non-compact
dimensions.  We found a configuration of circular exotic branes (with
anti-branes and/or image charges) in asymptotically flat
$\bbR^{3,1}\times S^1$.  As long as we keep away from the exotic brane,
the spacetime remains asymptotically flat but, if we go through the
brane, we go to a different spacetime with different asymptotics.  This
configuration does not satisfy the equation of motion and therefore is
not an actual solution of the theory.  However, it certainly gives
evidence for the existence of spacetimes with property (ii) at the
beginning of this section.
It would be very interesting to find an actual solution of string theory
with property (ii).

\section{Exotic branes and black holes}
\label{sec:exotic_branes_and_black_holes}

The supertube effect is a phenomenon in which some particular
combination of branes, when put together, produces a new type of brane.
One notable situation in string theory where multiple branes are put
together is the black hole.  In this section, we discuss the supertube
effect in the context of black holes.  In particular, we argue that, via
a multi-stage supertube effect, or ``double bubbling'', exotic branes
play an important role in black hole physics.

So far we discussed the supertube effect in which two types of brane are
put together, producing one new type of brane.  However, we can put
together three types of brane. As an example, take the 3-charge M2
system \cite{Bena:2004de} which is a well studied configuration in the
context of 5D black hole microstate counting \cite{Strominger:1996sh}.
In this case, the puffing up is known to occur as
\begin{align}
 \begin{array}{l}  
  \rm M2(56)\\
  \rm M2(78)\\
  \rm M2(9\ten)
 \end{array}
 \to
 \begin{array}{l}
  \rm M5(\psi 789\ten)\\
  \rm M5(\psi 569\ten)\\
  \rm M5(\psi 5678)
 \end{array},
 \label{puffup:M2M2M20}
\end{align}
where ``$\ten$'' denotes the $x^{10}$ direction.  Namely, the puffing up
occurs pairwise, producing three daughter branes which extend along an
arbitrary \emph{curve} parametrized by $\psi$ in $\bbR^4_{1234}$.  The
black ring solution \cite{Elvang:2004ds, Gauntlett:2004wh, Bena:2004de}
is the manifestation of this puffing up actually occurring.

Actually, the daughter charges on the right of \eqref{puffup:M2M2M20}
include combinations of charges which can pairwise puff up again, in
principle.  Such a second puffing-up would be:
\begin{align}
 \begin{array}{l}  
  \rm M2(56)\\
  \rm M2(78)\\
  \rm M2(9\ten)
 \end{array}
 \to
 \begin{array}{l}
  \rm M5(\psi 789\ten)\\
  \rm M5(\psi 569\ten)\\
  \rm M5(\psi 5678)
 \end{array}
 \to
 \begin{array}{l} 
  \rm 5^3(\phi 789\ten,\psi 56)\\
  \rm 5^3(\phi 569\ten,\psi 78)\\
  \rm 5^3(\phi 5678,\psi 9\ten)
 \end{array}
 \to\cdots
 \label{puffup:M2M2M2}
\end{align}
Namely, the system can polarize into exotic $5^3$ branes extended along
a \emph{surface} parametrized by $\psi,\phi$ in $\bbR^{4}_{1234}$. This
is the multi-stage supertube effect whose possibility was first pointed
out by the current authors in \cite{deBoer:2010ud}.  In
\cite{Bena:2011uw}, a supersymmetry argument was given in strong support
for such multi-stage supertube effect being indeed possible.  The
conjectured supersymmetric object along an arbitrary surface was dubbed
\emph{superstratum} in \cite{Bena:2011uw}.

Furthermore, as the dots in \eqref{puffup:M2M2M2} indicate, the
``granddaughter'' $5^3$-branes again form a combination that can combine
to produce still more charges.  Therefore, it appears that this
puffing-up process can in principle continue indefinitely, producing all
kinds of exotic charges appearing in Table
\ref{table:exotic_states}.\footnote{This process can in principle
produce objects with a codimension less than two.  However, their
relevance to black hole physics is more speculative and therefore we
focus on codimension-2 branes in this paper.}  If this is the case, then
the final state will be a complicated configuration of exotic
superstrata.  Conversely, it is also possible that this process happens
only a finite number of times for some dynamical reason. For example, whether 
supertubes provide a good (i.e. weakly coupled)
description of bound states depends on the string coupling and on the
other parameters of the theory, so presumably a similar statement is
valid for more complicated types of transitions. 
Either way, it
would be very interesting to investigate such processes can indeed
happen or not.

Another well-known example is the D0-D4-D4-D4 system, which has been
intensively studied in the context of 4D black hole microstate counting
\cite{Maldacena:1997de}, and involves four stacks of branes: D0,
D4(6789), D4(4589), D4(4567).  If we bring these four stacks together,
each pair is expected to puff up into a new brane stack:
\begin{align}
 {\rm D0}~~
 \begin{array}{l}
  \rm D4(6789)\\
  \rm D4(4589)\\
  \rm D4(4567)
 \end{array}
 ~~
 \to
 ~~
 \begin{array}{l}
  \rm NS5(6789\psi)\\
  \rm NS5(4589\psi)\\
  \rm NS5(4567\psi)
 \end{array}
 \begin{array}{l}
  \rm 5^2_2(6789,45\psi)\\
  \rm 5^2_2(4589,67\psi)\\
  \rm 5^2_2(4567,89\psi)
  \end{array}
 \label{D0D4D4D4puffup}
\end{align}
Interestingly, exotic branes make their appearance even at the first
stage of puffing up in this system.  The $\rm D4+D4\to 5^2_2$ supertube
we studied in section \ref{sec:supertube_effect_and_exotic_branes} in detail
is only a part of this process.  Therefore, a gravity description of
black hole physics in this frame inevitably involves exotic branes. It
would be interesting to find a solution corresponding to the right hand
side of \eqref{D0D4D4D4puffup}.  Also in this case, the daughter branes
on the right form a combination that can combine and produce new branes.
Again, the final state would be exotic superstrata along complicated
surfaces.

In the 2-charge system \cite{Lunin:2001jy}, the microscopic entropy
comes from the Higgs branch of the worldvolume theory associated with
the intersection of two stacks of branes.  In gravity, the same entropy
is explained by the degrees of freedom coming from the fluctuations of
the one-dimensional geometric object which is the result of puffing up
the intersection \cite{Rychkov:2005ji}.  In the 3-charge system, the
triple intersection of three stacks of branes leads to a more
complicated Higgs branch and larger microscopic entropy. It is
conceivable that the fluctuations of the above 2-dimensional exotic
superstratum that naturally appears, with its larger number of degrees
of freedom, account for the entropy of the 3-charge system.
It would hence be very interesting to construct non-geometric solutions
involving such exotic charges to see if they can really reproduce the
expected entropy.
The fact that the 3-charge supergravity microstates constructed thus far
(see {\it e.g.}\ \cite{Bena:2005va, deBoer:2009un}) do
not seem enough to account for the entropy of the 3-charge black hole
\cite{deBoer:2009un} may be related to the non-geometric nature of
exotic branes that have been overlooked.

In the explicitly known examples of microstate geometries for large
supersymmetric black holes (see {\it e.g.}\ \cite{Bena:2005va,
Berglund:2005vb, deBoer:2009un}) such exotic charges do not appear, and
it has been argued that purely geometric microstates alone cannot
account for their entropy \cite{deBoer:2009un}. It would be very
interesting to study and classify microstate geometries involving exotic
dipole charges, to see whether they can provide the missing microstate
geometries, and study their implications for black hole physics and the
fuzzball conjecture \cite{Mathur:2005zp}.\footnote{However, note that
more recent work \cite{Bena:2012hf, Lee:2012sc} shows that almost all
entropy of 4D black holes comes from ``pure-Higgs states'' which have no
angular momentum. Because configurations of branes along arbitrary
surface are expected to carry non-vanishing angular momentum in general,
it seems non-trivial to reproduce the degeneracy of such pure-Higgs
states.}

The multi-stage supertube effect, if it does occur, is special to
systems with 3 or more branes.  Because the daughter charge of a
supertube effect are not such that can combine with the parent charges
to produce a new charge, one needs more than two daughter charges to
produce a granddaughter.  This requires more than three parent charges
to begin with.  It is a curious coincidence that one needs at least
three stacks of branes to have a black hole with finite horizon in
string theory.

Also, it makes sense to construct black hole microstates using
codimension-2 objects, because they cannot be ``fattened''.  Namely,
supergravity solutions representing branes with codimension three or
larger can always be made non-extremal with a finite horizon, but there
is no non-extremal supergravity solution for codimension-2 branes.  This
is easy to see.  The metric produced by black D$p$-branes is
\cite{Duff:1994an}, by setting $q\equiv 7-p$,
\begin{align}
  ds_{\text{10,str}}^2&=H^{-1/2}(-Kdt^2+dx_1^2+\dots+dx_p^2)+H^{1/2}(K^{-1}dr^2+r^2d\Omega_{q+1}^2),\notag\\
 H&=h_0+\zeta\left({r_0\over r}\right)^{q},\qquad 
 K=1-\left({r_H\over r}\right)^{q},\\
 r_0^q&=c_q g_s N l_s^q,\quad
 c_q=(2\sqrt{\pi})^{q-2}\,\Gamma\!\left(\frac{q}{2}\right),\quad
 \zeta=\sqrt{1+\left[{1\over 2}\left({r_H\over r_0}\right)^{q}\right]^2}
 -{1\over 2}\left({r_H\over r_0}\right)^{q},\notag
\end{align}
where $N$ is the number of D$p$-branes, and $r_H$ is the position of the
horizon. The constant $h_0$ can be set to one by rescaling of
coordinates, but we leave it arbitrary.  We only displayed the metric,
but other fields are also excited.  In the extremal limit, $r_H\to
0,\zeta\to 1,K\to 1$.  If we further take the limit $p\to 7$ ($q\to 0$),
\begin{align}
 H=h_0+c_q g_s N
 \left({l_s\over r}\right)^{q}
 ~\xrightarrow{q\to 0}~
 {g_s N\over 2\pi q}+h_0+{g_sN\over 2\pi}\log{l_s\over r}
 +\CO(q).
\end{align}
Although the first term on the right is divergent as $q\to 0$, it can be
made finite by absorbing it into $h_0$. This way we can obtain the
metric for extremal D7-branes.  Let us ask if we can turn on
non-extremality $r_H>0$ keeping fields finite.  If we set
$r_H^q=a(q)\rho_H^q$, where $a(q)$ is some function of $q$,
\begin{align}
 K\xrightarrow{q\to 0} 1-a(q)-a(q)q\log\left({\rho_H\over r}\right).
\end{align}
In order to have a horizon, we need to keep the log so that $K=0$ has a
solution. This means that we need to take $a(q)\sim q^{-1}$ but this
will make $K$ diverge as $q\to 0$.  Therefore, there does not exist a
non-extremal D7-brane solution.
One may wonder if one can find codimension-2 black holes in the
blackfold approach \cite{Emparan:2009cs, Emparan:2009at}.  However, this
approach is applicable only when there are two widely separated scales
in the problem (thickness and curvature of the horizon), which is not
the case for codimension-2 objects.  Therefore, it is also consistent
with non-existence of codimension-2 black holes.
The fact that there is no ``fat'' codimension-2 object is good because,
if a microstate solution could be made non-extremal, or fattened, then
we would need some other objects to explain its Bekenstein--Hawking
entropy!

Although generic superstrata are expected to be exotic and thus
non-geometric, there can be special cases in which superstrata are completely
geometric.  Such geometric superstrata might not be sufficient for the
ultimate goal of reproducing black hole entropy, they are very
interesting in their own right and should also help us understand
non-geometric superstrata.  Geometric superstrata should be describable
within usual supergravity.  For recent progress toward constructing such
geometric superstrata in the context of 6D supergravity, see
\cite{Bena:2011dd, Niehoff:2012wu}.

In Eq.\ \eqref{ickz17Jun12}, we showed that the BPS bound state of two
kinds of brane with mass $g_s^{-a}$ and $g_s^{-b}$ is a supertube, which
is made of a brane with tension $T\sim g_s^{-(a+b)}$ and which is
spreading over distance $R\sim g_s^{(a+b)/2}$.  Then it is tempting to
conjecture that the BPS bound state of three kinds of brane with mass
$g_s^{-a},g_s^{-b},g_s^{-c}$ is a superstratum with tension $T\sim
g_s^{-(a+b+c)}$, typically spreading over distance $R\sim
g_s^{(a+b+c)/2}$, and that the bound state of four kinds of brane with
mass $g_s^{-a},g_s^{-b},g_s^{-c},g_s^{-d}$ is a superstratum with
tension $g_s^{-(a+b+c+d)}$ spreading over distance $R\sim
g_s^{(a+b+c+d)/2}$.  Based on this, we can make an interesting
observation, as follows.  Let us further assume that
\emph{non-supersymmetric} black holes are also made of polarized branes
spreading over distance $R\sim g_s^{(a+b+c)/2}$ or $R\sim
g_s^{(a+b+c+d)/2}$, and require that this $R$ reproduce the horizon
radius of the Schwarzschild black hole in $D$ dimensions,
\begin{align}
 r_H\sim (G_D)^{{1\over D-3}}\sim (g_s)^{2\over D-3},\label{fyig13Sep12}
\end{align}
where $G_D\sim g_s^2$ is the $D$-dimensional Newton constant.
The relation \eqref{fyig13Sep12} gives, for example,
\begin{align}
 D=6:~ r_H\sim g_s^{2/ 3},\qquad
 D=5:~ r_H\sim g_s,\qquad
 D=4:~ r_H\sim g_s^2,
\end{align}
which is to be equal to
$R\sim g_s^{(a+b+c)/2}$ or $R\sim
g_s^{(a+b+c+d)/2}$.
Because the mass of branes always come in integral powers of $1/g_s$, it
is impossible to have $R\sim r_H$ in $D=6$. On the other hand,
for $D=4,5$, it is possible to have brane charges satisfying $R\sim
r_H$.  For example,
\begin{align}
  \text{D1-D5-P system in $D=5$}   &: &\quad (a,b,c) &=(1,1,0), & T&\sim g_s^{-2},& R\sim g_s,\label{muhn11Jul12}\\
 \text{D0-D4-D4-D4 system in $D=4$}&: &\quad (a,b,c,d)&=(1,1,1,1),& T&\sim g_s^{-4},& R\sim g_s^2.\label{muhq11Jul12}
\end{align}
The first line \eqref{muhn11Jul12} suggests that the microstates of the
D1-D5-P black hole are described by polarized branes with tension $T\sim
g_s^{-2}$. In \cite{Bena:2011dd, Niehoff:2012wu}, it was argued that the
microstates of the \emph{supersymmetric} D1-D5-P black hole include
\emph{geometric} superstrata made of wiggly D1-D5 branes puffed up into
wiggly KK monopoles, which have tension $\sim g_s^{-2}$.  The above
seems to give support to such geometric superstrata actually being the
typical microstates of this system.  On the other hand,
\eqref{muhq11Jul12} suggests that the microstates of the 4D black hole
involves very exotic objects with tension $g_s^{-4}$ and its description
in terms of (multi-valued) metric in 10D is physically more
questionable, if not totally nonsensical.  One either has to go to 11D
or use the metric only as a qualitative guide.
The above discussion is very crude and not intended to be a rigorous
argument, equating the size of the supersymmetric superstrata and the
horizon radius of the Schwarzschild black hole.  However, we deem this
as an interesting and suggestive observation.

If a black hole microstate is made of some kind of
supertube/superstratum of exotic branes, one can imagine an object
freely falling into it.  The exotic supertube has some duality monodromy
around it.  Therefore, when the object goes into the microstate, it is
expected to generically go through the exotic superstratum and undergo a
duality transformation.  After exploring the complicated structure of
the superstratum, the object will eventually come back out (after a long
time which goes to infinity in the classical limit $G_N\to 0$).  For an
observer sitting at infinity, it appears that the object is no longer
the same entity but has turned into a $U$-dual version.  On the other
hand, from the point of view of the object, it appears that it has
returned to a different world---the $U$-dual version of the original
world it started from.  Of course, these two pictures are consistent
because they are only different (dual) descriptions of the same physics.
This is reminiscent of the still controversial black hole
complementarity \cite{Susskind:1993if, Stephens:1993an} which claims
that the two pictures of an object fallen into a black hole, in one of
which the object gets returned in the form of Hawking radiation and in
the other of which the object crosses the horizon unscathed, are two
different descriptions of the same physics.
Here we only point out this amusing similarity and leave further
investigations for future research.

In this paper, we have been arguing that exotic branes are generic
objects, and hence that, if we consider generic situations in string
theory, they will inevitably show up.  If we think this way, it is
natural that black holes must involve exotic branes.  Very near the
horizon of a black hole, large redshift allows us to have excitations
with very large proper energy, even non-perturbative ones, at very low
energy cost as measured at infinity.  So, near the horizon, all kinds of
objects in string theory can be pair created and therefore exotic
branes---the generic objects---will be the main ingredients floating
around near the horizon.  Namely, it seems natural that black holes are
made of non-geometric exotic branes.  This argument is valid even for
non-supersymmetric black holes, including Schwarzschild.

\section{Discussion and future directions}
\label{sec:discussion_and_future_directions}

Exotic branes are essential ingredients of string theory whose
significance has been long unappreciated.  In the current paper, we only
initiated the exploration of this {\it terra incognita\/} by studying
basic aspects of exotic branes, such as their monodromic nature,
supergravity solutions, and their implications for black hole physics.
Being fundamental elements in string theory, exotic branes are expected
to connect various areas in string theory and further investigation
should reveal its intriguing physics.

In this last section, we would like to discuss some of possible
directions for future research related to exotic branes, which we
particularly find interesting.

\medskip

{\it Non-abelian anyons.} Being codimension-2 objects and carrying
non-abelian monodromy charge, exotic branes have all the features to be
identified with non-abelian anyons.  Abelian anyons are particles in 2+1
dimensions whose wavefunction gets multiplied by $e^{i\theta}$ under
exchange of two particles, with general $\theta\in\bbR$ (bosons
correspond to $\theta=0$ and fermions to $\theta=\pi$).  Non-abelian
anyons are generalizations for which exchange of particles induces
multiplication by non-commutative matrices of the
wavefunction. Non-abelian anyons have attracted considerable attention
recently in the context of topological quantum computers
\cite{ToplQuantumComp} and it would be interesting to study exotic
branes as non-abelian anyons in this light.  It is possible, though
highly speculative, that exotic
branes are of some use for problems in quantum computation.

{\it Spacetimes with exotic monodromies.}  In a sense, exotic branes are
generalizations of F-theory $(p,q)$ 7-branes \cite{Greene:1989ya,
Vafa:1996xn}. The idea of F-theory as non-trivial fibrations of
$SL(2,\bbZ)$ monodromy allowed us to study non-perturbative aspects of
type IIB string theory and led to various applications, such as
realization of gauge theories with exceptional gauge groups
\cite{Johansen:1996am, Gaberdiel:1997ud, Gaberdiel:1998mv,
DeWolfe:1998eu, DeWolfe:1998pr, Bergshoeff:2009th} and string theory
realizations of phenomenological models \cite{Beasley:2008dc,
Beasley:2008kw}.
It is interesting to generalize the F-theory construction of fibering
duality groups to more general monodromies realized by exotic branes.
This direction was already pursued in \cite{Kumar:1996zx, Liu:1997mb,
Curio:1998bv, Leung:1997tw, Lu:1998sx, Vegh:2008jn} and recently there has been
recurring interest in this subject. 
One particularly interesting setting is type IIB superstring
compactified on K3, whose low energy description is given by
$d=6,\CN=4b$ supergravity.  This theory has $SO(5,m)$ duality group with
$m=21$,\footnote{Type IIB superstring compactified on $T^4$ can be
truncated to $\CN=4b$ supergravity with $m=5$, and can be studied in
this framework as well.} and there must be exotic branes with
$SO(5,m)$ monodromies around them.  It is interesting to explicitly
construct solutions with $SO(5,m)$ monodromies in this supergravity
theory.
Recently, Refs.~\cite{Chiodaroli:2012vc, Chiodaroli:2011nr} studied
supersymmetric solutions in this 6D theory preserving $AdS_2\times S^2$
symmetry and found interesting solutions with $SO(5,m)$ monodromies.
Their solutions correspond to self-dual strings ending on 3-branes,
whose 10D lift is the D1-D5 system (or the dual thereof) ending on a
D7-brane (or a dual thereof), giving 2D CFT with boundaries.  As one
goes around the D7-brane, one undergoes $SO(5,m)$ duality transformation
and the CFT becomes dualized as well.  Therefore, this gives an
interesting setup in which one can study properties of exotic branes in
terms of CFT, and {\it vice versa.}
Another recent paper \cite{Martucci:2012jk} is a more direct
generalization of F-theory configurations with $SL(2,\bbZ)$ monodromies
to $SO(5,m)$.  They assumed $\bbR^{1,3}$ isometry and the resulting
solutions include fibrations of a genus-2 Riemann surface over $\bbP^1$,
instead of torus fibrations over $\bbP^1$ as in F-theory.
It would be interesting to study the properties of such monodromic
spacetimes and also to construct more general solutions.  In particular,
this will help us understand what kind of monodromies are possible in
string theory, along the line of \cite{Bergshoeff:2006jj}.

A particularly interesting case to study is the classification of all
supersymmetric solutions of $\cN=16$ supergravity in three dimensions
\cite{Marcus:1983hb}. Preliminary work \cite{usnew} shows that these all
need to be of the form Riemann surface $\times$ time, but the analysis
of allowed Riemann surfaces with punctures and prescribed holonomies
appears to be very complicated. We hope to report more on this in the
future.

{\it Doubled geometry, double field theory and non-geometric
compactification.}  The spacetime around exotic branes involves
$U$-duality monodromy and is non-geometric.  The simplest case of
$U$-folds is the $T$-fold, in which spacetime is non-geometrically glued
together by $T$-duality.  $T$-duality is a perturbative duality in
string perturbation theory and worldsheet techniques are expected to be
useful for analyzing exotic brane of $T$-fold type, {\it i.e.}\ the
$5^2_2$-brane.
Doubled geometry \cite{Hull:2004in, Hull:2006va,
ReidEdwards:2008rd,Hull:2009sg,Nibbelink:2012jb} is a worldsheet
formalism which allows for studies of $T$-folds by doubling the
worldsheet fields in the compact torus directions.  This formalism is
expected to be useful to study the properties of $5^2_2$-branes, such as
worldsheet instanton effects on the $T$-duality among NS5-branes, KKM
backgrounds, and $5^2_2$-branes \cite{Tong:2002rq, Harvey:2005ab,
Jensen:2011jn}.

Double field theory (DFT) \cite{Hohm:2011gs, Zwiebach:2011rg} and its
$U$-duality generalizations \cite{Hillmann:2009ci, Berman:2010is,
Berman:2011pe, Berman:2011jh, Berman:2011cg} are frameworks in which the
usual coordinates $x^i$ and dual coordinates $\tilde{x}_i$ are on the
same footing and provide a natural setting in which to study
non-geometric exotic branes.  It would be interesting to study how
exotic branes are described in this new framework and study if it allows
to study aspects of exotic branes otherwise difficult to study.
The current formulation of DFT requires that we impose the so-called
strong constraint \cite{Hohm:2010jy, Hohm:2010pp} on the
$x^i,\tilde{x}_i$ dependence of fields (see however \cite{Hull:2009mi}).
This constraint allows the fields to depend only on $x^i$,
$\tilde{x}_i$, or their duals \cite{Hohm:2010jy}.  As a result, the
allowed configurations are ordinary geometric spacetimes locally, with
possible duality identifications globally.  This class of configuration
can locally be studied by the ordinary supergravity and the machinery of
DFT is not really needed.  For a more complete treatment of exotic
branes within DFT, it is necessary that the framework is developed
further so that we can relax the strong constraint.

However, more recently, it was shown \cite{Aldazabal:2011nj,
Geissbuhler:2011mx, Grana:2012rr, Berman:2012uy} (see also \cite{Andriot:2011uh, Andriot:2012wx, Andriot:2012an}) that one can
generalize DFT in a way such that its non-geometric Scherk--Schwarz
compactification reproduces the constraint equations in the low
dimensional theory expected for non-geometric compactifications.  It would be interesting to see if DFT can be used
to see if there is a relation between the non-geometric internal
manifolds appearing in the context of non-geometric compactifications
and the non-geometric spacetime produced by exotic branes.  The relation
would be similar to the geometric transition relating the geometry with
wrapped branes and the one in which the branes have disappeared and been
replaced by fluxes.
In other interesting recent work \cite{Chatzistavrakidis:2012qj}, a
Matrix Theory \cite{Banks:1996vh} description of non-geometric
compactifications was discussed.  It would be interesting to study how
exotic branes can be described within Matrix Theory and its $T$-dual
cousins.  In particular, it is interesting to see how the non-geometric
nature of spacetime is encoded in matrix configurations.

{\it Instanton corrections.}  It is well-known \cite{Green:1997tv} that
brane instantons induce higher curvature corrections to the low energy
effective action.  Such correction terms can be determined by requiring
$U$-duality symmetry, supersymmetry, and physical conditions on the
boundary of the moduli space \cite{Green:1998by}.  They are in general
proportional to the Eisenstein series which are eigenfunctions of the
Laplace operators on the moduli space, and provide valuable information
about non-perturbative physics of string theory.  Being codimension-2
objects, exotic brane instantons are expected to contribute to those
correction terms in $D\le 2$ dimensions, where the $U$-duality groups
are conjectured to be the infinite Kac--Moody groups
$E_{9},E_{10},E_{11}$.  It is interesting to examine how exotic brane
instantons are encoded in low energy effective action in $D\le 2$.  Recent
work on non-perturbative higher curvature corrections in $D\le 2$
includes \cite{Fleig:2012xa}.

{\it Exotic branes and $E_{11}$.}  It has been argued
\cite{Bergshoeff:2011se} that exotic branes are directly related to the
``mixed symmetry fields'' in 11 dimensions predicted by $E_{11}$
symmetry, which has been claimed to underlie M-theory
\cite{West:2001as}.  At this point, this is little more than numerology,
but it is certainly interesting to pursue this direction to better
understand the mysterious $E_{11}$ symmetry.  Recent papers discussing
the exotic branes in the context of $E_{11}$ include
\cite{Englert:2007qb, Cook:2009ri}.

In section \ref{sec:monodromies_and_charge_conservation}, we took some
explicit examples and demonstrated that charges are conserved in the
presence of non-trivial monodromies of exotic branes, if we use the
appropriate definition of charge.  It would be interesting to show this
for general cases in a more systematic formulation.  This basically
means to find the expression for Page charge that transforms covariantly
under $U$-duality.
In section \ref{sec:toward_truly_non-geometric_config}, we argued that
there exist configurations that are non-geometric even in lower
dimensions.  However, we could construct only off-shell configurations,
not solutions that satisfy the equation of motion.  It is desirable to
show that actual solutions exist which are non-geometric in lower
dimensions.

\section*{Acknowledgments}

We thank Iosif Bena, David Berman, Eric Bergshoeff, Bernard de Wit,
Roberto Emparan, Mariana Gra\~na, Shinji Hirano, Yosuke Imamura, Toru
Kikuchi, Tetsuji Kimura, Juan Maldacena, Don Marolf, David R.~Mayerson,
Paul McFadden, Sameer Murthy, Niels Obers, Takashi Okada, Kyriakos
Papadodimas, Boris Pioline, Joris Raeymaekers, Mukund Rangamani, Yuho
Sakatani, Savdeep Sethi, Tadashi Takayanagi, and Nick Warner for
fruitful discussions.  This work of JdB was supported in part by the
Foundation of Fundamental Research on Matter (FOM) and by an NWO Spinoza
grant.  The work of MS was supported in part by Grant-in-Aid for Young
Scientists (B) 24740159 from the Japan Society for the Promotion of
Science (JSPS)\@.

\appendix

\section{Conventions}
\label{app:conv}

When necessary, we display the rank of a $p$-form as a superscript as
$\omega^{(p)}$.  When we explicitly write the components of a $p$-from
$\omega^{(p)}$, we often omit the superscript;  for example,
\begin{align}
 \omega^{(p)}={1\over p!}\omega_{i_1\dots i_p}dx^{i_1}\wedge \cdots \wedge dx^{i_p}.
\end{align}

We define the Hodge dual of a $p$-form $\omega^{(p)}$ in $d$ dimensions as
\begin{align}
    (\ast \, \omega)_{i_1 \cdots i_{d-p}} = \frac{1}{p!}\epsilon_{i_1
  \cdots i_{d-p}}{}^{ j_1 \cdots j_p} \omega_{j_1 \cdots j_p},
\end{align}
where
\begin{align}
 \epsilon_{01 \cdots (d-1)} = -\sqrt{-g},\qquad
 \epsilon^{01 \cdots (d-1)} = +{1\over \sqrt{-g}}.\label{etgg14Apr12}
\end{align}
This means that
\begin{align}
   *(dx^{j_1}\wedge\cdots\wedge dx^{j_p})
 = \frac{1}{(d-p)!}\epsilon_{i_1 \cdots i_{d-p}}{}^{ j_1 \cdots j_p} 
 dx^{i_1}\wedge\cdots\wedge dx^{i_{d-p}}.
\end{align}
Note that
\begin{align}
 \epsilon^{i_1\dots i_p j_1\dots j_{d-p}}
 \epsilon_{i_1\dots i_p k_1\dots k_{d-p}}
 =-p!\,(d-p)!\,\delta^{[j_1}_{k_1}\cdots\delta^{j_{d-p}]}_{k_{d-p}}.
\end{align}
The minus sign comes from the Lorentzian signature.  Also note that, on
a $p$-form, $*^2=(-1)^{p(d-p)+1}$, where $+1$ comes from the Lorentzian
signature.

With this convention, we have
\begin{align}
 dx^{i_1}\wedge\cdots\wedge dx^{i_d}=\epsilon^{i_1\cdots i_d}\, \sqrt{-g}\,d^d x.
\end{align}
For $p$-forms $\omega,\eta$,
\begin{align}
 *\omega \wedge \eta
 =*\eta \wedge \omega
 &=-{1\over p!}\omega_{i_1\cdots i_p}\eta^{i_1\cdots i_p}\sqrt{-g}\,d^d x\\
 &=(-1)^{p(d-p)}\omega \wedge *\eta
 =(-1)^{p(d-p)}\eta \wedge *\omega
\end{align}

\bigskip
The $\Gamma$ matrices satisfy
\begin{align}
 \{\Gamma^{\hat{A}},\Gamma^{\hat{B}}\}&=2\eta^{\hat{A}\hat{B}},\qquad
 \eta^{\hat{A}\hat{B}}={\rm diag}(-++\dots +),
\end{align}
where hatted index such as $\hat{A},\hat{B}$ denote the local Lorentz
frame indices.  The $\Gamma$ matrices with spacetime indices are defined
by $\Gamma^M=e^M_{\hat A}\Gamma^{\hat A}$ where $e^M_{\hat A}$ is the
vielbein.  We also define $\Gamma^{MN\dots
P}\equiv\Gamma^{[M}\Gamma^N\cdots \Gamma^{P]}$, where antisymmetrization
is taken with weight one; for example, $\Gamma^{MN}\equiv {1\over
2}(\Gamma^{M}\Gamma^N-\Gamma^N\Gamma^M)$.  For a $p$-form $A^{(p)}={1\over
p!}A^{(p)}_{M_1\dots M_p}dx^{M_1}\wedge \cdots \wedge dx^{M_p}$, we define
\begin{align}
 \slash{A}^{(p)}\equiv {1\over p!}\Gamma^{M_1\dots M_p}A_{M_1\dots M_p}.
\end{align}
The covariant derivative of a spinor $\epsilon$ is defined by
\begin{align}
 \nabla_M\epsilon=\partial_M \epsilon +{1\over 4}\omega_{M\hat A\hat B}\Gamma^{\hat A\hat B}\epsilon,
\end{align}
where $\omega_{M\hat A\hat B}$ is the spin connection.

\subsection{Duality rules}
\label{app:duality_rules}

Let us first consider $T$-duality \cite{Imamura}.  Let the 10D metric in
string frame in type IIA be
\begin{align}
 ds_{\text{IIA}}^2=g_{\mu\nu}^{(9)}dx^\mu dx^\nu+e^{2\sigma}(\widetilde{dx^9})^2,\qquad
 \widetilde{dx^9}=dx^9+v_1,
\end{align}
where as picked $x^9$ as special direction, and $\mu,\nu=0,\dots,8$.  Let the NSNS
$B$-field be
\begin{align}
 B_2=b_2+b_1\wedge\widetilde{dx^9},
\end{align}
and define the 9D dilaton by
\begin{align}
 \varphi_9=\Phi-\half \sigma
\end{align}
where $\Phi$ is the 10D dilaton.  
Performing a $T$-duality along $x^9$, 
the  metric, $B$-field, dilaton in type IIB are given by
\begin{align}
 ds_{\text{IIB}}^2&=g_{\mu\nu}'^{(9)}dx^\mu dx^\nu+e^{2\sigma}(\widehat{dx^9})^2,\qquad
 \widehat{dx^9}=dx^9+v_1',\\
 B_2'&=b_2'+b_1'\wedge\widehat{dx^9},\qquad
 \varphi_9'=\Phi'-\half \sigma'
\end{align}
with
\begin{align}
 g_{\mu\nu}'^{(9)}=g_{\mu\nu}^{(9)},\qquad
 v_1'=-b_1,\qquad b_1'=-v_1,\qquad
 b_2'=b_2+b_1\wedge v_1,\qquad
 \sigma'=-\sigma,\qquad
 \varphi'=\varphi.
\end{align}
For the RR potentials, if decompose them in type IIA as
\begin{align}
 C_{\text{odd}}=c_{\text{odd}}+c_{\text{even}}\wedge \widetilde{dx^9},
\end{align}
the type IIB  ones are given by
\begin{align}
 C_{\text{even}}'=c_{\text{even}}+c_{\text{odd}}\wedge \widetilde{dx^9}.
\end{align}
Here, $C_{\text{odd}},C_{\text{even}}'$ are formal sums of RR potentials
with various rank, as defined in \eqref{ffo12Apr10},
\eqref{gpca13Sep12}.

The $S$-duality transformation rules are
\begin{align}
 ds_{\rm 10,str}'^2=e^{-\Phi}ds_{\rm 10,str}^2,\quad \Phi'=-\Phi,\quad
 B_2'=C_2, \quad C_2'=-B_2, \quad \tilde C_4'=\tilde C_4,
\end{align}
where $\tilde C_4=C_4-\half B_2\wedge C_2$.

The relation between the M-theory fields and type IIA ones is
\begin{align}
 ds_{11}^2&=e^{-{2\over 3}\Phi}ds_{\rm 10,str}^2
 +e^{{4\over 3}\Phi}(dx^\ten+C_1)^2,\qquad
 A_3=C_3+B_2\wedge dx^\ten,
\end{align}
where $x^\ten$ is the eleventh direction.

\section{10D lift of 8D T-duality on spinors}

\label{app:10D-8D}

\subsection{Relation between 10D and 8D spinors}


If we denote the 8D Gamma matrices by $\gamma^{\hat{\mu}}$,
$\mu=0,\dots,7$, which are $16\times 16$ matrices, the 10D Gamma
matrices $\Gamma^{\hat{M}}$, $M=0,\dots,9$, which are $32\times 32$
matrices, can be written as
\begin{align}
\label{fheg30Jul12}
\begin{split}
 \Gamma^{\hat{\mu}}&=\gamma^{\hat{\mu}}\otimes (-\tau^3), \qquad \mu=0,\dots,7,\\
 \Gamma^{\hat{8}}&=\1\otimes \tau^1,\qquad \Gamma^{\hat{9}}=\1\otimes \tau^2,
\end{split}
\end{align}
where $\tau^i$ are the Pauli matrices.  Here we are following the
convention of \cite{Polchinski:1998rr}.
The 8D chirality matrix
$\gamma^{9}\equiv -i\gamma^{\hat{0}}\cdots\gamma^{\hat{7}}$ is related
to the 10D one by
\begin{align}
 \Gamma^{11}&=\gamma^9\otimes \tau^3.
\end{align}
A 8D Majorana spinor $\zeta_8$ is defined by 
\begin{align}
 \zeta_8^*=B_{8}\zeta_8,\qquad
 B_{8}=\gamma^{\hat{3}}\gamma^{\hat{5}}\gamma^{\hat{7}},
\end{align}
while a 10D Majorana spinor $\zeta_{10}$ is defined by
\begin{align}
 \zeta^*_{10}=B_{10}\zeta_{10},\qquad
 B_{10}=\Gamma^{\hat{3}}\Gamma^{\hat{5}}\Gamma^{\hat{7}}\Gamma^{\hat{9}}
 =B_8\otimes i\tau^1.
\end{align}
One can take $B_8,B_{10}$ to satisfy
\begin{align}
 B_8^*=-B_8,\qquad B_{10}^*=B_{10}.
\end{align}

Using the above, it is easy to show that a 10D Majorana--Weyl spinor
$\zeta_{10\pm}$ with positive (negative) chirality can be decomposed as
\begin{align}
 \label{ndwz29Jul12}
\zeta_{10\pm}=\zeta_{8\pm}\otimes (\begin{smallmatrix}1\\0\end{smallmatrix})
 +iB_8 \zeta_{8\pm}^*\otimes (\begin{smallmatrix}0\\1\end{smallmatrix})
\end{align}
where $\zeta_{8\pm}$ is a 8D Weyl spinor with positive (negative)
chirality.  Note that a 10D Majorana--Weyl spinor and a 8D Weyl spinor
have the same number of independent (16 real) components.
Or, equivalently,
\begin{align}
\label{femp30Jul12}
 \zeta_{10\pm}=iB_8(\zeta_{8\mp}')^*\otimes (\begin{smallmatrix}1\\0\end{smallmatrix})
 +\zeta_{8\mp}'\otimes (\begin{smallmatrix}0\\1\end{smallmatrix}).
\end{align}
where $\zeta'_{8\pm}$ is a 8D Weyl spinor with positive (negative)
chirality.

\subsection{10D lift of 8D T-duality action on spinors}

In section \ref{ss:susy}, we discussed how the 8D spinor $\eta_A$ transforms
under the 8D duality transformation $q\in SL(2,\bbZ)_\rho$, and how the
transformation law can be rewritten in terms of the 10D spinor
$\epsilon$ as in \eqref{msqq29Jul12}.  Here, we derive this relation.

The 10D $\CN=2$ supergravity has two gravitinos and their supersymmetry
transformation law is parametrized by two 10D Majorana--Weyl spinors
$\epsilon_{1},\epsilon_2$, whose chirality depends on whether we are
considering type IIA or type IIB supergravity as follows:
\begin{align}
\begin{split}
 \text{IIA}:&\qquad \Gamma^{11}\epsilon_1=+\epsilon_1,\qquad
 \Gamma^{11}\epsilon_2=-\epsilon_2\\
 \text{IIB}:&\qquad \Gamma^{11}\epsilon_1=-\epsilon_1,\qquad
 \Gamma^{11}\epsilon_2=-\epsilon_2.
\end{split}
\end{align}
When compactified on a 2-torus $T^2_{89}$, this theory reduces to 8D
$\CN=2$ supergravity whose supersymmetry transformation law is
parametrized by two 8D spinors $\eta_A$, $A=1,2$, both of which are
positive chirality Weyl spinors ($\gamma^9\eta_A=+\eta_A$).  This theory
has $U$-duality group $SL(3)_U\times SL(2)_U$ whose maximal compact
subgroup is the $R$-symmetry group $SU(2)_R\times U(1)_R$.  If one
performs an $SL(3)_U\times SL(2)_U$ duality transformation, $\eta_A$
transforms under $SU(2)_R\times U(1)_R$ in the ${\bf 2}_{\half}$
representation as one can see from Table 16 of
\cite{Andrianopoli:1996ve}.  As usual
\cite{Schwarz:1983qr, PopeKaluzaKlein}, this is induced by the
compensating gauge transformation.

The 8D $T$-duality group $SO(2,2)=SL(2)_\tau\times SL(2)_\rho$ is a
subgroup of the 8D $U$-duality group $SL(3)_U\times SL(2)_U$. 
The relation between the two is  \cite{Obers:1998fb}
\begin{align}
\begin{split}
 \text{IIA}:&\qquad SL(2)_\tau\subset SL(3)_U,\qquad SL(2)_\rho=SL(2)_U,\\
 \text{IIB}:&\qquad SL(2)_\tau    =   SL(2)_U,\qquad SL(2)_\rho\subset SL(3)_U.
\end{split}
\end{align}
This means that,
if we do an $SL(2)_\rho$ transformation
\begin{align}
 q=\begin{pmatrix}a&b\\c&d\end{pmatrix}\in SL(2)_\rho,\qquad ad-bc=1,
\end{align}
the 8D spinor $\eta_A$, being in ${\bf 2}_{\half}$ of $SU(2)_R\times
U(1)_R\subset SL(3)_U\times SL(2)_U$, transforms as
\begin{align}
\label{ndov29Jul12}
 \eta_A\to 
 \begin{cases}
  e^{{i\over 2}\arg(c\rho+d)}\eta_A                   &\qquad \text{(IIA)},\\[1ex]
  [e^{{i\over 2}\arg(c\rho+d)\sigma^3}]_A{}^B\,\eta_B &\qquad \text{(IIB)},
 \end{cases}
\end{align}
which is \eqref{ndez29Jul12}.

We would like to rewrite \eqref{ndov29Jul12} in terms of the 10D spinor
$\epsilon=\left(\begin{smallmatrix}\epsilon_1\\
\epsilon_2\end{smallmatrix}\right)$.  Considering the chirality of
$\epsilon_{1,2},\eta_{A}$ and using
\eqref{ndwz29Jul12},\eqref{femp30Jul12}, the reduction rules are
\begin{align}
 \text{IIA}:&& \epsilon_1&=\eta_{1}\otimes (\begin{smallmatrix}1\\0\end{smallmatrix})
 +iB_8 \eta_{1}^*\otimes (\begin{smallmatrix}0\\1\end{smallmatrix}),
 &
 \epsilon_2&=
\eta_2\otimes (\begin{smallmatrix}0\\1\end{smallmatrix})
+iB_8\eta_2^*\otimes (\begin{smallmatrix}1\\0\end{smallmatrix})
 \\
 \text{IIB}:&&
 \epsilon_1&=\eta_{1}\otimes (\begin{smallmatrix}0\\1\end{smallmatrix})
 +iB_8 \eta_{1}^*\otimes (\begin{smallmatrix}1\\0\end{smallmatrix}),
 &
 \epsilon_2&=\eta_{2}\otimes (\begin{smallmatrix}0\\1\end{smallmatrix})
 +iB_8 \eta_{2}^*\otimes (\begin{smallmatrix}1\\0\end{smallmatrix})
\end{align}
Using this, it is easy to see that the transformation law
\eqref{ndov29Jul12} can be written as
\begin{align}
 \epsilon_1 \to e^{\pm {i\over 2}\tau_3 \arg(c\rho+d)}\epsilon_1,\qquad
 \epsilon_2 \to e^{\mp {i\over 2}\tau_3 \arg(c\rho+d)}\epsilon_2,
\end{align}
where the $\pm,\mp$ signs are for IIA/IIB\@.
In terms of the doublet $\epsilon=\left(\begin{smallmatrix}\epsilon_1\\
\epsilon_2\end{smallmatrix}\right)$, this is
\begin{align}
 \epsilon\to e^{\pm {i\over 2}\tau_3\sigma_3\arg(c\rho+d)}\epsilon
 =e^{\pm {1\over 2}\Gamma^{\hat{8}\hat{9}}\sigma_3\arg(c\rho+d)}\epsilon,
\end{align}
where in the last equality we used \eqref{fheg30Jul12}.
This equation was used in \eqref{msqq29Jul12}.

\section{Derivation of (\protect\ref{ickd12Jun12}), (\protect\ref{icrg12Jun12})}
\label{app:derive_int_db,int_dgamma}

From \eqref{jtti16Feb10},
\begin{align}
 \int_{\Sigma}d\beta_I =\int_{\Sigma}{*_3d}f_I
 =\int_{B^3}d{*_3df_I},
\end{align}
where $B^3$ is a 3-manifold such that $\partial B^3=\Sigma$.
Here,
\begin{align}
 d{*_3df_1}
 ={Q_1\over L}\int_0^L 
 dv\,\,d{*_3 d}\!\left({{1\over |{\bf x}-{\bf F}(v)|}}\right)
 =
 -{4\pi Q_1\over L}\int_0^L 
 dv\,\,\delta^{(3)}({\bf x}-{\bf F}(v))d^3x,
\end{align}
where $d^3x=dx^1\wedge dx^2\wedge dx^3$.  Thus we arrive at
\eqref{icrg12Jun12} for $I=1$ as follows:
\begin{align}
 \int_{\Sigma}d\beta_1
 =-{4\pi Q_1\over L}\int_0^L dv\int_{B^3}
 d^3x\,\delta^{(3)}({\bf x}-{\bf F}(v))
 =-{4\pi Q_1\over L}\int_0^L dv
 = -{4\pi Q_1},
\end{align}
where in the second equality we used the fact that the entire profile is
inside $B^3$. The derivation for $I=2$ is similar.

On the other hand,
\begin{align}
 \int_{c}d\gamma&=\int_{c}*_3dA
 =-{Q_1\over L}\int_0^L dv \int_c {*_3  d}
 \biggl({\dot{F}_i(v)dx^i\over |{\bf x}-{\bf F}(v)|}\biggr)
 =-{Q_1\over L}\epsilon_{ijk}\int_0^L dv \int_c
 {\dot{F}_i (x-F)_j dx^k\over |{\bf x}-{\bf F}(v)|^3}.\label{iwse13Jun12}
\end{align}
Because $d^2\gamma=0$, this integral is invariant under homotopic
deformation of $c$.  So, let us take it to be a very small circle going
around a point on the profile. Near that point, take a coordinate system
so that
\begin{align}
 F_1(v)=F_2(v)=0,\qquad F_3(v)=fv,\qquad
 x_1+ix_2=re^{i\theta}.
\end{align}
Because the integrand of \eqref{iwse13Jun12} is $\sim |x_3|^{-2}$,
the only contribution comes near that point, and
\begin{align}
 \int_{c}d\gamma
 &={nfQ_1\over L}\int_0^L dv \int
 {r^2 d\theta \over [r^2+(x_3-fv)^2]^{3/2}}
 ={4\pi n Q_1\over L},
\end{align}
where $n$ is the number of times the profile is going through the 
point. This is \eqref{ickd12Jun12}.

\section[Page charge for D-branes]{Page charge for D-branes\protect\footnote{We thank S.~Hirano and D.~Marolf for helpful discussions on this subject.}}
\label{app:page_chg}

In the presence of Chern--Simons terms or modified Bianchi
identities,\footnote{Here we are restricting to Abelian gauge fields.}
it is possible for gauge fields to carry charge and consequently there
are more than one possible notions of charge one can naturally define
\cite{Marolf:2000cb}.  Brane source charge is gauge-invariant and
localized but not conserved or quantized. Maxwell charge is
gauge-invariant and conserved but not localized or quantized.  Page
charge is conserved, localized, and quantized but is gauge-invariant
only up to small gauge transformation; it will change under large gauge
transformation.

In supergravity that is the low energy description of superstring
theory, such Chern--Simons terms are present and we must use an
appropriate notion of charge depending on the physical purpose of the
analysis.
Here we study different notions of D-brane charge in string theory.  We
first recall the equations of motion of type IIA and IIB supergravity
without sources.  Then, we introduce D-brane sources that appear on the
right hand side of the supergravity equations of motion and study how to
define brane source charge and Page charge for D-branes.  The result for
D-brane Page charge is used in the main text where we are interested in
charge conservation in exotic backgrounds.

The Page charge for D-branes was studied in \cite{Marolf:2000cb} in the
absence of NS5-branes. A partial analysis of D-brane Page charge in the
presence of NS5-branes was done in \cite{Aharony:2011yc} (see also
\cite{Aharony:2009fc}) based on explicit examples but no complete
analysis has appeared in the literature; in particular, it was not clear
how to define conserved Page charge in the presence of both D-branes and
NS5-branes.  We fill this gap and describe how to define Page charge in
the presence of both types of brane, although we consider only the
Wess--Zumino terms and not the DBI term.  We expect that this result has
a wide range of possible application.

In this Appendix, we will display the rank of a form as a subscript; for
example, $C_{p+1}$ is a $(p+1)$-form.  We will often omit the exterior
product symbol $\wedge$ to avoid clutter.  For our convention for
differential forms and the Hodge star $*$, see Appendix
\ref{app:conv}\@.

\subsection{Supergravity equations without sources}

\medskip
\subsubsection*{10D type IIA}

The bosonic part of the action for 10D IIA supergravity is
\begin{align}
 {2\kappa_{10}^2}S^{\text{IIA}}
 &=
 \int d^{10}x \sqrt{-g}\biggl[e^{-2\Phi}\left(R+4(\partial \Phi)^2-{1\over 2\cdot 3!}H_{\mu\nu\rho}H^{\mu\nu\rho}\right)
\notag\\
 & \qquad\qquad\qquad\qquad
 -{1\over 2\cdot 2}G_{\mu\nu}G^{\mu\nu}-{1\over 2\cdot 4}G_{\mu\nu\rho\sigma}G^{\mu\nu\rho\sigma}\biggr]-{1\over 2}\int B_2\wedge dC_3\wedge dC_3\notag\\
 &=
 \int \biggl[e^{-2\Phi}\left(-\,{*R}-4\,{*d\Phi}\wedge d\Phi+{1\over 2}{*H_3}\wedge H_3\right)
 \notag\\
 &\qquad\qquad\qquad\qquad
 +{1\over 2}{*G_2}\wedge G_2+{1\over 2}{*G_4}\wedge G_4
  -{1\over 2} B_2\wedge dC_3\wedge dC_3\biggr],\label{cgz12Apr10}
\end{align}
where
\begin{align}
2\kappa_{10}^2\equiv (2\pi)^7 l_{s}^8,\qquad
 H_3\equiv dB_2,\qquad 
G_2\equiv dC_1,\qquad G_4\equiv dC_3-H_3\wedge C_1.
\end{align}
The  bosonic equations of motion derived from this action and Bianchi
identities for form fields can be written as
\begin{subequations}
 \begin{align}
 dH_3 &= 0, &  d(*e^{-2\Phi}H_3)  &=-G_2\wedge *G_4 +\frac{1}{2}G_4\wedge G_4\label{klyi31Jan12}\\
 dG_2 &= 0, &  d* G_2 &= -H_3\wedge *G_4\label{jits9Aug12} \\
 dG_4 &= H_3 \wedge G_2,   & d*G_4 &= H_3\wedge G_4.
  \label{jitz9Aug12}
 \end{align}
\end{subequations}
If we define $G_6,G_8$ by
\begin{align}
  G_6=+*G_4,\qquad G_8=-*G_2,\qquad
  G_4=-*G_6,\qquad G_2=+*G_8,
\end{align}
or more concisely by \cite{Bergshoeff:2001pv}
\begin{align}
\begin{split}
  *G_p&=(-1)^{p(p-1)/2}G_{10-p}
 =(-1)^{\floor{p\over 2}}G_{10-p},\\
 G_p&=(-1)^{p(p+1)/2+1}\,{*G}_{10-p}
 =(-1)^{\floor{p\over 2}+p+1}\,{*G}_{10-p}
\end{split}
\label{mqkq18Apr12}
\end{align}
where $\floor{k}$ (the floor function) is the largest integer less than
or equal to $k$, then the equations \eqref{jits9Aug12},
\eqref{jitz9Aug12} can be written uniformly as
\begin{align}
  dG_p - H_3 \wedge G_{p-2}=0,\qquad p=2,4,6,8.
 \label{eqn12Apr10}
\end{align}
Some useful formulas involving $\floor{k}$ are, for $n\in\bbZ$,
\begin{align}
 (-1)^{\floor{n\over 2}}
 &=(-1)^{\floor{-n+1\over 2}}
 =(-1)^{n(n-1)/2},
 &
 (-1)^{\floor{-{n\over 2}}}
 &=(-1)^{\floor{n+1\over 2}}
 =(-1)^{\floor{n\over 2}+n},
 \notag\\
 (-1)^{\floor{n-1\over 2}} 
 &=(-1)^{\floor{n\over 2}+n+1},
 &
 (-1)^{\floor{-n-1\over 2}}
 &=(-1)^{\floor{n\over 2}+1}.\label{niqn11Aug12}
\end{align}

If we define the formal sum
\begin{align}
 G_{\text{even}}=\sum_{\text{$p$:\,even}}G_p
\end{align}
then the equations of motion \eqref{eqn12Apr10} can be written
collectively as \cite{Bergshoeff:2001pv}
\begin{align}
  dG_{\text{even}} -H_3 \wedge G_{\text{even}}=0.\label{hztl12Apr12}
\end{align}
and the relation \eqref{mqkq18Apr12} translates into the
``anti-self-duality'' of $G_{\text{even}}$,
\begin{align}
 *G_{\text{even}}+\CT G_{\text{even}}=0.
\end{align}
Here, transpose $\CT$ is defined by \cite{Imamura}
\begin{align}
 \CT dx^{i_1}\wedge\cdots \wedge dx^{i_p}
 = dx^{i_p}\wedge\cdots \wedge dx^{i_1}.
\end{align}
In other words, on a $p$-form, $\CT=(-1)^{p(p-1)/2}=(-1)^{\floor{p\over
2}}$.  The equation of motion for $B_2$ can be written as
\begin{align}
\label{xdn12Apr12}
\begin{split}
  dH_7&=-G_2\wedge G_6+{1\over 2}G_4\wedge G_4
 =\half\sum_n (-1)^n G_{2n}\wedge G_{8-2n}
 ={1\over 2}(G_{\text{even}}\wedge \CT G_{\text{even}})_8,\\
 H_7&\equiv e^{-2\Phi}{*H}_3,
\end{split}
\end{align}
where $()_8$ means the 8-form part.

The $H_3$ equation \eqref{klyi31Jan12} says that we can write $H_3$ in terms
of the potential $B_2$ as
\begin{align}
 H_3=dB_2.
\end{align}
The $G_{\text{even}}$ equation \eqref{hztl12Apr12} means that we can
write it in terms of the potential $C_{\text{odd}}$ as
\begin{align}
\label{ffo12Apr10}
\begin{split}
  G_{\text{even}}
 &=dC_{\text{odd}}-H_3\wedge C_{\text{odd}}.
\end{split} 
\end{align}
where $C_{\text{odd}}=\sum_{\text{$p$:\,odd}}C_{p}$.  The field
strengths $H_3$ and $G_{\text{even}}$ are invariant under the gauge
transformation
\begin{align}
 \delta B_2&=d\zeta_1,\qquad
 \delta C_{\text{odd}}=d\lambda_{\text{even}}-H_3\wedge \lambda_{\text{even}},
\label{idaq12Apr12}
\end{align}
where $\zeta_1$ is a 1-form while $\lambda_{\text{even}}$ is a sum of
even forms, $\lambda_{\text{even}}=\sum_{\text{$p$:\,even}}\lambda_p$.

Likewise, the $H_7$ equation \eqref{xdn12Apr12} can be solved by
\begin{align}
 H_7&=dB_6+{1\over 2}(-G_2C_5+G_4C_3-G_6C_1)
 =dB_6+\half\sum_n(-1)^n G_{2n}C_{7-2n}\notag\\
&=dB_6+{1\over 2}(\cT G_{\text{even}}\wedge C_{\text{odd}})_7.\label{ffwy18Apr12}
\end{align}
This is invariant
under the gauge transformation
\begin{align}
 \delta B_6
 &=d\zeta_5+{1\over 2}(G_2\lambda_4-G_4\lambda_2+G_6\lambda_0)
 =d\zeta_5+{1\over 2}\sum_n (-1)^{n-1} G_{2n}\lambda_{6-2n}\notag\\
 &=d\zeta_5-{1\over 2}(\cT G_{\text{even}}\wedge \lambda_{\text{even}})_6.
 \label{icci12Apr12}
\end{align}

\subsubsection*{10D type IIB}

In type IIB supergravity in 10D, the bosonic equations of motion and Bianchi
identities for the forms can be written as
\begin{align}
 dH_3 &= 0, & d(e^{-2\Phi}*H_3) &= G_1\wedge *G_3+G_3\wedge *G_5,\label{klvk31Jan12}  \\
 dG_1 &=0, & d* G_1  &= - H_3\wedge * G_3, \\
 dG_3 &= H_3 \wedge G_1, & d* G_3  &= -H_3\wedge * G_5,\\
 dG_5 &= H_3 \wedge G_3, & *G_5&=G_5
\end{align}
If we define $G_7,G_9$ by
\begin{align}
  G_7=-*G_3,\qquad G_9=+*G_1,\qquad
  G_3=-*G_7,\qquad G_1=+*G_9,
\end{align}
or more concisely by \eqref{mqkq18Apr12}, then the form equations of
motion can be written as
\begin{align}
  dG_p = H_3 \wedge G_{p-2}.
\end{align}
If we define the formal sum
\begin{align}
 G_{\text{odd}}&=\sum_{\text{$p$:\,odd}}G_p
\end{align}
then the equation of motion is simply
\begin{align}
  G_{\text{odd}} = H_3 \wedge G_{\text{odd}}.\label{eiz12Apr10}
\end{align}
and $G_{\text{odd}}$ satisfies ``self-duality''
\begin{align}
 *G_{\text{odd}}-\CT G_{\text{odd}}=0.
\end{align}
The equation of motion for $B_2$ is
\begin{align}
 dH_7&=-G_1\wedge G_7+G_3\wedge G_5
 =\half\sum_n (-1)^{n+1}G_{2n+1}\wedge G_{7-2n}
 ={1\over 2}(G_{\text{odd}}\wedge      \CT G_{\text{odd}})_8,\notag\\
 H_7&=*e^{-2\Phi}H_3.\label{xot12Apr12}
\end{align}

The equations of motion \eqref{eiz12Apr10} mean that we can write the
field strengths in terms of potentials as
\begin{align}
 G_{\text{odd}}
 =dC_{\text{even}}-H_3\wedge C_{\text{even}},
\qquad C_{\text{even}}=\sum_{\text{$p$:\,even}}C_p.\label{gpca13Sep12}
\end{align}
On the other hand, the $H_7$ equation \eqref{xot12Apr12} can be solved by
\begin{align}
 H_7&=dB_6+{1\over 2}(G_1C_6-G_3C_4+G_5C_2-G_7C_0)
 =dB_6+\half\sum_n(-1)^n G_{2n+1}C_{6-2n}\notag\\
 &=dB_6+{1\over 2}(\cT G_{\text{odd}}\wedge C_{\text{even}})_7.\label{ffzb18Apr12}
\end{align}

We have the gauge symmetry
\begin{align}
 \delta B_2&=d\zeta_1,\qquad
 \delta C_{\text{even}}=d\lambda_{\text{odd}}-H_3\wedge \lambda_{\text{odd}},
 \qquad \lambda_{\text{odd}}=\sum_{\text{$p$:\,odd}}\lambda_p,
 \label{ibzv12Apr12}
\end{align}
under which the field strengths $G_{\text{odd}},H_3$ are invariant.
Also, it is easy to show that $H_7$ is invariant under
\begin{align}
 \delta B_6
 &=d\zeta_5+{1\over 2}(G_1\lambda_5-G_3\lambda_3+G_5\lambda_1)
 =d\zeta_5+{1\over 2}\sum_p (-1)^{n} G_{2n+1}\lambda_{5-2n}\notag\\
 &=d\zeta_5+{1\over 2}(\cT G_{\text{odd}}\wedge \lambda_{\text{odd}})_6.
\end{align}

\subsection{Inclusion of sources}

In the above, we considered equations of motion of type IIA/IIB
supergravity in 10D without charge source.  Now, let us introduce
D-brane sources and study how the equations of motion get modified.  By
looking at the structure of the equations, we will define conserved Page
charge for D-branes.

\subsubsection{D-brane sources}

First, let us begin with the case without NS5-brane source and hence
$dH_3=0$.  However, $B_2,H_3$ can be non-vanishing.  Here we consider
both type IIA/B at the same time and represent $C_{\text{odd}}$ and
$C_{\text{even}}$ both by $C$\@.  Similarly, $G=G_{\text{even/odd}}$ and
$\lambda=\lambda_{\text{even/odd}}$.

Since the D$p$-brane is electrically coupled to $C_{p+1}$, the D-brane
action includes $\int_{p+1} C$, where $\int_{p+1}$ denotes the integral
over the $(p+1)$-dimensional worldvolume of the D$p$-brane.  However,
this minimal coupling $\int_{p+1} C$ is not invariant under the $C$
gauge transformation
\begin{align}
 \delta C=d\lambda-H_3\lambda.\label{fuvg10Aug12}
\end{align}
Therefore, the minimal coupling is not sufficient and should be modified
to the well-known Wess--Zumino (WZ) action $\int_{p+1} e^{-B_2}C$, which
is invariant under \eqref{fuvg10Aug12} because its variation is
a total derivative:
\begin{align}
 \delta (e^{-B_2} C)=e^{-B_2}(d\lambda-H_3\lambda)
 =d(e^{-B_2}\lambda ).\label{jwxb18Apr12}
\end{align}

This is not the end of the story;
in order to make the WZ term invariant also under the $B_2$ gauge
transformation
\begin{align}
 \delta B_2=d\zeta_1,
\end{align}
we should have a worldvolume 1-form gauge field $V_1$ with the
transformation rule
\begin{align}
 \delta V_1=-{1\over 2\pi \ap}\zeta_1,
\end{align}
so that
\begin{align}
 2\pi \ap\cF_2\equiv B_2+2\pi \ap F_2\equiv B_2+2\pi \ap dV_1
\end{align}
is invariant.  This is how we find the WZ action for a D$p$-brane to be
\begin{align}
 S_{\rm WZ}^{\text{D$p$}}=
 {1\over (2\pi)^pl_s^{p+1}}\int_{p+1}e^{-2\pi \ap\cF_2}C
 = {1\over (2\pi)^pl_s^{p+1}}\int_{10} e^{-2\pi \ap\cF_2}\, C \, 
 \delta^{\text{D$p$}}_{9-p},
 \label{inyw18Apr12}
\end{align}
where $\int_{10}$ denotes the integral over the entire 10-dimensional
spacetime. $\delta^{\text{D$p$}}_{9-p}$ is a $(9-p)$-form whose support
is a delta function along the brane worldvolume and whose form
components are transverse to the brane worldvolume.  More precisely,
$\delta^{\text{D$p$}}_{9-p}$ satisfies, for any $(p+1)$-form
$\omega_{p+1}$,
\begin{align}
 \int_{p+1}\omega_{p+1}
 = \int_{10} \omega_{p+1} \wedge \delta^{\text{D$p$}}_{9-p}.
\end{align}
In other words, $\delta^{\text{D$p$}}$ is the Poincar\'e dual of brane worldvolume.
For example, if a D$p$-brane is extending along $x^0,\dots,x^p$ and
sitting at $x^{p+1}=\cdots=x^9=0$, then
$\delta^{\text{D$p$}}_{9-p}=\pm\delta(x^{p+1})\cdots \delta(x^9)\,
dx^{p+1}\wedge \cdots \wedge dx^{9}$, where the $\pm$ signs correspond
to two possible orientations of the brane.  If the  D-brane has no
endpoint, then $d\delta^{\text{D$p$}}_{9-p}=0$.  When there are multiple
D-branes, the WZ action for the total D-brane system is
\begin{align}
 S_{\rm WZ}^{\text{D}}
 &= \sum_{\text{$p$:\,even/odd}} {1\over (2\pi)^pl_s^{p+1}}\int_{10} e^{-2\pi \ap\cF_2}\, C \, \delta^{\text{D$p$}}_{9-p},\label{ffgq10Aug12}
\end{align}
where $p$ is even (odd) for type IIA (IIB)\@.  We will write such a sum
just as $\sum_p$ henceforth; whether $p$ is even or odd must be clear
from the summand.  For later convenience, let us write
\eqref{ffgq10Aug12} as
\begin{align}
 {2\kappa_{10}^2}\,S_{\rm WZ}^{\text{D}}
 &= \int_{10} e^{-2\pi \ap\cF_2}\, C\, \delta^{\rm D},
 \qquad
 \delta^{\rm D}\equiv \sum_p (2\pi l_s)^{7-p} \delta^{\text{D$p$}}_{9-p}.
\label{isir9Aug12}
\end{align}

Henceforth, we assume that the D-branes do not have endpoints and
therefore
\begin{align}
 d\delta^\text{D$p$}_{9-p}=d\delta^{\rm D}=0.\label{jgpa10Aug12}
\end{align}
Generalization to the case with endpoints is straightforward but we do not
consider it for simplicity of the discussion.

How do the $C$ equations of motion, \eqref{hztl12Apr12} and
\eqref{eiz12Apr10}, change in the presence of the D-brane source
\eqref{isir9Aug12}?  Because \eqref{ffgq10Aug12} involves not only
$C_{p<4}$ but also their duals $C_{p>4}$, which are not independent
degrees of freedom,\footnote{For type IIB, $C_4$ has self-dual field
strength and only a half of its components, up to gauge transformation,
are independent degrees of freedom.} care must be taken when deriving the equations of
motion for $C$.  However, it is known that one can
derive the correct equations of motion by taking the following
``democratic'' action \cite{Fukuma:1999jt, Bergshoeff:2001pv}
\begin{align}
  2\kappa_{10}^2\,
 S^{\text{bulk,RR}}= {1\over 4}\sum_p\int_{10} *G_{p+2}\wedge G_{p+2}
 = {1\over 4}\int_{10} *G\wedge G,
 \label{jznt18Apr12}
\end{align}
where summation is over all $p$; namely, $p=0,2,4,6$ for type IIA and
$p=-1,1,3,5,7$ for type IIB\@. Here, $G_{p+2}$ are defined by
$G_{p+2}=dC_{p+1}-H_3C_{p-1}$. We treat all $C_{p+1}$'s independent when
deriving the equation of motion, and then impose the duality condition
\eqref{mqkq18Apr12} afterward.  Explicitly, the variation of the action
\eqref{jznt18Apr12} with respect to $C_{p+1}$ is
\begin{align}
  2\kappa_{10}^2\,
 \delta S^{\text{bulk,RR}}&= {1\over 2}\sum_p\int_{10} {*G}_{p+2}\wedge (d\delta C_{p+1}-H_3\wedge \delta C_{p-1})\notag\\
 &= {1\over 2}\sum_p (-1)^{p+1}\int_{10} ( d{*G}_{p+2}+H_3\wedge {*G}_{p+4})\wedge  \delta C_{p+1},
\label{ncgm18Apr12}
\end{align}
On the
other hand, the variation of the WZ action \eqref{isir9Aug12} is
\begin{align}
 {2\kappa_{10}^2}\,\delta S_{\rm WZ}^{\text{D}}
 =  \half\sum_p
 \int_{10}e^{-2\pi\ap\cF_2}\,\delta C_{p+1}\, \delta^{\rm D}
 =  \half\sum_p (-1)^{p+1}
 \int_{10}e^{-2\pi\ap\cF_2}\,\delta^{\rm D}\,\delta C_{p+1}.\label{fhtm10Aug12}
\end{align}
Note that, in the democratic formulation, one must divide the
interaction \eqref{isir9Aug12} by two to get the correct result
\cite{Freed:2000tt, Belov:2006xj}.  Namely, we obtained the variation \eqref{fhtm10Aug12} by varying $1/2$ of \eqref{isir9Aug12}.
Also, the $(-1)^{p+1}$ sign in the last expression
came from commuting $\delta C_{p+1}$ and $\delta^{\rm D}$ through.
Combining the two,\footnote{Note that the DBI action for D-branes does
not contribute to the $C$ equation of motion.} the $C$ equation of
motion in the presence of D-brane sources is found to be
\begin{align}
  d{*G}_{p+2}+H_3\wedge {*G}_{p+4}
 +e^{-2\pi\ap\cF_2}\,\delta^{\rm D}=0.\label{lkfo10Aug12}
\end{align}
By using the duality relation \eqref{mqkq18Apr12}, we arrive at
\begin{align}
 dG_{8-p}-H_3\, G_{6-p}=
 (-1)^{\floor{p\over 2}}
 \bigl( e^{-2\pi \ap\cF_2} \delta^{\rm D} \bigr)_{9-p}
 \equiv *j^{\text{D$p$,bs}}_{p+1}.
 \label{hpcm9Aug12}
\end{align}
This equation defines the brane source current $(p+1)$-form
$j^{\text{D$p$,bs}}_{p+1}$, which is clearly gauge invariant because the
left hand side is.  It is also localized at the position of the D-brane.
By summing  over $p$, we
can write \eqref{hpcm9Aug12} more concisely as
\begin{align}
 dG-H_3G
 =e^{+2\pi \ap\cF_2}\deltabD= *j^{\text{D,bs}},
 \label{nkac11Aug12}
\end{align}
where we defined
\begin{align}
 \deltabD
 \equiv 
 \sum_p (-1)^{\floor{p\over 2}}(2\pi l_s)^{7-p} \delta_{9-p}^{\text{D$p$}}
 =
 \cT\delta^{\rm D},\qquad
 j^{\text{D,bs}}\equiv \sum_p j^{\text{D$p$,bs}}_{p+1}.
\end{align}
The sign in front of $\cF_2$ flipped in going from \eqref{hpcm9Aug12} to
\eqref{nkac11Aug12} because of the following manipulations:
\begin{align}
 \sum_p (-1)^{\floor{p\over 2}}(e^{-2\pi \ap \cF_2}\delta^D)_{9-p}
 &=\sum_{p,n} (-1)^{\floor{p\over 2}}{(-2\pi \ap \cF_2)^n\over n!}
 (2\pi l_s)^{7-p-2n}\delta^{\text{D$(p+2n)$}}_{9-p-2n}\notag\\
 &=\sum_{q,n} (-1)^{\floor{q\over 2}-n}{(-2\pi \ap \cF_2)^n\over n!}
 (2\pi l_s)^{7-q}\delta^{\text{D$q$}}_{9-q} \qquad (q=p+2n)\notag\\
 &=\sum_{q,n} (-1)^{\floor{q\over 2}}{(2\pi \ap \cF_2)^n\over n!}
 (2\pi l_s)^{7-q}\delta^{\text{D$q$}}_{9-q}
 =e^{2\pi \ap \cF_2}\deltabD.
\end{align}

If we insert the definition $G=dC-H_3C$ into \eqref{nkac11Aug12}, we obtain
\begin{align}
 d(dC-H_3C)-H_3(dC-H_3 C)
 =d^2C=*j^{\text{D,bs}},
\end{align}
where we used $dH_3=0$.  This is just right, because the violation of
the Bianchi identity $d^2C=0$ is directly related to the existence
of  brane sources coupled to $C$.  So, it is appropriate
to call the quantity $j^{\text{D,bs}}$ brane source current.
Explicitly written, the expression for D$p$-brane source current
$*j^{\text{D$p$,bs}}_{p+1}$ is
\begin{align}
\begin{split}
  *j^{\text{D$0$,bs}}_{1}
 & =
 (2\pi l_s)^{7}
 \left[\delta^{\text{D$0$}}_9
 -{\cF_2\over 2\pi}\delta^{\text{D$2$}}_7
 +{1\over 2}\left({\cF_2\over 2\pi}\right)^2\delta^{\text{D$4$}}_5
 -{1\over 3!}\left({\cF_2\over 2\pi}\right)^3\delta^{\text{D$6$}}_3
 \right],\\
 *j^{\text{D$2$,bs}}_{3}
 & =
 (2\pi l_s)^5\left[-\delta^{\text{D$2$}}_7
 +{\cF_2\over 2\pi}\delta^{\text{D$4$}}_5
 -{1\over 2}\left({\cF_2\over 2\pi}\right)^2\delta^{\text{D$6$}}_3
 \right],\qquad \dots
\end{split}\label{gasb13Aug12}
\end{align}
Namely, we have the standard expression of the combination
$2\pi\ap\cF_2=B_2+2\pi \ap F_2$ inducing lower D-brane charges on
D-brane worldvolume.

\bigskip
However, a problem with the brane source current $j^{\text{D,bs}}$ is
that it is not conserved; by taking exterior derivative of
\eqref{nkac11Aug12}, we obtain
\begin{align}
  d\,{*j}^{\text{D,bs}}=
 d^2G
 +dH_3\,G
 -H_3\, {*j}^{\text{D,bs}}\neq 0.
\end{align}
The first term, which superficially vanishes, does not necessarily
vanish in the presence of singular sources.  The second term vanishes if
there is no NS5-brane.  The last term is non-vanishing in general.

If we can rewrite the $H_3G$ on the left hand side of
\eqref{nkac11Aug12} in terms of a total derivative $d(...)$, one can
define a current that is conserved.  Such conserved current is called
Page current \cite{Page:1984qv, Marolf:2000cb}.  There are two ways to
do that; namely, \eqref{nkac11Aug12} can be rewritten as
\begin{align}
 d(e^{-B_2}G)
 &= e^{-B_2}e^{2\pi \ap\CF_2}\deltabD
 =e^{2\pi \ap F_2}\deltabD ,\label{lsbu3Jul12}\\
 d(G+H_3\wedge C)
 &= e^{2\pi \ap\CF_2}\deltabD
 = e^{B_2+2\pi \ap F_2}\deltabD .\label{lseo3Jul12}
\end{align}
In deriving the  second one we used that $dH_3=0$ in the absence of
NS5 source. Therefore, one may think that there are two possible ways to
define conserved Page current $j^{\text{D,Page}}$, using the two
expressions on the right hand side, because the left hand side is
written as $d(...)$ which apparently vanishes by acting by $d$, implying
that the Page current is conserved, $d{*j}^{\text{D,Page}}=0$.  However,
this is too quick because $C,G$ are singular at the point of the
source and they are not necessarily annihilated by $d^2$.  Actually, it
is clear, from the explicit expression on the right hand side, that
\eqref{lsbu3Jul12} is annihilated by $d$ but \eqref{lseo3Jul12} is not.
Namely,
\begin{align}
 d(e^{2\pi \ap F_2}\deltabD )=0,\qquad
 d(e^{B_2+2\pi \ap F_2}\deltabD )=H_3\, e^{B_2+2\pi \ap F_2}\deltabD \neq 0,
\end{align}
where we used $d\deltabD=0$ because of \eqref{jgpa10Aug12}.  Therefore,
the correct choice is \eqref{lsbu3Jul12}; the conserved Page current
$j^{\text{D,Page}}$ is
\begin{align}
 d(e^{-B_2}G)=e^{2\pi \ap F_2}\deltabD 
 \equiv *j^{\text{D,Page}},\qquad d\,{*j}^{\text{D,Page}}=0.\label{gape12Aug12}
\end{align}
Because the current $j^{\text{D,Page}}$ is conserved, we can define the
conserved Page charge for D$p$-brane contained in a $(9-p)$-volume
$V^{9-p}$ by
\begin{align}
 Q^{\text{D$p$,Page}}
 \equiv {1\over (2\pi l_s)^{7-p}}\int_{V^{9-p}}*j^{\text{D,Page}}
 ={1\over (2\pi l_s)^{7-p}}\int_{S^{8-p}}e^{-B_2}G,\label{gmcy13Aug12}
\end{align}
where $S^{8-p}=\partial V^{9-p}$ is an $(8-p)$-surface.

By comparing \eqref{gape12Aug12} with \eqref{nkac11Aug12}, we see that
the Page current $*j^{\text{D,Page}}$ is obtained by subtracting from
the D-brane source current $*j^{\text{D,bs}}$ the D-brane charges
induced by the spacetime field $B_2$.  Namely, Page current
$*j^{\text{D,Page}}$ is induced only by the D-brane worldvolume field
$F_2$.  The explicit expression for $*j^{\text{D$p$,Page}}_{p+1}$ is
obtained from the
one for $*j^{\text{D$p$,bs}}_{p+1}$,
\eqref{gasb13Aug12}, by replacing $\cF_2$ by $F_2$.


\subsubsection{NS5-brane source}
\label{ss:NS5source}

Up until now, we assumed that there is no NS5-brane source and therefore
$dH_3=0$.  Now let us consider NS5-brane source so that $dH_3\neq 0$.

Note that it appears that the RR $C$ fields on NS5 worldvolume induce
D-brane charges.  This can be motivated by dualities as follows. For
example, $(B_2)_{12}$ on D5(12345) induces D3(345).  The $S$-dual of this
statement is that $(C_2)_{12}$ on NS5(12345) induces D3(345), and a
further $T$-duality along $x^3$ says that $(C_3)_{12}$ on NS5(12345)
induces D2(45). However, the way this works must be not so simple.  In
order that $(C_3)_{12}$ on NS5(12345) induces D2(45), one naively thinks
that the NS5-brane WZ terms contains $\int_6 C_3\wedge C_3$, but
this identically vanishes.  We will discuss below in what sense D-brane
source charges are induced on NS5 worldvolume.

We can derive the WZ coupling for the NS5-brane by following the logic
around \eqref{jwxb18Apr12} where we derived the WZ term for D-branes by
gauge invariance.  This is the strategy taken in
\cite{Bergshoeff:2011zk}.  The NS5-brane is electrically coupled to
$B_6$, but the minimal coupling $\int_6 B_6$ is not gauge invariant (we
defined $B_6$ in \eqref{ffwy18Apr12} for IIA and in \eqref{ffzb18Apr12}
for IIB)\@.  Just like we defined the gauge invariant field
$2\pi\ap\CF_2 \equiv B_2+2\pi\ap F_2$ on D-brane worldvolume, we can define
gauge invariant fields on the NS5 worldvolume by introducing worldvolume
gauge fields  \cite{Bergshoeff:2011zk}.  In type IIA, we define
\begin{align}
 \cG_1 \equiv C_1+dc_0,\qquad
 \cG_3 \equiv C_3+dc_2-H_3c_0,\qquad
 \cG_5 \equiv C_5+dc_4-H_3c_2,
\end{align}
where $c_{0,2,4}$ are worldvolume fields with the gauge transformation
rule
\begin{align}
 \delta c_0=-\lambda_0,\qquad
 \delta c_2=-\lambda_2,\qquad
 \delta c_4=-\lambda_4.
\end{align}
Then it is not difficult to find a combination of fields which
transforms into a total derivative:
\begin{align}
\delta\left[ B_6+{1\over 2}(-\cG_5 C_1+\cG_3 C_3-\cG_1 C_5)\right]
 =d\left[\zeta_5+{1\over 2}(\cG_5 \lambda_0-\cG_3 \lambda_2+\cG_1 \lambda_4)\right].
\end{align}
Therefore, the WZ term for IIA NS5 is
\begin{align}
 2\kappa_{10}^2 S_{\text{WZ}}^{\text{NS5A}}
 =(2\pi l_s)^2
 \int_{6}
 \left[ B_6+{1\over 2}(-\cG_5 C_1+\cG_3 C_3-\cG_1 C_5)\right].
\end{align}
This is pretty much the same as eq.\ (3.22) of \cite{Bergshoeff:2011zk}.
Explicitly written, this is
\begin{align}
 2\kappa_{10}^2 S_{\text{WZ}}^{\text{NS5A}}
 =(2\pi l_s)^2
 \int_{6}
 \left[ B_6+{1\over 2}\Bigl(-(dc_4-H_3c_2) C_1+(dc_2-H_3c_0) C_3-dc_0 C_5\Bigr)\right].\label{neod18Apr12}
\end{align}
Note that quadratic terms in $C$ canceled out; for example, as mentioned
above, there is no term $C_3\wedge C_3$. By adding this WZ to the bulk
action \eqref{jznt18Apr12} and taking variation with respect to
$C_{1,3,5}$, assuming that we can still use the democratic formulation
in the presence of NS5 source, we obtain the equations of motion
\begin{align}
\begin{split}
  dG_8-H_3G_6&=(2\pi l_s)^2 (dc_4-H_3c_2)\,\delta_4^{\text{NS5}},\\
  dG_6-H_3G_4&=(2\pi l_s)^2 (dc_2-H_3c_0)\,\delta_4^{\text{NS5}},\\
  dG_4-H_3G_2&=(2\pi l_s)^2  dc_0        \,\delta_4^{\text{NS5}}.
\end{split}\label{hbuu13Sep12}
\end{align}
Note that, in contrast to the D-brane WZ action \eqref{isir9Aug12}, we
do not divide \eqref{neod18Apr12} by two, since we are not using a
democratic formulation for $B_2$ and $B_6$.  We see from
\eqref{hbuu13Sep12} that the spacetime RR potentials $C$ do not induce
D-brane source charges via the WZ coupling \eqref{neod18Apr12}.

Similarly, in type IIB, we can define
\begin{align}
 \cG_0=C_0,\qquad
 \cG_2=C_2+dc_1,\qquad
 \cG_4=C_4+dc_3-H_3c_1,\qquad
 \cG_6=C_6+dc_5-H_3c_3,
\end{align}
where $c_{1,3,5}$ are worldvolume fields with the gauge transformation
rule
\begin{align}
 \delta c_p=-\lambda_p.
\end{align}
Then we can show
\begin{align}
\delta\left[ B_6+{1\over 2}(\cG_0 C_6-\cG_2 C_4+\cG_4 C_2-\cG_6 C_0)\right]
 =d\left[\zeta_5+{1\over 2}(\cG_0 \lambda_0-\cG_2 \lambda_2+\cG_4 \lambda_2-\cG_6\lambda_0)\right].
\end{align}
Therefore, the WZ term for the IIB NS5 is
\begin{align}
 2\kappa_{10}^2 S_{\text{WZ}}^{\text{NS5B}}
 &= (2\pi l_s)^2
 \int_{6}
 \left[ B_6+{1\over 2}(-\cG_6 C_0+\cG_4 C_2-\cG_2 C_4+\cG_0 C_6)\right]
 \notag\\
 &= (2\pi l_s)^2
 \int_{6}
 \left[ B_6+{1\over 2}\Bigl(-(dc_5-H_3c_3)C_0+(dc_3-H_3c_1) C_2-dc_1 C_4\Bigr)\right].
\end{align}
The equations of motion derived from this are
\begin{align}
\begin{split}
  dG_9-H_3G_7&=(2\pi l_s)^2 (dc_5-H_3c_3)\,\delta_4^{\text{NS5}},\\
  dG_7-H_3G_5&=(2\pi l_s)^2 (dc_3-H_3c_1)\,\delta_4^{\text{NS5}},\\
  dG_5-H_3G_3&=(2\pi l_s)^2 dc_1         \,\delta_4^{\text{NS5}}.
\end{split}
\end{align}
Again, the spacetime RR potentials $C$ do not induce D-brane
source charges via the WZ coupling.

This result can be stated as follows: both in type IIA and IIB, the
equations of motion in the presence of NS5-branes are
\begin{align}
  dG-H_3G&=(2\pi l_s)^2(dc-H_3c)\delta_4^{\text{NS5}},\qquad c=\sum_p c_p.
 \label{ngeh15May12}
\end{align}
If we include D-branes as well, we have the following equation of motion:
\begin{align}
  dG-H_3G&=e^{2\pi \ap \cF_2}\deltabD
 +(2\pi l_s)^2(dc-H_3c)\delta_4^{\text{NS5}}.\label{kiqs18Apr12}
\end{align}
Actually, this equation \eqref{kiqs18Apr12} is not gauge invariant in
this form.  This is because the NS5-brane is the magnetic source of
$B_2$ and therefore the Bianchi identity $dH_3=d^2B_2=0$ is modified to
\begin{align}
 dH_3=*j^{\text{NS5,bs}}_6
 =(2\pi l_s)^2\delta^{\text{NS5}}_4\neq 0,\label{kidh18Apr12}
\end{align}
where $j^{\text{NS5,bs}}_6$ is the NS5-brane source current.  We assume
that there is no induced NS5 current on other branes and thus
$*j^{\text{NS5,bs}}_6=(2\pi l_s)^2\delta^{\text{NS5}}_4$ (for example,
$C_2$ on $7_3$ induces NS5, but we assume that there is no such induced NS5). Eq.\
\eqref{kidh18Apr12} means that the combination
\begin{align}
 G=dC-H_3C\label{kjhb18Apr12}
\end{align}
is no longer gauge invariant under the $C$ gauge transformation
\begin{align}
 \delta C=d\lambda-H_3\lambda.\label{irfw18Apr12}
\end{align}
Explicitly, gauge invariance is violated by
\begin{align}
 \delta G&=d(d\lambda-H_3\lambda)-H_3 (d\lambda-H_3\lambda)
 =-dH_3\,\lambda+H_3\,d\lambda-H_3\, d\lambda\notag\\
 &=-(2\pi l_s)^2\delta_4^{\text{NS5}}\lambda\neq 0.\label{hblo2May12}
\end{align}
where we assumed $d^2\lambda=0$.  Therefore, neither side of the
equation of motion \eqref{kiqs18Apr12} is gauge invariant.

We can rectify this problem by defining an improved field strength
$\hat{G}$ by
\begin{align}
 \hat G&\equiv dC-H_3C-(2\pi l_s)^2\delta_4^{\text{NS5}}c\label{nfsd15May12}
\end{align}
which can be easily shown to be gauge invariant.  This is defined in
spacetime, although it involves $c$ defined only on the NS5 worldvolume,
because of the factor $\delta_4^{\text{NS5}}$.  Note that the left hand
side of \eqref{kiqs18Apr12} can be written as
\begin{align}
 dG-H_3G=d\hat{G}-H_3\hat{G}+(2\pi l_s)^2(dc-H_3c)\delta_4^{\text{NS5}}.
\end{align}
If we substitute this into \eqref{kiqs18Apr12}, the terms involving $c$
cancel between the two sides and we finally obtain the following gauge
invariant equation of motion:
\begin{align}
 d\hat{G}-H_3\hat{G}=e^{2\pi \ap \cF_2}\deltabD.\label{gmfs13Aug12}
\end{align}
The NS5 charge does not appear in this equation. Formally, this equation
has the same form as in the case without NS5-brane source,
\eqref{nkac11Aug12}, if we replace $G\to \hat{G}$.

Just as in the case without NS5-branes, \eqref{gmfs13Aug12} can be rewritten as
\begin{align}
 d(e^{-B_2}\hat{G})=e^{2\pi \ap F_2}\deltabD\equiv *\hat\jmath^{\text{D,Page}}.
\end{align}
The D-brane Page current $\hat\jmath^{\text{D,Page}}$ defined through
this equation is manifestly closed and thus conserved.  Even in the
presence of NS5-branes, the D-brane Page current
$\hat\jmath^{\text{D,Page}}$ is induced only by $F_2$, the D-brane
worldvolume field; namely, neither spacetime fields $B_2,C$ nor the NS
worldvolume field $c$ induces D-brane Page current on NS5.
Just as in \eqref{gmcy13Aug12}, we can define the charge contained in a
$(9-p)$-volume $V^{9-p}$ as
\begin{align}
 \hat{Q}^{\text{D$p$,Page}}
 \equiv {1\over (2\pi l_s)^{7-p}}\int_{V^{9-p}}*\hat\jmath^{\text{D,Page}}
 ={1\over (2\pi l_s)^{7-p}}\int_{S^{8-p}}e^{B_2}\hat{G}.
\end{align}

Note that the RR potentials $C$ do not induce D-brane  charges on
the NS5 in the naive way we expected.  However, actually,
D-brane charges are indeed induced, although rather secretly.  
If we plug the definition of $G$, \eqref{kjhb18Apr12}, into the
left hand side of the equation of motion \eqref{kiqs18Apr12}, we obtain
\begin{align}
 dG-H_3 G=d^2C- (2\pi l_s)^2 C\, \delta^{\text{NS5}}_4.
\end{align}
By equating the last expression with
the right hand side of \eqref{kiqs18Apr12}, we obtain
\begin{align}
 d^2C 
 &= e^{2\pi \ap \cF_2}\deltabD+
 (2\pi l_s)^2\cG \delta_4^{\text{NS5}}
 \equiv *\hat{\jmath}^{\text{D,bs}}.
 \label{nhml15May12}
\end{align}
This is exactly as expected, because the violation of the Bianchi
identity $d^2C=0$ is directly related to the existence of charges that
$C$ couples to.  The first term of \eqref{nhml15May12} is the D-brane
charges induced on the D-brane worldvolume, while the second term is the
D-brane charges induced on the NS5-brane worldvolume.
It is appropriate to call this quantity the D-brane source current,
$*\hat{\jmath}^{\text{D,bs}}$, in the presence of NS5-brane source, as
we already did in \eqref{nhml15May12}. This quantity is gauge invariant
and localized but not conserved.

At the beginning of \ref{ss:NS5source}, we raised a puzzle of how
D-brane charges can be induced on NS5-brane worldvolume by RR $C$
fields, even though for example $C_3\wedge C_3$ identically vanishes.
We can now see how this puzzle is resolved: the induced D-brane charge on the
NS5-brane comes \emph{not} from the NS5 WZ action but secretly from the
bulk supergravity action \eqref{jznt18Apr12}.  From \eqref{ncgm18Apr12},
\begin{align}
 2\kappa_{10}^2\,\delta S^{\text{bulk,RR}}
 &=\half \sum_p (-1)^{\floor{p\over 2}}\int_{10} (dG_{9-p}-H_3\wedge G_{7-p})\wedge \delta C_{p},
\end{align}
where we already used the duality relation \eqref{mqkq18Apr12} at this
point.  If we rewrite $dG_{9-p}-H_3\wedge G_{7-p}$ in terms of $C$ using
\eqref{kjhb18Apr12}, we get
\begin{align}
 2\kappa_{10}^2\,\delta S^{\text{bulk,RR}}
 &=\half \sum_p (-1)^{\floor{p\over 2}}\int_{10} (d^2C_{8-p}-dH_3\wedge C_{6-p})\wedge \delta C_{p}.
\end{align}
The second term in the integrand means that, at the point where
$dH_3\neq 0$, there is induced D-brane charge proportional to $C$.

Let us conclude with remarks on the limitation of the results obtained
in this section.  First, we only took into account D-brane charges
induced by WZ coupling of NS5-branes.  The complete NS5-brane action
includes the DBI part, which involves RR potentials \cite{Eyras:1998hn,
Bandos:2000az} and can induce D-brane charges.  A more complete analysis
requires that we take that also into account.  Secondly, we assumed that
there is no induced NS5-brane charge.  For example, $C_2$ on $7_3$-brane
can induce NS5 (this is the $S$-dual statement of $B_2$ on D7 inducing
D5) and, in more general situations, we should take such induced
NS5-branes into account.  Because an induced NS5/brane can have further induced
charges in it, we expect much richer structure arising in such more
general situations. We should also point out that we have neglected 
various topological issues related to the fact that the B-field is really
a two-gerbe, D-brane charges are properly computed using K-theory, etc.
Finally, we left the F1 charge completely out of
the discussion.  We leave a fuller analysis of brane charges for  very
interesting future research.


\end{document}